%% file: Mbh-Lbt_I.tex
\documentclass[12pt,usenatbib,usegraphicx]{emulateapj}

\usepackage{epsfig}
\usepackage{latexsym,amssymb} 
\usepackage{amsmath}
\usepackage{textcomp}
\usepackage{dcolumn}
\usepackage{booktabs}

\include{Mbh-Lbt_abbrev} 

\lefthead{L\"asker et al.}
\righthead{Bulge luminosities from dedicated NIR data}
\slugcomment{Version \today; accepted by {\it The Astrophysical Journal}.}

\begin{document}
 
\title{Supermassive Black Holes and Their Host Galaxies -- \\ \textrm{I}. Bulge luminosities from dedicated near-infrared data}

\author{Ronald L\"{a}sker$^{1,2}$\footnote{Based on observations obtained with WIRCam, a joint project of CFHT, Taiwan, Korea, Canada, France, at the Canada-France-Hawaii Telescope (CFHT) which is operated by the National Research Council (NRC) of Canada, the Institute National des Sciences de l'Univers of the Centre National de la Recherche Scientifique of France, and the University of Hawaii.}, Laura Ferrarese$^2$, Glenn van de Ven$^1$}
\affil{${}^1$Max-Planck Institut f\"ur Astronomie, K\"onigstuhl 17, D-69117, Heidelberg, Germany; e-mail:laesker@mpia.de}
\affil{${}^2$NRC Herzberg Institute of Astrophysics, 5071 West Saanich Road, Victoria, BC V9E2E7, Canada}

\begin{abstract}
In an effort to secure, refine and supplement the  relation between central Supermassive Black Hole masses, $\mbh$, and the bulge luminosities of their host galaxies, $\lbul$, we obtained deep, high spatial resolution $K$-band images of 35 nearby galaxies with securely measured $\mbh$, using the wide-field WIRCam imager at the Canada-France-Hawaii-Telescope (CFHT). A dedicated data reduction and sky subtraction strategy was adopted to estimate the brightness and structure of the sky, a critical step when tracing the light distribution of extended objects in the near-infrared. From the final image product, bulge and total magnitudes were extracted via two-dimensional profile fitting. As a first order approximation, all galaxies were modeled using a simple \sersic-bulge + exponential-disk decomposition. However, we found that such models did not adequately describe the structure that we observe in a large fraction of our sample galaxies which often include cores, bars, nuclei, inner disks, spiral arms, rings and envelopes. In such cases, we adopted profile modifications and/or more complex models with additional components. The derived bulge magnitudes are very sensitive to the details and number of components used in the models, although total magnitudes remain almost unaffected. Usually, but not always, the luminosities and sizes of the bulges are overestimated when a simple bulge+disk decomposition is adopted in lieu of a more complex model. Furthermore we found that some spheroids are not well fit when the ellipticity of the \sersic\ model is held fixed. This paper presents the details of the image processing and analysis, while in a companion paper we discuss how model-induced biases and systematics in bulge magnitudes impact the $\mbh-\lbul$ relation.
\end{abstract}

\keywords{galaxies: bulges, galaxies: photometry, galaxies: structure, methods: observational, techniques: photometric}

\section{Introduction}
\label{sec:intro}

The correlation between Supermassive Black Hole (\smbh) masses, $\mbh$, and the luminosity of their host galaxies' bulges, $\lbul$, is noteworthy for at least two reasons.
First, it is believed to hold important clues regarding the origin of \smbh s and the evolution of galaxies \citep[e.g., ][]{SilkRees98, Granato_etal04, Hopkins_etal06, Croton_etal06, VolNatGul11b, JahnkeMaccio11}. Second, it allows one to infer \smbh\ masses -- which are notoriously difficult to measure based on resolved kinematics -- from a simple estimate of the bulge luminosity. This, in turn, enables detailed studies of \smbh\ demographics, both in the local and high redshift Universe \citep[e.g.,][]{McD04,Marconi_etal04,Shankar_etal04,Tundo_etal07}. There are, therefore, very good reasons to pursue the calibration of the $\mbh-\lbul$ relation in a precise and unbiased manner. Of particular interest is the relation at near-infrared (NIR) wavelengths, not only because dust obscuration is a lesser concern, compared to optical bands, but also because the NIR luminosity is a better tracer of the underlying stellar mass, due to the fact that the stellar mass-to-light ratio is a weaker function of stellar population (age, metallicity) in the NIR than in the optical \citep[][]{BelldeJong01,Cole_etal01}. If the $\mbh-\lbul$ relation is reflecting a more fundamental relation with bulge mass, its scatter is therefore expected to decrease when moving from optical to NIR bands \citep[][]{MH03,HR04}.

The $\mbh-L_\mathrm{bul,NIR}$ relation was first investigated using \twomass\ $J$, $H$ and $K$ data by \cite{MH03} [hereafter MH03]. Although the details of the photometric analysis were not included in MH03,  the limited depth and resolution of the \twomass\ data, combined with uncertainties in the sky subtraction, is a challenge when when trying to produce a reliable bulge/disk decomposition. 

These limitations were addressed by \cite{V12} [hereafter V12], who used data from the \ukidss\ survey for a sample of 25 galaxies with reliable $\mbh$ measurement. Thanks to the improved depth and spatial resolution of the \ukidss\ data ( $\approx 2\mg$ deeper and with $\approx 3\times$ better resolution than \twomass\ data), \citeauthor{V12} were able to include nuclei, bars and cores in the decompositions, although proper modeling of the sky background remained a concern in their analysis.

Our work represents the next step forward. Exploiting the superior image quality at Mauna Kea, we used the $20' \times 20'$  WIRCam imager at the Canada-France-Hawaii Telescope (CFHT) to obtain deep, wide-field K-band images for 35 galaxies with secure $\mbh$ detections. Our data are approximately $2\mg$ deeper than the \ukidss\ data, and $4\mg$ deeper than \twomass\ data, and represent an improvement of a factor 2-4 in spatial resolution. In addition, they benefit from a dedicated observational strategy and data reduction pipeline designed to produce a reliable map of the spatial and temporal variation of the NIR background. We exploit the superior quality of the data, as well as the flexibility offered by the latest \galfit\ profile fitting code \citep[][]{GF3}, to perform 2D-decompositions that extend well beyond ``standard'' bulge+disk models. We find additional components and profile modifications to be justified and necessary \textit{in almost all galaxies harboring disks}, and demonstrate the large impact on the resulting bulge magnitudes. Model-based total luminosities are supplemented by parameter-free estimates. We use these to derive the, hitherto unpublished, correlation between $\mbh$ and \textit{total} $K$-band luminosity of the host, $\ltot$.

This paper is organized as follows: In \S\ref{sec:data} we present the \smbh\ host galaxy sample and describe in detail the data characteristics, sky-subtraction strategy, and data reduction pipeline. The photometric analysis is described in \S\ref{sec:imageana}, including the decomposition technique, the shortcomings of bulge+disk models, and the improved decompositions. The resulting parameters are given in \S\ref{sec:results}, where we also present a comparison between several derived magnitudes and literature values. We discuss and summarize our findings in \S\ref{sec:discussion} and \S\ref{sec:summary}. The appendix supplies descriptions of the observed galaxy properties and the image modeling process individually for every galaxy, including plots of observed and model profiles as well as residual images.

The \smbh\ scaling relations derived from this data are presented and discussed in a companion paper (L\"asker et al. 2013b, hereafter Paper-II).

\section{NIR Imaging Data}
\label{sec:data}

Our sample of \smbh\ host galaxies, listed in Table \ref{tab:obs}, comprises 35 galaxies with securely measured\footnote{based on dynamical modeling of spatially resolved
stellar or gas kinematics} $\mbh$. We did not observe nine of the galaxies included in the MH03 sample, since their \smbh\ mass estimate is deemed uncertain (M31, M81, NGC1068, NGC4459, NGC4594, NGC4596 and NGC4742) or due to declination constraints (NGC5128). On the other hand, we include in our sample seven galaxies for which a \smbh\ mass measurement became available after the MH03 study was published (IC4296, NGC1300, NGC1399, NGC2748, NGC3227, NGC3998, and PGC49940). A further five galaxies, for which a BH mass was published after our data were obtained, are included in V12, although their sample does not include 21 galaxies that are included in this paper. For details on the sample selection criteria and applied distances, please see Section 2 in Paper-II.

\subsection{Data Acquisition}
\label{subsec:data}

As mentioned in the introduction,  all existing scaling relations between bulge luminosities and \smbh\ masses in the NIR rely on \twomass\ (MH03) or \ukidss\ (V12) data. The shallow depth and limited spatial resolution of the \twomass\ data, as well as difficulties in background subtraction, especially for large galaxies, pose a challenge for reliable decomposition. An incorrect estimate of the background, or failing to properly account for spatial variations in the sky, can lead to large photometric error (random and systematic), especially in studies that rely on an accurate characterization of the extended ``wings" of giant elliptical galaxies. Likewise, limited depth prevents identifying and tracing low surface brightness features, while limited spatial resolution hinders the identification and fitting of small or low-surface brightness morphological components (nuclei, spiral arms, nuclear bars, etc.). These shortcomings were partially addressed in V12, but uncertainty about the reliability of the background subtraction remained.

Our homogeneous imaging data set has sub-arcsecond resolution (median $0\farcs8$ in the final image stack, compared to $2$ to $3\arcsec$ of the \twomass\ data), and reaches a signal-to-noise ratio of S/N=1 at $24\magarcsec$, a factor of 40 ($4\mg$) deeper than \twomass\, and $2\mg$ deeper than the \ukidss\ data used in V12. Moreover, we developed and applied an optimized dithering strategy and data-reduction pipeline that reduces both random and systematic uncertainty in the modeling and removal of the sky background.

All observations were carried out using WIRCam\footnote{\cite{WIRCam}} at the CFHT, in order to benefit from the excellent image quality and reduced NIR sky emissions of the Mauna Kea site. The large ($20\arcmin\times20\arcmin$) FOV of the instrument can accommodate even the largest of our targets, and the smaller galaxies do not require time-consuming off-target nodding otherwise necessary to monitor the sky background. WIRCam's FOV is composed of an array of 4 detectors, each consisting of $2048\times2048$ pixels, with a pixel scale of $0\farcs3$ that comfortably samples the point spread function (PSF). The $45\arcsec$-gaps between the detectors, as well as several bad pixel areas located mainly in the detectors' corners, can be properly sampled by dithering the exposures. WIRCam's read noise ($30\,\mathrm{e}^-$) is small compared to the background flux noise ($\approx 180\,\mathrm{e}^-$ on a typical 20s-exposure at average sky brightness), and the dark current ($\approx0.05\,\mathrm{e}^-/\mathrm{s}$) is negligible compared to the background flux ($\approx1000\,\mathrm{e}^-/\mathrm{s}$). Typical total exposure times range from $\approx500\,\mathrm{s}$ to $1000\,\mathrm{s}$, divided into $\approx24$ to 48 single exposures of $\approx20\,\mathrm{s}$ duration to avoid saturation (the background flux fills half the electron well after $\approx25\,\mathrm{s}$).

For all galaxies, the observing strategy consisted of a sequence of a large followed by a small dithering pattern. Each sequence started with the galaxy centered on one of the four detectors; after one $\approx20\,\mathrm{s}$ exposure, the telescope was slewed by $\approx10\arcmin$ (1/2 of WIRCam's FOV), thus centering the target on an adjacent detector. Once a series of four exposures (each with the galaxy centered on one of the  4 detectors) was completed, the pointing was changed according to a small ($1\farcs5$) dithering pattern, and the large dithering pattern was repeated. The entire  large/small dithering sequence was repeated between 4 and 28 times, depending on the galaxy, until the final total exposure time was reached (see Table 1).

The small dithering pattern was designed to permit correction for detector artifacts and removal of small sources when building the sky frame.  Due to the large-scale dithering, about half of our target galaxies are entirely imaged within a single detector, allowing the remaining 3 detectors to be used for background \textit{structure} determination (see \S\ref{subsec:reduction}). For the remaining galaxies (identified in column (6) of Table \ref{tab:obs}), separate sky exposures were acquired before and after each series of 4 large-dithered exposures by slewing the camera by $\approx 1.5\deg$. For these these galaxies the average background \textit{level} (although not its spatial variations) in a given exposure can still be measured from uncontaminated parts of the on-target image. Only in the case of the four largest galaxies, namely NGC221 (M32), NGC4258 (M106), NGC4347 (M84) and NGC4486 (M87), both the sky level and the spatial variations need to be measured in off-target exposures: for these galaxies sky exposures were therefore obtained after every science frame.

Finally, for every galaxy, shorter exposures (2.5s) were obtained to recover the centers of galaxies that saturate in the long exposures. 

\begin{table*}
 \centering
 \caption{Targets and image quality}
 \newcolumntype{d}[1]{D{#1}{#1}{-1}}
 \begin{tabular}{lcccccr@{ = }lcccc}
  \toprule
  Galaxy & Hubble & our class. & $m-M$ & ref. & $A_K$ & \multicolumn{2}{c}{exposure time} & sky & $\sigma_{b}$ & FWHM & $\beta$ \\
   & type (RC3) & & & & & \multicolumn{2}{c}{} & & $[\magarcsec\,]$ & $[\,\mathrm{arcsec}\,]$ & \\
  (1) & (2) & (3) & (4) & (5) & (6) & \multicolumn{2}{c}{(7)} & (8) & (9) & (10) & (11) \\
  \midrule
  \input{table_obs}
  \bottomrule
 \end{tabular}
\tablecomments{Columns (1) and (2) give the name of our targets, and Hubble type taken from the RC3 catalog. For comparison, column (3) is based on our images and analysis. Apart from giving a basic classifcation into Elliptical (single component), Lenticular (disk present) and Spiral systems, it indicates the presence of a bar. Distance moduli and corresponding references are given in columns (3) and (4) respectively. The code in column (4) is as follows 1: \protect\cite{Tonry_etal01}, 2: redshift distances (NED), 4: \protect\cite{Herrnstein_etal99}, 5: \protect\cite{Mei_etal07}, 7: \protect\cite{Jensen_etal03}, and 8: \protect\cite{Blakeslee_etal09}. Distance moduli are based on Surface Brightness Fluctuation when available. For further details on the distances, see Section 2 in Paper-II. Column (5) lists $K$-band galactic foreground extinctions as taken from NED. Column (6) gives the number of regular (long) exposures, their individual and total exposure times. Not listed here are short exposures that recover saturated galaxy centers, and intermittent off-target (``sky'') exposures to monitor the background. However, the presence of sky exposures is indicated in column (7). See Section \ref{subsec:data} for the observing strategy, dithering pattern and sky subtraction procedure. Column (8) gives the surface brightness uncertainty $\sigma_b$ in the background on the final image stack. Column (9) and (10) respectively give the full-width at half-maximum (FWHM) and asymptotic powerlaw of the point-spread function, as modeled by a Moffat profile (see Section \ref{subsec:metadata}).}
 \label{tab:obs}
\end{table*}

\subsection{Data reduction pipeline and background subtraction}
\label{subsec:reduction}

Data reduction was performed using \iraf\ routines unless otherwise stated. We start by inspecting all detrended frames (as provided by CFHT) and reject a few exposures that show abnormally high background levels or have erroneous pointing. All remaining exposures are bias-subtracted and rescaled to a common zero point, according to the standard-star zero points given in the image headers.

Before the images can be co-added, the background needs to be subtracted. Although it is possible in the 2D modeling analysis described in \S3 to treat and model the sky as a separate component, doing so can lead to significant degeneracies, especially for galaxies with extended low-surface-brightness wings. Sky subtraction for NIR exposures is a challenging task, since the typical sky surface brightness in the $K$-band is $\mu_b\sim13.5~\text{to}~14.5\magarcsec$, $\sim10$ magnitudes brighter than the galaxy surface brightness limit we wish to reach. Additionally, the background is modulated by a spatial pattern (structure) as well as temporal variability which, if not properly modeled, can lead to severe biases and systematics in the final galaxy photometry.

In what follows, we first describe the NIR background characteristics as we observe them in our data. Afterwards, we present our background subtraction strategy.

In WIRCam images, the background can be properly characterized as the sum of two independent components. The first component arises from the sky. It is highly time variable, but exhibits relatively little spatial structure and may therefore be described almost entirely by a its \textit{level}, $\mu_b$. This component varies by $\sim 1\%\times\mu_b$ ($\sim 19\,\magarcsec$) on a 30-second timescale (the length of an individual exposure plus readout/dithering overhead), and $\sim 2.5\%\times\mu_b$ ($\sim 18\magarcsec$) on a 80-second timescale (the typical time between subsequent exposures when slewing to the sky). The sky spatial structure, by contrast, is virtually time-independent: the r.m.s. of the difference between subsequent images, after correction for the time-independent detector signature (as described below) is $\sim23\magarcsec$, i.e. 40 to 100 times fainter than the temporal variation in the sky level.

The second component is characteristic to each detector and is due to emission from the instrument and its housing, illumination effects, deviant pixels and flatfield residuals. This component is virtually time-independent. Its spatial structure is however very pronounced, with an amplitude of $\sim 5\%\times\mu_b$ (corresponding to an r.m.s. deviation of $\sim2\%\times\mu_b$). It is dominated by a smooth large-scale ($\approx 10\arcmin$, i.e. comparable to the detector size) pattern, but also includes bright streaks, smooth doughnut-like rings and sharp-edged patches, all approximately a few arcminutes in size, as well as some regions compromised by dark arcsecond-scale patches. Dark rows on the boundaries between blocks assigned to different amplifiers are also visible.

The background levels of the four detectors differ by a near-constant factor (up to $\sim10\%\times\mu_b\simeq 17\magarcsec$ \textit{after} scaling to a common zero-point as mentioned above), and therefore need to be measured and subtracted for each detector separately. Additionally, the background needs to be accurately removed \textit{before} co-adding individual science frames: due to detector gaps and dithering, pedestals would otherwise remain between different areas of the mosaiced co-added frame.

Background determination and subtraction is performed using two separate iterations of the same procedure. All objects (the target galaxy as well as foreground stars and other contaminants) need to be identified and masked in each exposure. The masks are produced by first running \sex\ \citep[][]{SEx} on each exposure, using a small (32 pixel) background grid that accommodates for small-scale variations in the background. Extended features and objects missed by \sex~ are additionally masked by hand. The mask thus obtained is combined with the bad pixel mask provided by the standard CFHT pipeline. Once all objects and bad pixels are masked, we measure the median sky level (a single number for each detector) on all detectors that are sufficiently unaffected by flux from extended sources, adopting a maximum allowed mask fraction of 10\%. This criterion automatically excludes the detector that contains the target galaxy. Suitable detector frames are normalized by their median (including those obtained on off-target exposures) and then median-combined to obtain a map of the time-independent component of the background (large-scale pattern, rings, streaks, etc, as described above); in the process unidentified hot/dead pixels are detected and added to the bad pixel mask. The resulting detector-specific background map is subtracted from each individual frame (before normalization), producing images that are largely corrected for the time-independent background component. This way, the images become much more suitable for reliable source detection (masking) and measurement of the time-variable sky level.

The procedure is then repeated: a new mask is created, this time by running \sex\ while masking the objects detected in the first pass, and with a wider (128 pixels) background grid, thereby improving the detection of extended sources (e.g., in the ``wings'' of stellar profiles). The median is measured again, detector-by-detector, in combination with the improved masks. This time, the time-variable sky level is estimated also on detectors containing the target galaxy, by extrapolating the median background levels measured on adjacent detectors. The background levels thus determined are subtracted from each frame. 

At this stage, the frames are ready to be co-added. Although an astrometric solution is given in the headers of the detrended images provided by CFHT, we found it to be too imprecise for our purposes. A new astrometric solution is therefore computed and all frames are corrected for field distortion using \scamp\ \citep[][]{Scamp} with a 4th-order polynomial. This ensures minimal degradation of the point-spread function (PSF). The resulting frames are then co-added using \swarp\ \citep[][]{SWarp}. This step is performed separately for the main (long) exposures, the off-target (``sky'') exposures and the short exposures. Short and long exposures are \textit{not} co-added at yet. 

Although the procedure described above produces good results, we make use of the fact that the co-added frame has a significantly higher signal-to-noise ratio than each individual frame used to create the masks. We therefore create a final mask from the deep co-added frame, again using \sex~ (with a 128 pixel, median-filtered background grid) combined with some masking by-hand. This significantly improves the masking of  low-surface brightness features (including but not limited to the target galaxy), which are most problematic for background characterization. The entire procedure as described above (from creation of the detector-specific map of the time-invariant background, to the final co-addition) is repeated once more using these ``deep" masks.

We note that although for the four largest galaxies sky exposures were obtained after \textit{every} on-target exposure (see \S\ref{subsec:data}), in practice the additional time it takes to perform a large slew to the sky (60s, compared to 10s for a regular dithering), is such that the background level changes quickly enough to introduce significant uncertainties when interpolating between bracketing sky exposures. For these galaxies, after extensive experimentation, we found it better to apply the same subtraction method as for all other galaxies, after very conservative masking. Only M32 requires additional manipulations, due to the overlap with the M31 disk, and is discussed separately in Appendix \ref{appA}.

Finally, the images are photometrically calibrated by cross-correlating point sources detected in the co-added images with the \twomass\ catalog. The centers of galaxies that saturate in the long exposures are replaced with data from the short exposure stacks, scaled by the average flux ratio measured on an annulus surrounding the saturated galaxy center. 

\section{2D image analysis}
\label{sec:imageana}

Our measurements of apparent magnitudes are based on two-dimensional (2D) image decomposition performed using \galfit\ \citep[][]{GF3}. Before describing the procedure in detail, we provide a general overview of our approach to the modeling.

We require each galaxy model to contain a bulge component, with radial surface brightness profile described by a \sersic\ law \citep[][]{Sersic63}: 
\begin{equation}
\Sigma(R)=\Sigma_e\exp\{b_n[1-(R/R_e)^{1/n}]\}~, \label{eqn:sersic}
\end{equation}
where $b_n$ is defined such that half of the total flux is enclosed within $R_e$, the effective radius. $R_e$, the \sersic\ index $n$, and the apparent magnitude $m=m(\Sigma_e,R_e,n)$ characterize the radial profile. Wherever it can be identified (see below), a ``disk" component with exponential profile, equivalent to a \sersic\ profile with $n \equiv 1$ and $R_e = b_1 R_s = 1.678 R_s$ replaced by the scale radius $R_s$, is added. Each component is further characterized by center ($x_0,y_0$), axial ratio $q$ and position angle ($PA$). We place no prior constraints on any parameter and, when fitting a bulge and a disk, allow parameters to be mutually independent. Such \sersic\ bulge (+ exponential disk) models have been applied in most previous studies that aim at bulge extraction, and we refer to them as ``standard models".

The apparent bulge magnitude in the standard model, $\mbstd$, is one of the \galfit's output parameters, and can be easily converted into absolute bulge magnitude $\Mbstd$ and $K$-band luminosity\footnote{The absolute $K$-band magnitude of the Sun, $M_{K,\odot}=3.28$ is taken from http://www.ucolick.org/\textasciitilde cnaw/sun.html} $\lbstd$ using the distance moduli and extinction corrections listed in Table 1. The disk component magnitude, $m_d$, is then added to yield the galaxy's total magnitude ($\mtstd$ and $\Mtstd$).

After fitting all images with standard models, and measuring the corresponding bulge and total magnitudes, most (30 out of 35) galaxies showed characteristic residuals in the model-subtracted images. While large residuals are expected for spiral galaxies, bulge(+disk) profile mismatches are observed in \textit{all} galaxies with a disk component, and even in some of the ellipticals. This leads us to perform more detailed and complex fits to account for additional components (usually bars, central point sources and spiral arms), necessary profile modifications, such as diski-/boxiness and truncations, and masking of giant ellipticals' cores (see \S 3.3). We refer to these more complex models as ``improved" throughout the remainder of this paper. We will retain the standard models' results for comparison with previous studies of the $\mbh-\lbul$ scaling relations, and present them along with magnitudes derived from improved models. 

In the following subsections, we detail the steps leading to our bulge magnitude measurements, and elaborate on the most common challenges in obtaining them in an accurate, yet consistent and systematic manner. The \galfit\ results (i.e., the best-fit parameters) are presented in Table \ref{tab:galfits}. 

\subsection{1D-profiles}
\label{subsec:1D-prof}

We extract one-dimensional semi-major axis (SMA) profiles for every galaxy before commencing the two-dimensional fits via \galfit. This is done using the \iraf\ task \textit{ellipse}. Extraction of 1D-profiles, as well as \galfit\ modeling, requires object masks; in both cases we use the final masks derived as part of the background estimation procedure (\S\ref{subsec:reduction}) after un-masking the target galaxy. \textit{Ellipse} produces SMA profiles for surface brightness $\Sigma$, ellipticity $\varepsilon$, position angle $PA$, and the higher-order harmonic amplitude $B_4$ that measures isophotal deviations from perfect ellipsoids (disky and boxy). 

The purpose of the 1D profiles is to help choosing the \galfit\ component configuration and suitable initial parameter values. For instance, while visual inspection of the images is usually enough to reveal the presence of a disk, the profiles confirm (or refute) the visual impression in a quantitative way. Maxima in ellipticity, especially when met in conjunction with maxima in $B_4$ (diskiness), are also a good indicator of (embedded) disks. Throughout the 2D-fitting process, the 1D information is also a tool to judge the quality of a particular fit and to assess subsequent fitting strategy. When fitting the improved models, the 1D-profiles aid in finding configurations that include additional components; for example, bars may be indicated and confirmed by means of extrema in position angle, and a nucleus by an inflection in the otherwise smoothly-curved brightness profile of the bulge.

\subsection{Input metadata to \galfit}
\label{subsec:metadata}

Beyond the science image, metadata supplied to \galfit\ comprise a mask (discussed in \S\ref{subsec:reduction} and \S\ref{subsec:metadata}), the point-spread function (PSF), and the noise image. All are indispensable for realistic modeling and may have a significant impact on the fit result. Therefore, care should be taken in order to construct them reliably.

\begin{figure*}
  \begin{center}
    \includegraphics[width=\linewidth]{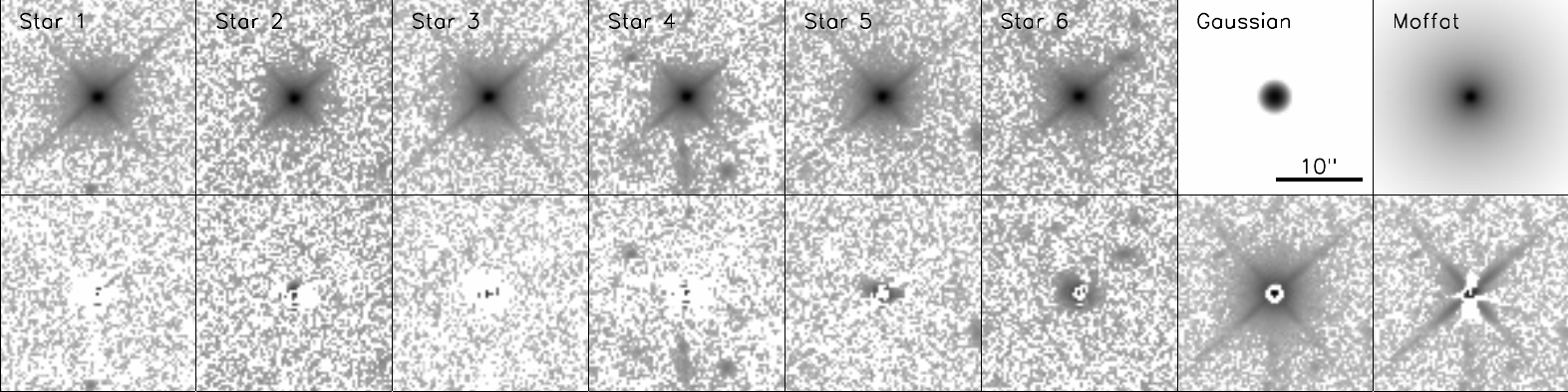}
  \end{center}
  \caption{Demonstration of the quality of the adopted PSF model and its advantage over analytic functions, using NGC1300 as a typical example. The first six columns show the images (top row) of individual stars that are summed to obtain the PSF model (image in Fig. \ref{fig:psf_pros}), and the residual after the PSF model has been subtracted (bottom row). The last two columns show the best-fitting Gaussian and Moffat approximation to the PSF (top) with residuals shown in the corresponding bottom panels. Although the Moffat profile reproduces the radial structure of the PSF reasonably well (see also Fig. \ref{fig:psf_pros}), it cannot account for the azimuthal structure. Each image is 75 pixels ($22\farcs5$) across.}
  \label{fig:psf_imgs}
\end{figure*}

\begin{figure}
  \begin{center}
    \includegraphics[width=\linewidth]{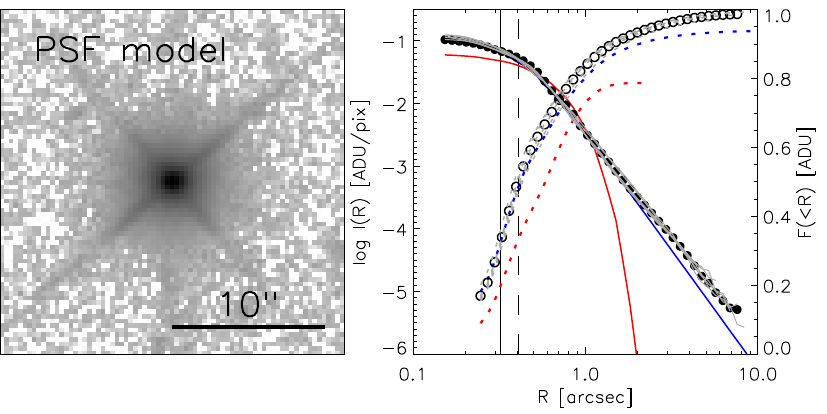}
  \end{center}
  \caption{Left panel: PSF model of NGC1300, the sum of the six star images shown in Fig. \ref{fig:psf_imgs}. Right panel: surface brightness profiles and enclosed flux of the same PSF model (circles), individual stars (grey lines) and analytic profile fits (Gaussian: red, Moffat: blue). The star profiles are difficult to discern, as they are very similar to each other and the PSF model. Vertical solid and dashed lines indicate the half-maximum and effective radius, respectively.}
  \label{fig:psf_pros}
\end{figure}

The PSF image is crucial since \galfit\ needs to convolve each model prior to fitting the images and computing $\chi^2$ and its derivatives. We extract the PSF individually for each stacked image by co-adding several (typically 5-15) cutout images of stars with high signal but without signs of interlopers or saturation. We prefer a PSF model derived this way over an analytic function, because the latter is generally not flexible enough. For example, the WIRCam PSF is neither Gaussian nor elliptical. We illustrate this, and the reliability of PSF model construction, in Figures \ref{fig:psf_imgs} and \ref{fig:psf_pros}. In particular, Figure \ref{fig:psf_pros} (left panel) shows the close agreement of the radial profiles of the PSF model and all stars used to construct it. This indicates that PSF variability across the FOV is low, 
background residuals are negligible, and degradation from centering errors are minimal. In contrast to a Gaussian, a Moffat function, $\Sigma(R)=\Sigma_0(1+(R/R_d)^2)^{-\beta}$, represents the \textit{radial} profile (but not the detailed \textit{shape}) of the PSF reasonably well. Hence, the Moffat function is suitable to quantify the image quality, by fitting it to our PSF model images. For each galaxy, the resulting best-fit full-width at half-maximum, $\mathrm{FWHM}=2 R_d\sqrt{2^{1/\beta}-1}$, and asymptotic powerlaw index ($\beta$) are given in Table \ref{tab:obs}.

The image of the local expected pixel noise (``sigma image'') directly enters the $\chi^2$ computation and optimization. We calculate the sigma image by measuring the background noise after applying complete source masking, modulate it by the local weight (image provided by \swarp, see \S\ref{subsec:reduction}), and add it in quadrature to the local signal noise (ADU $\times$ local gain, see the \galfit\ user's manual).

Aside from these metadata images, \galfit\ requires the fitted image region to be defined. Inclusion of (correctly background subtracted) image regions far outside the optical radius of the target galaxy is essential for accurate modeling and determination of parameters, including the magnitude \citep[see][]{GF3}. Therefore, the fitted image region ought to be as large as possible, ideally $\approx10$ effective radii or more. This is in contrast to the claim by \cite{Sani_etal11}, who advocate fitting only to a central region with high S/N and high $\chi^2$ \textit{per pixel}.

\subsection{Beyond Bulge+Disk models}
\label{subsec:beyond_b+d}

\subsubsection{General considerations}
\label{subsec:impmod_general}

Improved models are necessary to model the majority of our target galaxies, whose structural diversity is evidenced by the characteristic residuals seen when using standard (bulge or bulge+disk) models. The fitting code we employ (\gfthree) allows for considerable complexity, although we try to converge on the simplest models that produce adequate fits to the data. To decide whether to include a disk or additional components as well as profile modifications in our analysis, we take a multi-prong approach, based on the 1D profiles discussed in \S\ref{subsec:1D-prof}, and a visual analysis of the residual image produced by subtracting the best model from the original image. Additionally, we judge the quality of a model by how ``robustly exponential'' the disk is: we replace the exponential by a \sersic\ profile, and observe how far the best-fit \sersic\ index deviates from $n=1$. If it converges to a value outside $\sim[0.7,\,1.3]$ while at the same time the observed profile is clearly exponential, we take this as an indication (albeit not proof) that the model requires additional components or profile modifications. 

We also consider whether the convergence of a model to a given fit solution is well-behaved. This becomes relevant for some of the more complex multi-component models and models featuring a truncation or rotation function. The solution is considered well-behaved if changing the initial parameters by $\sim50\%$ produces fit results consistent to within $\sim0.01\mg$ for the bulge and $\sim1-5\%$ for other magnitude and shape parameters. We discard models that do not meet this criterium in favour of alternative models with less "noisy" convergence, while at the same time taking into account the overall ability of the model to reproduce morphologically significant features seen in the data. Thereby, requiring a well-behaved solution helps us to avoid degenerate models or "overfitting" of the data. For the same reasons, we do not consider the minimum $\chi^2$-value as the only criterion to judge the quality of a model. Instead, in preferring a model or particular fit solution over another, we are also guided by the morphological interpretation and the representation of recognizable galaxy components. However, in practice, the best morphologically motivated models are also those with the lowest $\chi^2$.

Despite the overall diversity, we found that our target galaxies broadly constitute several classes in terms of the characteristic residuals left after fitting the best-fit \sersic\ or Bulge+Disk model. The residual patterns, the galaxy type they are linked to, and our scheme to account for them, are described in the following subsections.

\subsubsection{Elliptical galaxies}
\label{subsec:impmod_ellipticals}

Elliptical galaxies are not always fit well by a two-dimensional \sersic\ profile. Several (IC1459, IC4296, NGC1399, NGC3379, NGC4261, NGC4291, NGC4374, NGC4486, NGC4649) exhibit a \textit{core}, a central light deficit with respect to a \sersic\ model (typically within a few arcseconds, see \citealt{Ferrarese_etal06b}). In the improved models, we correct for this mismatch and the incurred bias in the derived parameters by masking the core (typically, magnitude, effective radius and \sersic\ index are underestimated if cores are not masked). Even more common in ellipticals, including cored ellipticals, are residuals resulting from the assumption, which is implicit in \galfit, that each component must have constant ellipticity and position angle. This is rarely the case. Radial changes in ellipticity (e.g. NGC7052), deviations from ellipses ($B_4$ and gradients thereof, e.g. NGC4261), and isophotal twists (e.g. IC1459) are often seen. We do not account for these structures in our improved models both because they are generally mild and should therefore not strongly affect the derived magnitudes, and also because \galfit\ cannot reproduce radial variations in $\varepsilon$ and $B_4$ using  a single-component model\footnote{two components with coupled parameters, differing only in, e.g., ellipticity, may be joined by mirror-symmetric truncation, but we found that such models converge very slowly and usually still yield unsatisfactory residuals}. In some ellipticals we detect weak small-scale \textit{substructure} that often resembles embedded highly-flattened disks. If we cannot establish a robust model with uniquely interpretable bulge and disk (e.g. in NGC4473, NGC5845), we still fit those galaxies with a single-\sersic\ profile.

\subsubsection{Envelopes and embedded disks in lenticulars}
\label{subsec:impmod_envelopes}

There is a group of early-type galaxies (NGC821, NGC3115, NGC3377, NGC4342, NGC4697) harbouring a thin embedded disk with low flux fraction ($\sim10\%$). The dominant component, which has broadly spheroidal appearance, is not well represented by a single \sersic\ profile, and in order to model the thin disk, a separate \sersic\ component (in addition to the bulge and disk) must be included. Because of the disk's very low axis ratio, these galaxies are probably seen nearly edge-on. The ellipticity typically increases steadily from the center and, after peaking at intermediate radii, levels out to an intermediate value at large radii. In some cases, a second ellipticity peak is also seen, as well as one or more maxima in the diskiness parameter B4.

If the extra \sersic\ component is omitted (i.e if a bulge+disk model is fitted), residuals barely improve over a single-\sersic\ model: the fitted disk component is too flattened ($q\lesssim0.1$) and the disk is still clearly seen in the residuals. Strong residuals typically persist also in the centre and appear to originate in a mismatch of the bulge component. Moreover, bulge \sersic\ indices in the standard model ($n = 5.3$ to 8.7) are higher than generally seen in early-type galaxies of comparable magnitude \citep[e.g.][]{Ferrarese_etal06b}, likewise indicating that they might be biased by component(s) that are not properly accounted for. Finally, the initial disk parameters require fine tuning in order to converge to a shape and size that resembles the thin disk at least approximately. If initialized with too large axis ratio or too small size, the ``disk'' component will readily be fitted to a rounder shape, so that the overall model mostly corrects for mismatches in the centre.

None of these shortcomings change fundamentally if the disk is allowed to stray from exponential ($n=1$), in which case the best fit solution for the disk typically has $n\sim2$. If the \sersic\ ``disk'' component is initialized with parameters that lead it to be fitted with a round shape, overall residuals \textit{do} improve relative to forcing $n=1$, even if they still show the thin embedded disk.

However, if the third component with \sersic\ profile is included, all residuals improve significantly, the flattest component's size corresponds to the observed ellipticity peak at intermediate radii, and it's \sersic\ index decreases to $n\lesssim1.5$. We note that, while in spiral galaxies disks are observed to have closely exponential profiles, there is no reason why this should always be the case. Indeed, when fitting bulge+disk components in bulge-dominated galaxies, the bulge and disk parameters are often degenerate, and {\it assuming} an exponential profile for the disk is often required for convergence.

Our data, thanks to its depth, might provide evidence for the existence of disks (which we define here loosely as highly flattened components, with no assumption as to their kinematics) with non-exponential profiles. Interpretation of the extra component is however ambiguous: it may represent part of a bulge that is insufficiently modeled by a single-\sersic\ profile in two dimensions due to the ellipticity or position angle gradients, it could be a genuine component distinct from the bulge, such as a thick (``hot'') disk, or it could simply be an artifact introduced by the assumption of an exponential disk. This ambiguity is reflected in the ``minimal bulge" and ``spheroid'' magnitudes (see \S\ref{subsec:bulgemag}). We refer to the \sersic\ component with higher central surface brightness as ``bulge", and the second \sersic\  as ``envelope". The latter may have higher or lower axial ratio than the bulge, but is always less flattened than the disk.

\subsubsection{Bars and Nuclei}
\label{subsec:impmod_bars+nuclei}

The most common deviations from the bulge+disk morphology in lenticular and spiral galaxies are \textit{bars} and \textit{nuclei}. We identify and fit the former in NGC1023, NGC1300, NGC2778, NGC2787, NGC3227, NGC3245, NGC3384, NGC3998, NGC4258 and NGC7457, confirming their Hubble classification as given in deVaucouleurs' RC3 catalogue in most cases. We choose to represent bars by a \sersic\ profile with allowed boxiness (\galfit\ input file parameter $C_0>0$). We also attempted to fit bars using a modified Ferrer profile (one more parameter than the \sersic\ profile, see \citealt{GF3}), but found that in general this does not improve the fit. After fitting, we check that the bar component has $n\lesssim1$ before adopting the model. In all cases except NGC2778 ($n_\mathrm{bar}=0.1$), $n_\mathrm{bar}$ converged to a value between 0.3 and 0.8.

Nuclei are modeled as point sources and included for NGC1300, NGC2787, NGC3998, NGC4697 and NGC7457. We then test whether the nucleus is resolved by applying a \sersic\ or King profile in lieu of the PSF. In all cases, the characteristic radius of the component is well below our resolution limit, and the $\chi^2$ as well as other parameters of the model remain virtually unchanged. Therefore, modeling our nuclei as point sources is justified, independent of their physical nature (AGN versus nuclear star cluster). In some galaxies (e.g. NGC821, NGC1023, NGC3245, NGC3377) the putative nucleus is too faint to allow for accurate modeling, and is therefore not included. The nucleus in M87 is masked since the entire core region is masked and it hence does not effect the fit result.

\subsubsection{Spiral arms and profile modifications}
\label{subsec:impmod_modifications}

Another obvious additional component are spiral arms, observed in NGC1300, NGC2748, NGC3227 and NGC4258. They are modeled separately from the disk by a \sersic\ profile (not an exponential), modified by a \textit{rotation function}, as described in \cite{GF3}. An exception is NGC3227, where a significantly better fit could be achieved by applying the rotation to the outer parts of the disk component itself and leaving the (inner) spiral/bar component unmodified (see appendix).

In the case of NGC1300, the spiral arms component is further modified by an inner \textit{truncation} via multiplication with a tanh-function \citep[see][]{GF3}. We also found it necessary to introduce such a truncation for the disk component of NGC2787 and NGC3998 in order to account for the ring, and for NGC2787's bar which does not connect through the galaxy center. We generally do not truncate bulge components in order to maintain consistency with equation (\ref{eqn:sersic}) throughout.

\subsection{Model-independent magnitudes}
\label{subsec:growthcurves}

Considering the intricacies and potential biases involved in 2D-image modeling, we also derive non-parametric total magnitudes, $\miso$, using a curve-of-growth analysis. These are defined as $\miso=m_{ZP}-2.5\log F_{24}$, where $F_{24}=F(<R_{24})$ is the flux inside the radius at which the surface brightness drops below $24\magarcsec$.  We do not extrapolate the flux, in order to maintain independence of image-specific residual background fluctuations and (uncertain) assumptions about the galaxy profile in the outer parts. The surface-brightness limit was chosen to be as low as possible, but well above the background fluctuations we observe in our images.

Curve-of-growth magnitudes are derived within circular apertures, but we use the {\rm ellipse} analysis (\S\ref{subsec:1D-prof}) by replacing masked or saturated pixels with values from the \textit{\iraf.ellipse} model image. Curve-of-growth profiles are shown in appendix A.

\section{Results}
\label{sec:results}

\begin{table*}
 \centering
 \caption{\galfit\ results}
 \begin{tabular}{ll*{3}{r@{ / }l}*{2}{r@{ / }l}*{4}{l}} 
  \toprule
  Galaxy  & Hubble & \multicolumn{6}{c}{Bulge} & \multicolumn{4}{c}{Disk} & \multicolumn{4}{c}{Additional components} \\ 
  \cmidrule(lr){3-8}\cmidrule(lr){9-12}\cmidrule(lr){13-16}
  & type & \multicolumn{2}{c}{$m_b$} & \multicolumn{2}{c}{$R_e~[\,\arcsec\,]$} & \multicolumn{2}{c}{$n$} & \multicolumn{2}{c}{$m_d$} & \multicolumn{2}{c}{$R_s~[\,\arcsec\,]$} & $m_3$ & $m_4$ & $m_5$ & $m_6$ \\ 
  (1) & (2) & \multicolumn{2}{c}{(3)} & \multicolumn{2}{c}{(4)} & \multicolumn{2}{c}{(5)} & \multicolumn{2}{c}{(6)} & \multicolumn{2}{c}{(7)} & (8) & (9) & (10) & (11) \\
  \midrule
  \input{table_galfits}
  \bottomrule
 \end{tabular}
\tablecomments{Shown are the Hubble type according to RC3 in column (2); bulge magnitude, effective radius and \sersic\ index in columns (3-5) and, when fitted, disk magnitude and scale radius in columns (6,7). Each of columns (3-7) can have up to two entries: the first corresponds to the parameters derived for the ``standard" model (either a single-\sersic\ bulge or a \sersic\ bulge plus exponential disk, see \S\ref{sec:imageana}), while the second corresponds to the bulge and disk parameters derived when an elliptical galaxy's core has been masked, or when additional components are included (``improved" model, see \S\ref{subsec:beyond_b+d}). The magnitudes of additional components are listed in columns (8)-(11), with the component type shown in brackets: ``psf" for point source (point-spread function); ``cdisk" for central edge-on disk; ``bar" for bar (\sersic\ with $n\leq1$, boxy isophotes and higher flattening than bulge \textit{and} disk); ``spiral" for spiral arms (\sersic\ modified by rotation function, optionally Fourier and bending modes); ``env" for envelope.}
\label{tab:galfits}
\end{table*} 

Table \ref{tab:galfits} lists the most relevant parameters derived from the \galfit\ 2D-image analysis: the apparent magnitudes of the components (uncorrected for extinction), the effective radius ($R_e$) and \sersic\ index ($n$) of the bulge, as well as the scale radius ($R_s$) of the exponential disk. For each component, there are other fitted parameters, such as center position ($x_0,y_0$), axial ratio $q=1-\varepsilon$ and position angle, which are not listed.

From the component magnitudes, we derive model-based bulge and total magnitudes by summing component fluxes as needed according to their definition (Table \ref{tab:magdef}, Sections \ref{subsec:bulgemag} and \ref{subsec:totmag}). Absolute magnitudes are calculated from apparent magnitudes using the distance moduli and extinction corrections given in Table \ref{tab:obs}. Including the parameter-free estimate of total magnitude ($\Miso$), we thus obtain four estimates for the bulge magnitude, $\Mbul=\{\Mbstd,~\Mbmin,~\Mbmax,~\Msph\}$, and four estimates for the total magnitude, $\Mtot=\{\Mser,~\Mtstd,~\Mtimp,~\Miso\}$. Of these, $\Mbstd$ and $\Mtstd$ are derived using a ``standard" model (bulge plus, if needed, disk), while $\Mbmin$, $\Mbmax$, $\Msph$ and $\Mtimp$ are derived from improved models that include additional components. $\Mser$ is a measure of total luminosity that results from applying a single \sersic\ profile only, regardless of galaxy morphology. We present it merely for completeness and comparison with the more precise $\Mtstd$ and $\Mtimp$. All magnitudes based our WIRCam images are listed in Table \ref{tab:magabs}, along with magnitudes from the literature taken from MH03 (bulge) and the \twomass\ database (total magnitudes). A graphic comparison between standard, improved and literature magnitudes is given in Figures \ref{fig:magcomp_wc} and \ref{fig:magcomp_lit}.

\begin{table*}
 \centering
 \caption{Magnitudes and their definitions}
 \begin{tabular}{lll}
  \toprule
  Luminosity & Short name & Definition \\
  \midrule
  \multicolumn{3}{c}{this work:} \\
  $\Mbstd$ & ``standard bulge'' & bulge component of the standard \sersic-bulge(+exponential disk) model \\
  $\Mbmin$ & ``minimal bulge'' & bulge component of the improved (additional components or masked core) model  \\
  $\Mbmax$ & ``maximal bulge'' & all components except disk and (if present) spiral arms of the improved model \\
  $\Msph$ & ``spheroid'' & bulge component plus envelope (if present) of the improved model \\
  \midrule
  $\Mser$ & ``\sersic'' & magnitude of a single-\sersic\ model (all galaxies) \\
  $\Mtstd$ & ``standard total'' & sum of bulge and disk of the standard model \\
  $\Mtimp$ & ``improved total'' & sum of all components of the improved model \\
  $\Miso$ & ``isophotal'' & flux within aperture delimited by the $24\magarcsec$ isophote \\
  \midrule
  \multicolumn{3}{c}{previous studies:} \\
  $\Mmh$ & -- & bulge magnitude from \protect\cite{MH03}; data: \twomass, models: bulge+disk \\
  $\Mv$ & -- & bulge magnitude from \protect\cite{V12}; data: \ukidss, models: include bars, nuclei and cores \\
  $\Mtwom$ & -- & total magnitude from the \twomass\ database ($K_\mathrm{tot}$, includes extrapolated flux) \\
  \bottomrule
 \end{tabular}
 \tablecomments{Summary of the magnitudes used in this paper. Magnitudes derived in this work are listed first (eight lines) and based on our CFHT/WIRCam $K$-band imaging. For details on standard and improved image models, see Sections \ref{sec:imageana} and \ref{subsec:beyond_b+d}. For details on the magnitude definitions, see Sections \ref{subsec:bulgemag} and \ref{subsec:totmag}. The three bottom lines list magnitudes we take from the literature for comparison, all corrected for our adopted distances.}
\label{tab:magdef}
\end{table*} 

\begin{table*} 
 \caption{Resulting absolute magnitudes}
 \centering
 \begin{tabular}{l*{13}c}
 \toprule
   Galaxy & core & disk & imp & $\Mtwom$ & $\Miso$ & $\Mser$ & $\Mtstd$ & $\Mtimp$ & $\Mbstd$ & $\Mbmin$ & $\Mbmax$ & $\Msph$ & $\Mmh$ \\
  \midrule
  \input{table_magabs}
  \bottomrule
\end{tabular}
\tablecomments{All values have been derived from our WIRCam data, except for the total magnitude $\Mtwom$, which is taken from the \twomass\ database, and $\Mmh$, which is the bulge magnitude from MH03. Both are corrected for our distances and listed for comparison. Aside from $\Miso$ (the isophotal magnitude derived from the curve of growth analysis, see \S\ref{subsec:growthcurves}), all magnitudes are based on \galfit\ models (see Section \ref{sec:imageana}). \galfit-based magnitudes are derived from single-\sersic\ models ($\Mser$), bulge+exponential disk models ($\Mtstd$, $\Mbstd$), or improved models ($\Mtimp$, $\Mbmin$, $\Mbmax$, $\Msph$). Columns ``core'' and ``disk'' indicate whether a core or a disk was detected, and ``imp'' whether an improved model (masked core or additional components) was used to fit the data.}
\label{tab:magabs}
\end{table*}

\subsection{Bulge magnitudes}
\label{subsec:bulgemag}

In addition to the standard bulge magnitude ($\Mbstd$) from simple bulge+disk models, we derive three distinct bulge magnitudes for each galaxy with an improved model that includes, as needed, one or several additional components (nucleus, bar, spiral arms, inner (secondary) disk, envelope) or profile modifications (see Section \ref{subsec:beyond_b+d}). This effects 17 out of 35 galaxies, which are all those galaxies with a disk except NGC4564. The ``minimal'' bulge magnitude, $\Mbmin$, is the magnitude of the improved model's bulge component alone. The ``maximal'' bulge magnitude, $\Mbmax$, results from summing the flux of \textit{all} components \textit{except} the disk and, if present, the spiral arms. Therefore, $\Mbmax$ represents an upper limit for the bulge magnitude in all cases. $\Mbmin$ generally reflects the conventional definition of the bulge and represents its best approximation in most cases. The exceptions are the ``edge-on lenticulars" discussed in \S\ref{subsec:impmod_envelopes}. Here, the ``envelope" needed to fit the profiles might represent, all or in part, a distinct morphological component from the bulge, in which case $\Mbmin$ is gives rough lower limit to the bulge magnitude. It might, on the other hand, result from the inability of a single \sersic\ component to fit the spheroid. For this reason we introduce a third definition of bulge magnitude, $\Msph$, which includes the flux of both ''bulge'' and ``envelope'' components. For galaxies without an envelope, $\Msph=\Mbmin$. 

\begin{figure*}
  \begin{center}
    \includegraphics[width=\linewidth]{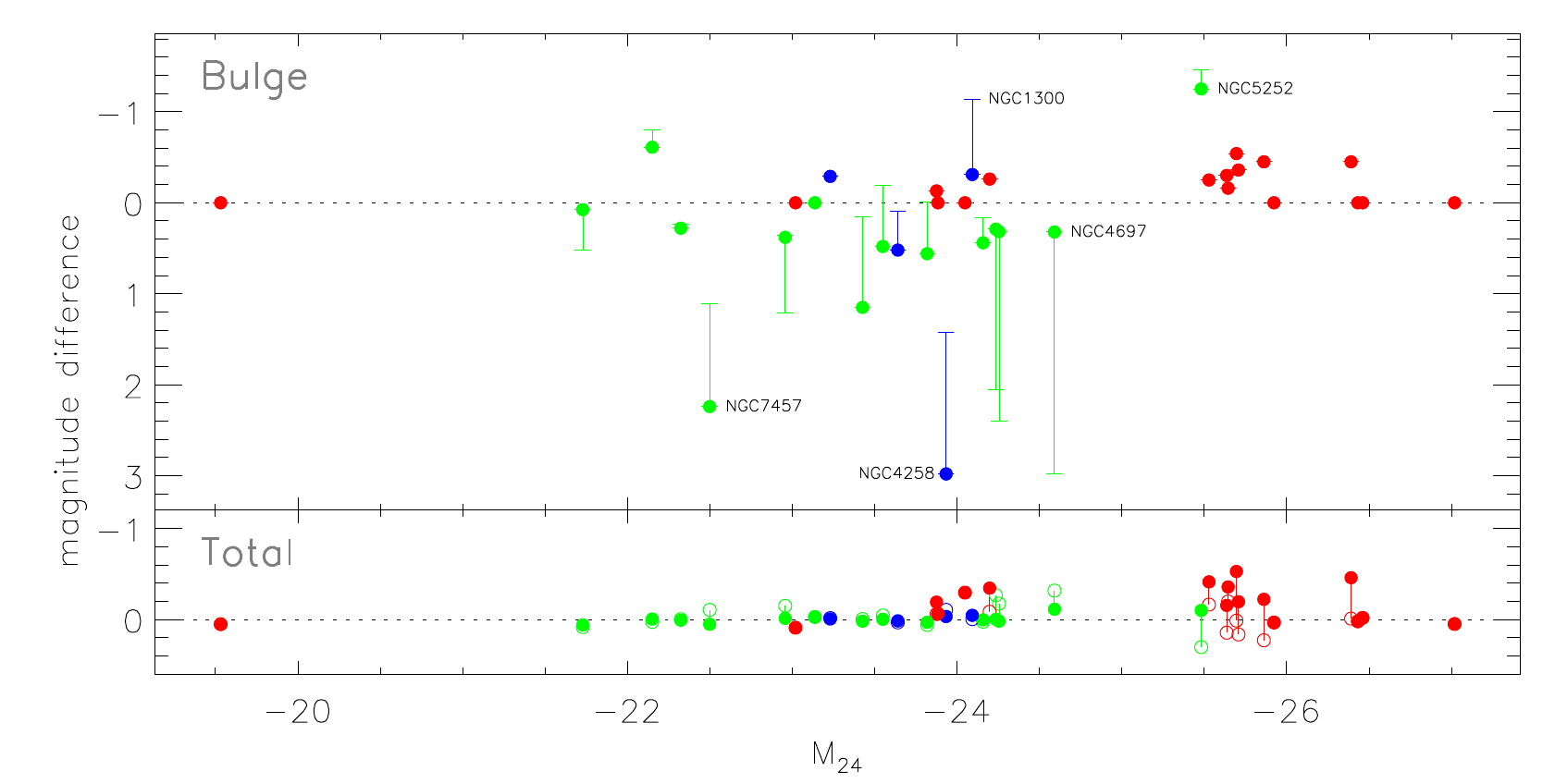}
  \end{center}
  \caption{Comparison of different types of bulge (top panel) and total (bottom panel) magnitudes, all derived from our WIRCam data. Colors indicate the galaxy type: elliptical (red), lenticular (green) and spiral (blue), as classified by us and listed in Table \ref{tab:obs}. Magnitude differences (y-axis) are plotted against total magnitude, $\Miso$ (\S\ref{subsec:growthcurves}). Top panel: differences between ``standard'' and ``improved'' bulge magnitudes. Filled circles: $\Msph-\Mbstd$; lower and upper bar limit: $\Mbmin-\Mbstd$ and $\Mbmax-\Mbstd$, respectively. Bottom panel: difference $\Mtstd-\Miso$ (open circles) and $\Mtimp-\Miso$ (filled circles) between \galfit-based and isophotal total magnitudes. Total magnitudes are relatively invariant with respect to the applied photometric method, while standard and improved bulge magnitudes differ by up to a few magnitudes. For magnitude definitions and discussion of labeled galaxies, see Table \ref{tab:magdef} and Section \ref{subsec:bulgemag}. Linear relation fits to the plotted values are given in Table \ref{tab:magcomp}.}
  \label{fig:magcomp_wc}
\end{figure*}

In Figure \ref{fig:magcomp_wc} (upper panel), we show the differences between improved and standard bulge magnitudes: $\Mbmin-\Mbstd$ and $\Mbmax-\Mbstd$ (lower and upper ''limits''), as well as $\Msph-\Mbstd$ (filled circles). All three are plotted jointly against total magnitude ($\Miso$) to allow for their direct visual comparison. $\Mbmin$ and $\Mbmax$ sometimes differ considerably from one another, reflecting the flux included in components other than bulge and disk (and spiral arms, if present). For galaxies without multi-component models (ellipticals and NGC4564), $\Mbmin$, $\Mbmax$, and consequently $\Msph$ are equal by definition. For most galaxies, improved bulge magnitudes differ significantly from $\Mbstd$ (represented by the horizontal dotted line), by $\sim0.5$ to $\sim1\mg$ on average (see Table \ref{tab:magcomp}). In case of ellipticals with a core, $\Mbstd$ always underestimates the magnitude, as anticipated (see Section \ref{subsec:impmod_ellipticals}). For multi-component models, however, the sign of individual differences with $\Mbstd$ is not always what one might naively expect. In a few cases the bulge magnitude derived in the standard model, $\Mbstd$ is \textit{fainter} than the conventional bulge magnitude in the improved model ($\Mbmin$), despite the the fact that improved model includes additional components that might be expected to ``absorb" some of the bulge flux. Conversely, in numerous cases (NGC821, NGC1023, NGC2778, NGC3115, NGC3998, NGC4342, NGC4564, NGC4697, NGC7457), $\Mbstd$ is {\it brighter} than $\Mbmax$, the total minus the disk luminosity of the improved model. This confirms that ignoring additional components (bars, spiral arms, nuclei, etc..), and forcing a single-\sersic\ or \sersic+exponential disk model to the data, can lead to severe (and unpredictable) biases in the derived bulge magnitude. We have labeled some of these cases in Fig.\ref{fig:magcomp_wc} and discuss them in more detail below. 

NGC5252 stands out as the galaxy with the largest negative difference $\Mbmin-\Mbstd$, i.e. the galaxy for which the bulge magnitude is most underestimated when using a standard bulge+disk model. The galaxy has a bright nucleus in addition to a bulge and a disk component (see Appendix A). Neglecting the nucleus (i.e. fitting a standard model to the profile) led to a best-fit \sersic\ model for the bulge with unreasonably high $n$, and therefore we decided to fix the parameter (somewhat arbitrarily) to $n = 4$ (i.e. a deVaucouleurs profile). When including the nucleus in the improved model, the bulge \sersic\ index converges to $n=5.0$, close to the (fixed) value in the standard model. In the improved model, however, the bulge is brighter. We attribute this to a bias towards a small effective radius (an opposite bias would occur if the \sersic\ index was not fixed) in the standard model, due to the luminous inner region. As can be seen in appendix \ref{appA}, the improved model, which includes the nucleus, produces vastly reduced residuals. Another galaxy for which  the bulge flux is \textit{under}estimated by the standard bulge+disk model is NGC1300, for which as many as four additional components might be present. Here though, the bias can be traced back to an unrealistic model of the exponential disk, which appears to fit the light of the large-scale bar. Further contribution to biased bulge parameters may come from the small, but bright, inner disk. For this galaxy, the \sersic\ index \textit{increases} considerably when components other than bulge+disk are included in the model.

NGC7457 is an example of the more common situation in which the bulge flux in the standard model is overestimated, and is in fact larger than the upper limit on the bulge magnitude in  the improved model. The galaxy has a bulge, a large scale disk, a bar and a nucleus. Ignoring the latter two components, and the bar in particular, leads to underestimate the disk's flux as well as its scale radius, and to overestimate the bulge $R_e$ (see data and model profile of NGC7457 in appendix \ref{appA}). Effectively, in the standard model, the bulge over-extends to fit the large scale profile, which is in reality dominated by the disk, thus causing its magnitude to be overestimated. A similar situation occurs in NGC4258, where the (main) disk, although displaying a partial truncation at intermediate radii, has a large scale radius. That this extension is not part of  the bulge may be inferred by the unrealistically high bulge \sersic\ index ($\approx 8$) derived in the standard model, but also by the low axial ratio at large radii. The overestimate of the bulge magnitude in the standard model is further compounded by the presence of a bright, small and highly inclined inner disk which, when not included in the model, biases the bulge profile to a high \sersic\ index and effective radius.

A different situation occurs in edge-on lenticulars like NGC4697. In the standard model, the best-fit disk (which is forced to have an exponential profile) is too flattened and its flux is therefore underestimated: as can be seen in Appendix A, the best fit standard model does not provide a good match to the data. As discussed  in   \$\ref{subsec:beyond_b+d}, an additional component (``envelope") needs to be included to provide a good fit: without, $\Mbstd$ is biased too high. In the particular case of NGC4697 (and NGC821), the inner \sersic\ component is flatter that the outer. Yet it is probably not an inner disc, as its \sersic\ index is $>2$, and its axial ratio significantly greater than that of the exponential disc component.

The cases mentioned above are only examples; similar situations are encountered in the majority of galaxies harboring a disk. For all of these, the inadequacy of a standard model is evident from a simple inspection of the residuals from the fits and comparison of the projected 1D model to the semi-major axis profiles of surface brightness, ellipticity, position angle and diskiness. However, the examples serve to illustrate the danger of applying a blind 2D decomposition to nearby galaxies: we found that inspection of the data and models, as well as  careful supervision while running the code, were needed to provide not only a good fit, but also a realistic physical description of our targets.

\begin{table}
 \centering
 \caption{Magnitude differences}
 \begin{tabular}{cr@{ $-$ }l*{4}{c}}
  \toprule
  {} & \multicolumn{2}{c}{$y=\Delta M$} & $\langle x=\Miso \rangle$ & $a=\langle y \rangle$ & $b$ & r.m.s. \\
  \midrule
  \input{table_magcomp}
  \bottomrule
 \end{tabular}
 \tablecomments{Trends of absolute magnitude differences $\Delta M$ as plotted in Figures \ref{fig:magcomp_wc} and \ref{fig:magcomp_lit}. Fitted are relations of the form $y=a+b(x-\langle x \rangle)$ with r.m.s. scatter in variable $y=\Delta M$. In all fits, $x=\Miso$, $\langle x \rangle $ its average, and therefore $a=\langle \Delta M \rangle$ is the average difference between two magnitude definitions (see Table \ref{tab:magdef}). Notable is the large scatter for bulge magnitudes (lines 1-3), as well as the offset between $\Msph$ and $\Mv$. The trends with total magnitude (slopes $b$) are relatively mild, $\sim0.1$.}
 \label{tab:magcomp}
\end{table} 

\subsection{Comparison of bulge magnitudes with literature values}
\label{subsec:magcomp}

\begin{figure*}
  \begin{center}
    \includegraphics[width=\linewidth]{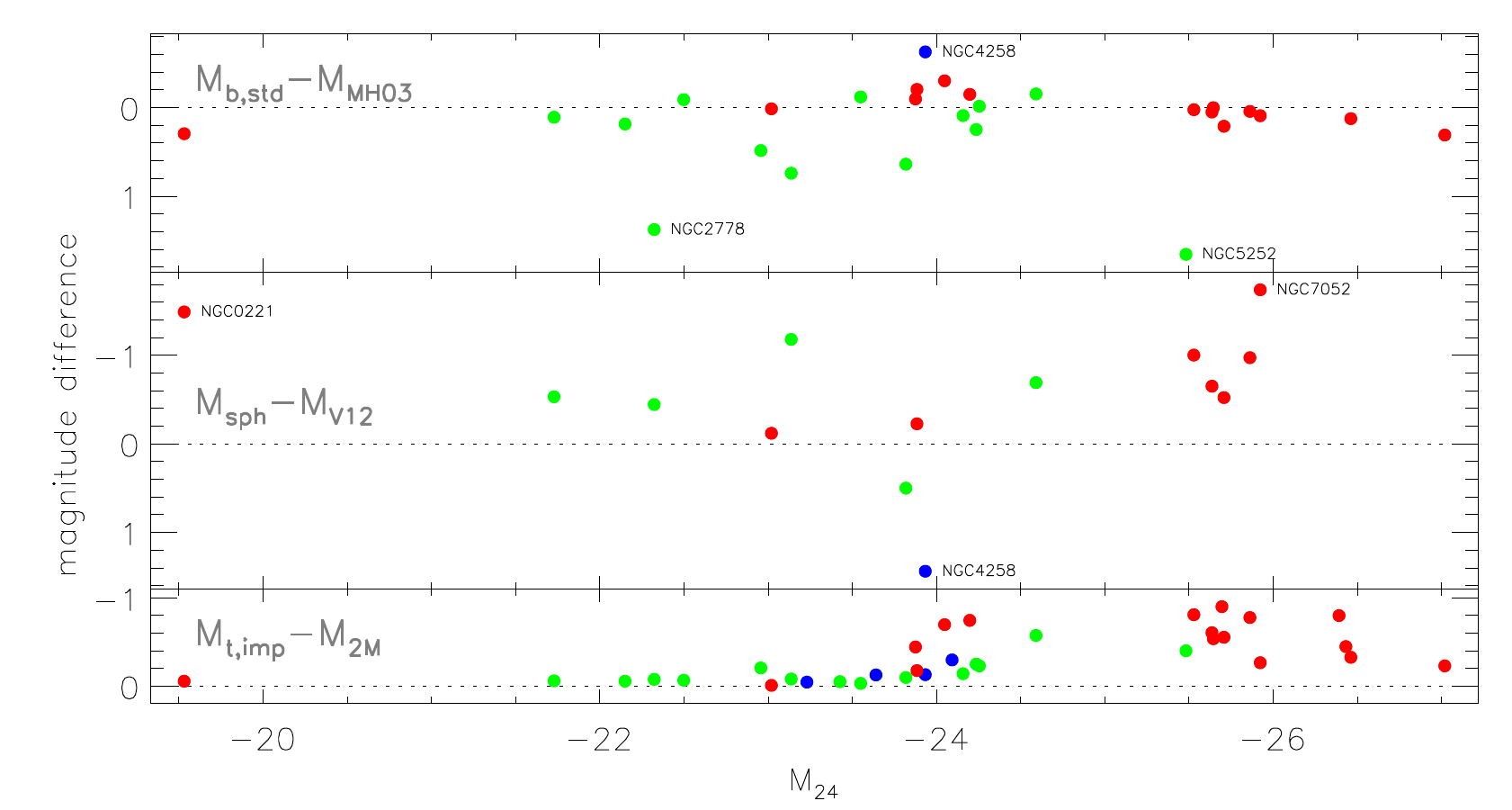}
  \end{center}
 \caption{Comparison of absolute $K$-band magnitudes derived from our WIRCam images with values available from the literature, rescaled to our adopted distances. Plotted are magnitude differences against total magnitude, $\Miso$. Colors are the same as in Figure \ref{fig:magcomp_wc}. Top panel: $\Mbstd-\Mmh$ difference between our ``standard'' bulge magnitudes and those derived in MH03. Both use the same decomposition technique (\galfit and bulge+disk models), but differences are often significant. Middle panel: $\Msph-\Mv$ difference between our ``improved'' spheroid magnitudes and those derived by V12. Although V12 account for nuclei, bars and and cores, the scatter in this plot is large, even when excluding the labeled galaxies (see \S\ref{subsec:magcomp}). Bottom panel: $\Mtimp-\Mtwom$ difference between our ''improved'' and \twomass\ total $K$-band magnitude. Compared to bulge magnitude differences, the scatter is reduced, but \twomass\ values appear to underestimate the flux systematically, particularly for elliptical galaxies. For magnitude definitions and further details, see Table \ref{tab:magdef} and Sections \ref{subsec:magcomp}. Linear relation fits to the plotted values are given in Table \ref{tab:magcomp}.}
 \label{fig:magcomp_lit}
 \vspace{20pt}
\end{figure*}

The upper panel of Figure \ref{fig:magcomp_lit} shows a comparison between our standard model bulge magnitudes, $\Mbstd$, and the bulge magnitudes given in \cite{MH03}. The latter were derived from \twomass\ data using the same analysis software and model configuration (bulge+disk) we used to derive $\Mbstd$. The magnitude differences scatter considerably (r.m.s. of $0.5\mg$ around the average, see Table \ref{tab:magcomp}), with no obvious trend with magnitude or galaxy type (i.e. whether a disk was included or not). Our standard bulges are on average $0.17\mg$ fainter than MH03's. As mentioned in the introduction, this is not unexpected given the the limited depth of the \twomass\ data and uncertainties in the background subtraction.

Cases where $\Mmh$ is significantly brighter than $\Mbstd$ from our WIRCam data are NGC5252, a lenticular galaxy with a bright nucleus, and NGC2778, a lenticular with a weak bar. The causes for these differences are not clear, but it is possible that NGC5252's very bright nucleus has biased the MH03 model to an unrealistically high \sersic\ index, and that the NGC2778 disk might not have been recognized in the shallower \twomass\ data. We note that the MH03 bulge magnitudes for NGC2778 is $\sim0.7\mg$ brighter than the {\it total} magnitude listed in the \twomass\ database, and $\sim0.6\mg$ brighter than the {\it total} magnitude we estimate for this galaxy\footnote{Our data and analysis, on the other hand, leads to good agreement between  $\Mtstd$, $\Mtimp$ and $\Miso$}. However, applying a single-\sersic\ model to the WIRCam data (i.e. omitting the disk) leads to a bulge (total) magnitude in reasonable agreement with MH03 (within $0.15\mg$), giving credibility to our explanation that the disk component was neglected in the MH03 analysis.

The negative outlier in $\Mbstd-\Mmh$ is NGC4258, a nucleated spiral galaxy. The galaxy hosts an extended, low surface brightness disk that, if unaccounted for, might have biased the sky estimate in the MH03 analysis, leading to oversubtraction. Notable is also CygA, which suffers from heavy stellar foreground contamination which, if insufficiently masked, could lead to overestimate the galaxy flux.

In the middle panel of Figure \ref{fig:magcomp_lit}, we compare the spheroid magnitudes ($\Msph$, bulge plus, if present, envelope flux, see \S\ref{subsec:bulgemag}) from our improved models, to the bulge magnitudes derived by V12, who also accounted for nuclei, bars and cores in their modeling. The comparison shows a significant bias -- with our spheroid flux being, on average, $\sim 0.5\mg$ brighter than reported by V12 -- as well as considerable scatter (see Table \ref{tab:magcomp}). This is only partly due to differences in the models adopted for specific galaxies:  V12 decompose NGC221 and NGC7052 into bulge+disk, while we do not find compelling evidence for a disk in either galaxy. Conversely, V12 do not account for the central disk of NGC4258, which probably leads them to overestimate the bulge flux, as discussed in \S\ref{subsec:bulgemag}. For most other galaxies, ellipticals in particular, V12 adopt the same components  we used in our improved models. Yet, even after excluding the outliers mentioned above, the substantial offset and scatter between our and V12's magnitudes persists. The small sample overlap (14 galaxies) precludes to establish a firm reason for the disagreement; however likely culprits are differences in the estimation of the background, uncertainty in bulge extraction resulting from differences in data quality (resolution, depth), as well as degeneracies between parameters when fitting multiple components.

\subsection{Total magnitudes}
\label{subsec:totmag}

Our analysis leads to three separate estimates of total magnitudes: $\Mtstd$ from the standard model (\S\ref{sec:imageana}), $\Mtimp$ from the improved model (\S\ref{subsec:beyond_b+d}) and $\Miso$ from a non-parametric curve-of-growth analysis (\S\ref{subsec:growthcurves}). These are compared in the bottom panel of Figure \ref{fig:magcomp_wc}. Differences between $\Mtstd$, $\Mtimp$ and $\Miso$ are small, relative to the variance between bulge magnitudes (\S\ref{subsec:bulgemag} and \S\ref{subsec:magcomp}). In galaxies with disks,  the total luminosity typically, but not always, decreases slightly when improved models are used. The most notable exception is NGC5252, whose bulge \sersic\ index needed to be fixed to $n=4$ in the standard model (see \S\ref{subsec:bulgemag} and the appendix). Magnitudes for elliptical galaxies (red symbols) are underestimated, by a few tenths of a magnitude, when cores are not masked while fitting a \sersic\ profile ($\Mtstd$, open circles) compared to $\Mtimp$ (filled circles). Finally, $\Mtimp$ is slightly brighter, on average, than $\Miso$. This is to be expected given that, while the former is extrapolated to infinity, the latter reflects the flux within the isophote at which the surface brightness equals $24\magarcsec$.

Comparison of $\Mtimp$ with \twomass\ total magnitudes (see Figure \ref{fig:magcomp_lit}), which are also derived by extrapolating the profile, shows small scatter ($0.2\mg$) relative to bulge magnitude comparisons (top and middle panel in the same figure). However, \twomass-based magnitudes appear to be systematically biased: they are fainter ($0.34\mg$ on average), and increasingly so for the brightest and most extended (giant elliptical) galaxies. This is likely due to background oversubtraction in the \twomass\ images.

\subsection{Magnitude uncertainties}
\label{sec:magerr}
                                                                                                                                                                                                                                                                      
The formal uncertainties of all our magnitudes measurements are very low -- typically below $0.001$ mag. In the case of magnitudes derived from \galfit, these uncertainties reflect the local change in $\chi^2$ corresponding to the $1\sigma$-confidence interval. The actual magnitude uncertainties are of course significantly larger \citep[see also the discussion in][]{GF3} and originate mainly in uncertainties in the appropriateness of the functional form of the profiles adopted for the various components, in the number of components used, and in the ambiguity in the interpretation of the components for some galaxies (see \S\ref{subsec:beyond_b+d}). Total magnitudes are more robust: they show little dependency on the details of the modeling, and agree closely with the non-parametric curve-of-growth values. Additional, but less dominant, sources of uncertainty are in the background determination/subtraction, in the PSF model, the noise map, and the masks.

Quantifying such systematic uncertainty is very difficult. However, an educated guess can be gathered from the difference between the values derived using the standard and improved models and, in the latter case, between minimal and maximal values (see \S\ref{subsec:bulgemag}). The effect of magnitude uncertainties in the parametrization of the $\mbh-\lbul$ relation will be discussed in Paper-II. Here, we emphasize that the magnitudes we derive hold only under the condition that the \textit{adopted} model (profile, number and types of components, and metadata) represent a valid physical description of the data. 

In the case of isophotal magnitudes, we recognize that they necessarily represent lower limits, as they only include the flux within the $\mu=24\magarcsec$ isophote. The uncertainty here originates in the unknown fraction of omitted flux, which in turn depends on the outer profile of a given galaxy. A general rough estimate, using the fundamental plane of Elliptical galaxies and assuming de Vaucouleurs' profile, indicates that the ``missing flux" should be $\lesssim0.1\mg$ even for giant elliptical galaxies ($M_K=-24\mg$). This agrees with the small systematic difference we observe between $\Miso$ and $\Mtimp$ (\S\ref{subsec:totmag} and Table \ref{tab:magcomp}).

\section{Discussion}
\label{sec:discussion}

\subsection{Impact of additional components}
\label{subsec:impact}

Most of our non-elliptical target galaxies deviate significantly from the canonical \sersic-bulge plus exponential-disk morphology. We emphasize that these deviations are not merely seen in the residual images once the best fit bulge+disk model is subtracted from the data, but are often noticeable in the original images and are reflected in the complexity of the one-dimensional surface brightness profiles. It is worth noticing that dust obscuration is not a significant source of contamination in our analysis: except in case of the spiral galaxy NGC4258, where a small region near the galaxy center is partially obscured, dust lanes/patches are not visible in any of our galaxies.

While the morphological complexity of the galaxies is not surprising, as significant substructure is commonly seen in early-type galaxies with intermediate luminosity \citep[see, for instance,][]{Ferrarese_etal06b}, what is noteworthy is the fact that ignoring such components, and adopting an overly simplified 2-D model, can significantly bias the derived bulge magnitudes. Depth and spatial resolution are critical to accurately discern and model stellar nuclei, small nuclear disks, bars and the outer, low-surface brightness regions of galaxies. It follows that bulge parameters derived from data that do not permit an accurate characterization of the morphological diversity in nearby galaxies should be treated with caution. Case-by-case descriptions of the structure seen in our targets can be found in the appendix; here we summarize the most commonly observed consequences of imposing \sersic\ bulge (+ exponential disk) models on galaxies that deviate from such a simple morphology.

\textit{Cores in ellipticals}: When a core, i.e. a depletion of light relative to the inner extrapolation of the  \sersic\ law that best fits the galaxy profile beyond a few arcseconds, is present and not accounted for (e.g. by masking), the resulting parameters of the \sersic\ profile are biased such that luminosity, effective radius and \sersic\ index $n$ are all underestimated, sometimes drastically (up to a factor of 2-3, e.g. IC4296, NGC1399, M87, M60). The change in $n$ is not surprising, as a model with lower \sersic\ index $n$ features a shallower profile at small radii compared to a high-$n$ model. Underestimating $n$ causes the total flux to decrease since a profile with low $n$ is steeper in the outer parts than a profile with high $n$. For example, for a  \sersic\ model with $n=10$, $\sim 8\%$ of the total flux is at radii $r>20 R_e$, whereas for $n=4$ (de Vaucouleurs) this fraction is less than $1\%$.

\textit{Nuclear (point-)sources}: Here the same mechanisms as for cores is at play, albeit with opposite effect: bright nuclei, if neglected, lead to \textit{overestimate} the \sersic\ index of the bulge component. A profile with high $n$ is steeper in  the inner parts (thus providing a better fit for the nuclear component) but shallower (more extended) in the outer parts, leading to an artificially bright magnitude. This effect is compounded if a large-scale disk is present (e.g. NGC7457): in this case, the latter may be degenerate with the \sersic\ profile in the outer parts, and the best-fit solution may evolve to an entirely different configuration, in which, boosted by the artificially high $n$, the bulge component also dominates at the largest radii. 

\textit{Bars and Inner Disks}: If unaccounted for, these components may also bias the bulge parameters. Apart from contributing flux that should not be attributed to the bulge, they may bias $R_e$ and $n$ to either higher or lower values. The effect may become amplified due to the bulges' overlap, and therefore partial degeneracy, with the disk.

\begin{figure*}
  \begin{center}
    \includegraphics[width=\linewidth]{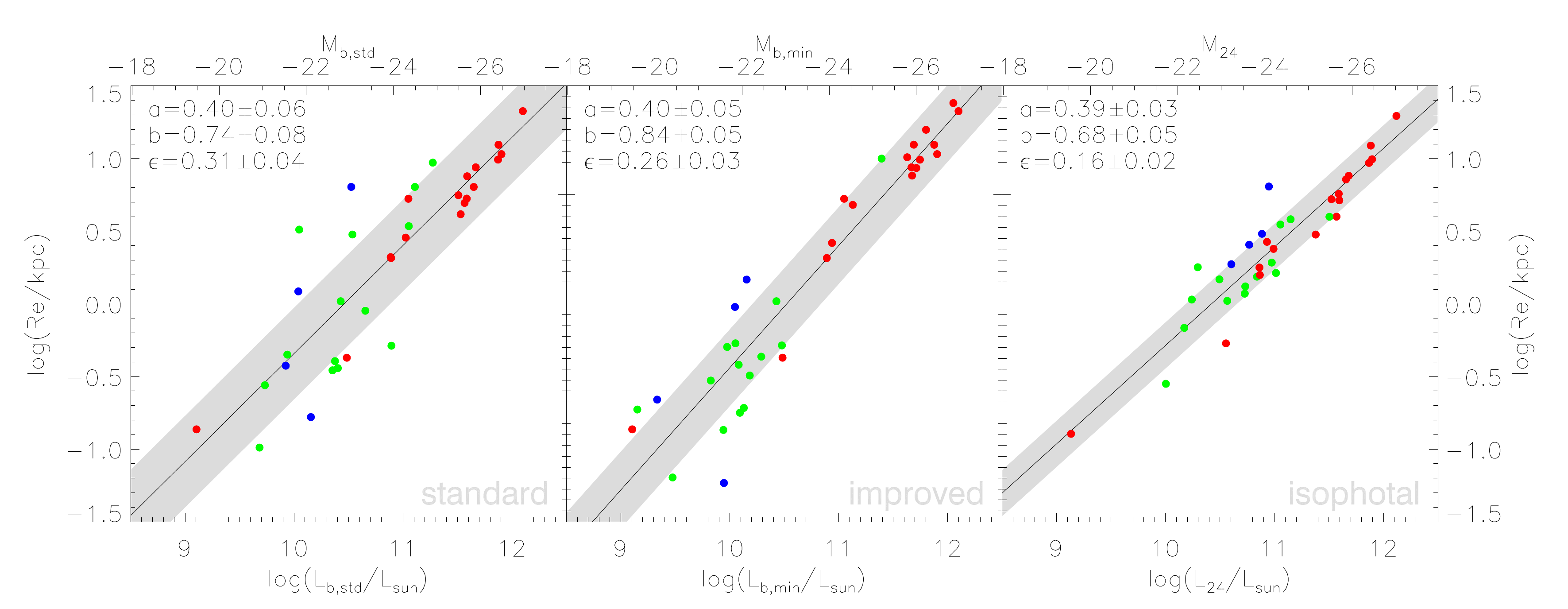}
  \end{center}
 \caption{Size-Luminosity relations derived from our WIRCam $K$-band images. In case of bulge parameters (left panel: standard bulge+disk fits; middle: improved models), the effective radius of the bulge component ($R_e$ of the \sersic\ profile) is plotted against its luminosity. In case of the total light distribution (right panel), the effective radius and luminosity ($\liso$) are derived from the curve-of-growth. Colored symbols represent elliptical (red), lenticular (green) and spiral galaxies (blue). The relations have been fitted by $\log(R_e/\,\mathrm{kpc})=a+b\log(L/10^{11}\lsun)$ including intrinsic scatter ($\eps$) in $R_e$ via a maximum likelihood method (described in the appendix of Paper-II). Neither $R_e$ nor luminosities have been assigned an uncertainty. The ensuing best-fit parameters are displayed along with their $1\sigma$-uncertainties, and represented by the solid line and the shaded area, which has a $2\eps$ width in $R_e$. It is evident that the Size-Luminosity relation tightens when improved decompositions are applied in lieu of bulge+disk models; the relation using total photometry is even tighter.}
 \label{fig:sizelum_comp}
 \vspace{20pt}

\end{figure*} 

\subsection{Reliability and Size-Luminosity relation}
\label{subsec:sizelum}

The Size-Luminosity relation of bulges provides indirect evidence of the advantage that improved decompositions have over bulge+disk models. Assuming that bulges broadly are kinematic and evolutionary analogues to elliptical galaxies, they should similarly follow a Size-Luminosity relation via projection of the fundamental plane of these systems. If so, then a careful decomposition and accounting for non-spheroidal features (apart from the disk) should tighten the Size-Luminosity relation with respect to a standard bulge+disk decomposition. This is indeed the case, as comparison of the left and middle panel of Figure \ref{fig:sizelum_comp} shows: improved models appear to recover the (true) bulge parameters, effective radius ($R_e$) and luminosity ($\lbul$), with higher precision. The steeper slope that results from the improved models is also more in line with the slope expected from the fundamental plane of elliptical galaxies \citep{JunIm08}. Likewise, the total luminosity is known to correlate with the radius enclosing half of the total light \citep{Shen_etal03}. In the right panel of Figure \ref{fig:sizelum_comp}, we plot the corresponding relation that results from our non-parametric curve-of-growth analysis of the WIRCam data. The ensuing correlation's slope and intrinsic scatter are consistent with the $r$-band relation of early-type galaxies derived by \cite{Shen_etal03}. Notably, it is also significantly tighter than the Size-Luminosity relation of bulges. This \textit{may} indicate that the measured bulge parameters are still relatively uncertain, even when improved decompositions are used to derive them; alternatively bulges may not, or not to high degree, represent a homologous family of dynamical stellar systems as, for instance, elliptical galaxies do.

\subsection{The role (or not) of pseudobulges}
\label{subsec:pseudobulges}

Finally, we comment here briefly on the possible distinction between classical and pseudo-bulges, in view of recent claims that black hole scaling relations might differ depending on whether the bulge belongs to one or the other class \citep[e.g.][but see also \citealt{Greene_etal10} for an alternative claim regarding the $\mbh-\lbul$ relation]{Hu08,GreeneHoBarth08,Nowak_etal10,Sani_etal11,K11}. Typically, pseudo-bulges are defined as  having low \sersic\ index ($n<2$) and for being associated with distinct morphological features, including nuclear bars, spiral structures, dust, and flattening similar to the disk \citep[][]{K11}. 

We apply those criteria to our imaging data. As mentioned at the beginning of this subsection, none of our targets exhibits dust lanes in the $K$-band. At our resolution of $\sim0\farcs8$, we observe neither spiral structures nor bars in the nuclear regions. All of our bulges appear less flattened than the respective disks. In a number of galaxies, we \textit{do} identify inner disks and model them with with \sersic\ components with flattening similar to the disk's and $n<2$ (in fact, all of our best fits have $n\lesssim1$). However, these inner disks are seen in {\it addition} to a bulge component, they do not {\it replace} it, and therefore are not likely to represent ``pseudo bulges". In other words, we fail to identify pseudo-bulges in our data based on morphological features alone. When we consider the \sersic\ index, if we restrict ourselves to the bulge+disk decomposition, three galaxies have bulges with $n<2$: NGC1300, NGC2787 and NGC3384. However, in all three cases, when improved models are used, the \sersic\ indices of the bulges increase above $n=2$ ($n=4.3$, $2.8$ and $2.5$ specifically). On the other hand, four galaxies for which $n>2$ when bulge+disk models are used, see $n$ decrease to $n_\mathrm{bul}<2$ when improved models are adopted (NGC3245, NGC3998, NGC4342 and NGC7457). For the first three galaxies, however, the improved model is not a perfect fit to the data, suggesting that the bulge parameters are likely quite uncertain.

In conclusion, with our data and using improved models that fare better at avoiding biased bulge parameters, there are only 4 candidate galaxies that may not harbor a ``classical" bulge. Yet, even in those cases the classification as pseudo-bulge is tentative and based only on the \sersic\ index being smaller than 2. All other galaxies feature bulges with $n>2$. Based on our sample, we conclude that a morphologically based classification of bulges into two separate classes is extremely subjective, and do not support it. 

\section{Summary and conclusions}
\label{sec:summary}

Using the wide field of view WIRCam imager at CFHT, we have obtained deep, high spatial resolution near-IR ($K$-band) images for 35 nearby galaxies  with securely measured Supermassive Black Hole masses $\mbh$. Our goal is to study and characterize the NIR $\mbh-\lbul$ and $\mbh-\ltot$ relations using  a homogeneous imaging data set that supersedes all $K$-band data previously available for our sample galaxies. In particular, we required 1) increased imaging depth to reduce component degeneracy and to allow for reliable bulge parameter estimates; 2) a dedicated dithering and data reduction strategy to improve subtraction of the strong and variable NIR-sky background; and 3) high spatial resolution to resolve and model small components, such as stellar nuclei or inner disks which, if unaccounted, can potentially bias the derived bulge parameters. These criteria are not met by \twomass\ data, which formed the basis of the first NIR $\mbh-\lbul$ relation (MH03). Likewise, the \ukidss\ $K$-band images used by \cite{V12} still suffer from residual background fluctuations and include only a fraction of our \smbh-host galaxy sample.

We described a dedicated data reduction procedure specifically designed to provide accurate modeling of the background, a task that is significantly aided by the wide field of view ($20'\times20'$) of WIRCam. Our iterative procedure exploits the fact that the background can be characterized as two independent components: a spatially invariant (on the scale of a single CCD), time dependent component, and a spatially complex, but time-invariant pattern. Using the 2D software \gfthree, we found that while 17 galaxies (all classified as ellipticals) can be adequately modeled by a single-\sersic\ profile, all other galaxies (18, including three classified as elliptical in the RC3) required the addition of (at least) a disk, which we assumed to be exponential. The resulting bulge magnitudes, $\Mbstd$, typically differ by several tenths of a magnitude from the values published by MH03 and based on \twomass. 

However, we found that such ``standard" bulge+disk models do not generally provide good fits to the data. Such discrepancies can be resolved by the inclusion of additional components, most commonly bars (8 galaxies), galactic nuclei (10 galaxies) and inner disks (6 galaxies). Making use of the flexibility offered by \galfit, we also model evident spiral arms (4 galaxies) and rings (2 galaxies). Moreover, in 5 early-type systems with highly inclined disks, we found it inevitable for an adequate fit to introduce, in addition to the bulge, a second large-scale component with \sersic\ profile ($n\gtrsim1$). It is unclear whether such ``envelope" is simply needed to account for deviations from a \sersic\ model in the outer part of the bulge, is a spurious component introduced by the assumption that the disk is described by an exponential profile, or whether it constitutes a real, morphologically separate stellar component.  Finally, we observe a central light deficit (measured relative to the inner extrapolation of the \sersic\ law that best fits the outer profile) in 9 elliptical galaxies: for these, the core region was masked when fitting the data.

While the total magnitudes we derived are largely independent of the details of the modeling (e.g. the number of components used), and agree well with the estimate obtained from a non-parametric curve-of-growth analysis, the bulge magnitudes vary significantly according to the specifics of the model used to represent the galaxy. For galaxies requiring additional components, bulge magnitudes are on average $0.36\mg$ fainter than derived using a simple bulge+disk decomposition. This serves as a warning that careful analysis and supervision must be applied when fitting nearby galaxies to avoid biases and systematics in the derived bulge magnitudes. 

In a companion paper (Paper-II) the bulge and total magnitudes presented in this contribution will be used to provide a detailed characterization of the NIR $\mbh-\lbul$ and $\mbh-\ltot$ relations for Supermassive Black Holes.


\section*{Acknowledgments}
\label{sec:acknowledgments}

This research has made use of the NASA/IPAC Extragalactic Database (NED) which is operated by the Jet Propulsion Laboratory, California Institute of Technology, under contract with the National Aeronautics and Space Administration. 


\bibliographystyle{apj}
\bibliography{Mbh-Lbt}


\appendix

\section{Individual galaxy decompositions}
\label{appA}

This appendix presents detailed notes and diagnostic plots pertaining to the galaxy image decompositions which led to the bulge and total magnitudes presented in Section \ref{sec:results}. We intend to demonstrate the intricacies and uncertainties often inherent when trying to extract the bulge light from a galaxy image, the importance of an accurate identification of distinct morphological components to obtain unbiased structural parameters, and the frequent inability of simple bulge plus exponential disk models in fitting the data. We describe the photometric characteristics and the corresponding decomposition strategies for each galaxy individually. For a summary description and general discussion, see Section \ref{sec:imageana}.

The notes and figures are ordered according to the galaxy name, as in Table \ref{tab:obs}. The figures include the original image, the residual image, and the radial surface brightness profile of the data and model. Individual components, when present, are plotted separately. For galaxies with disk, we also include the curve-of-growth, and relevant results from the 1D isophotal analysis (ellipticity, position angle and/or the ``B4'' parameter that describes whether the isophotes are disky or boxy). We note that all but one of the disk galaxies required an ``improved model'', i.e. additional components besides a bulge and a disk. These models are also shown in the Figures. In some cases of galaxies with improved models, we also show the single-\sersic\ model, to highlight how the addition of an exponential profile to the single-\sersic\ model is unable to account for the disk unless at least one additional component, which we term ``envelope'', is included. \\

\begin{figure*}
\begin{center}
\includegraphics[width=\linewidth]{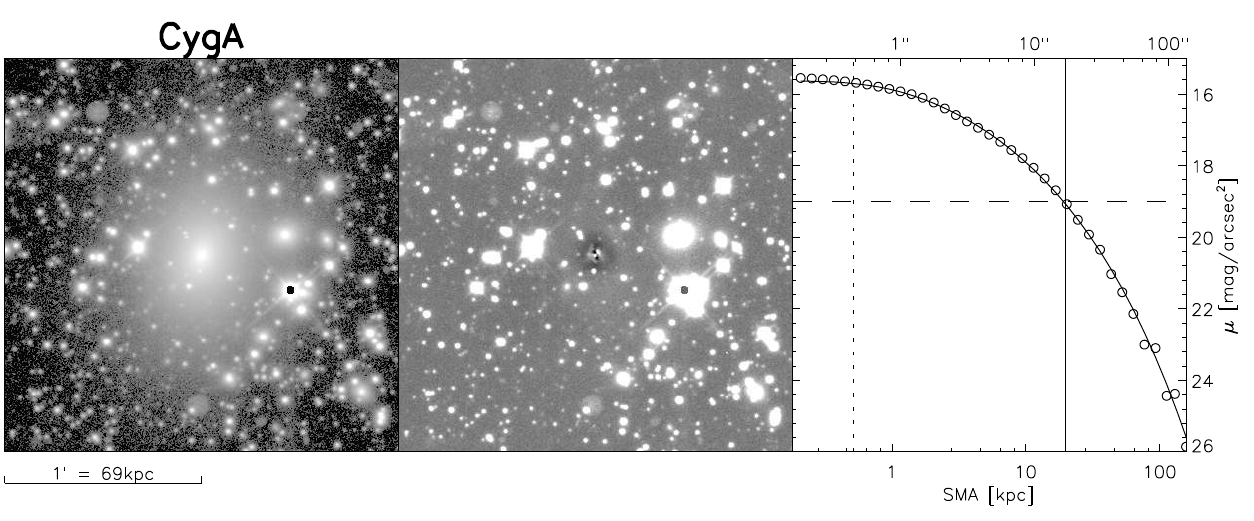}
\end{center}
\caption{Panels showing a cutout of the co-added image (left), the residuals from the single-\sersic\ model obtained by \galfit\ (middle), and the surface-brightness profiles (right) of the data (open circles) and model (solid line). The solid vertical line indicates the galaxy effective radius as determined from the curve-of-growth. The dotted vertical line corresponds to the FWHM of the image point-spread function. The dashed horizontal line indicates the level at which the grayscale in the residual image saturates (white/black for positive/negative residuals). While the grayscale in the residual image scales linearly with flux (gray corresponding to zero), the data image is displayed on a logarithmic scale.}
\label{fig:CygA}
\end{figure*}

{\bf CygA} (Fig. \ref{fig:CygA}) is the most luminous ($L_K\sim1.3\times10^{12}\lsun$) and intrinsically largest ($R_e\sim20\kpc$) object in our sample. Yet, it is also the most distant ($d\sim240\mpc$) target and consequently amongst those with the smallest apparent size ($R_{e,24}=17\arcsec$). Photometric measurements are complicated by the fact that the galaxy suffers from severe foreground contamination. For this galaxy, therefore, the resolution and depth of our WIRCam data are of vital importance, not only to resolve the object itself, but also to reliably mask the numerous stellar sources overlapping it. We found it essential to construct an ``2nd-pass'' object mask from the residual image, and to use it to mask interlopers in the final fit. Despite being a giant elliptical, we cannot identify a core in this galaxy, possibly because of contamination from the unresolved nuclear source. At all radii, a \sersic\ profile fits exceptionally well. \\

\begin{figure*}
\begin{center}
\includegraphics[width=\linewidth]{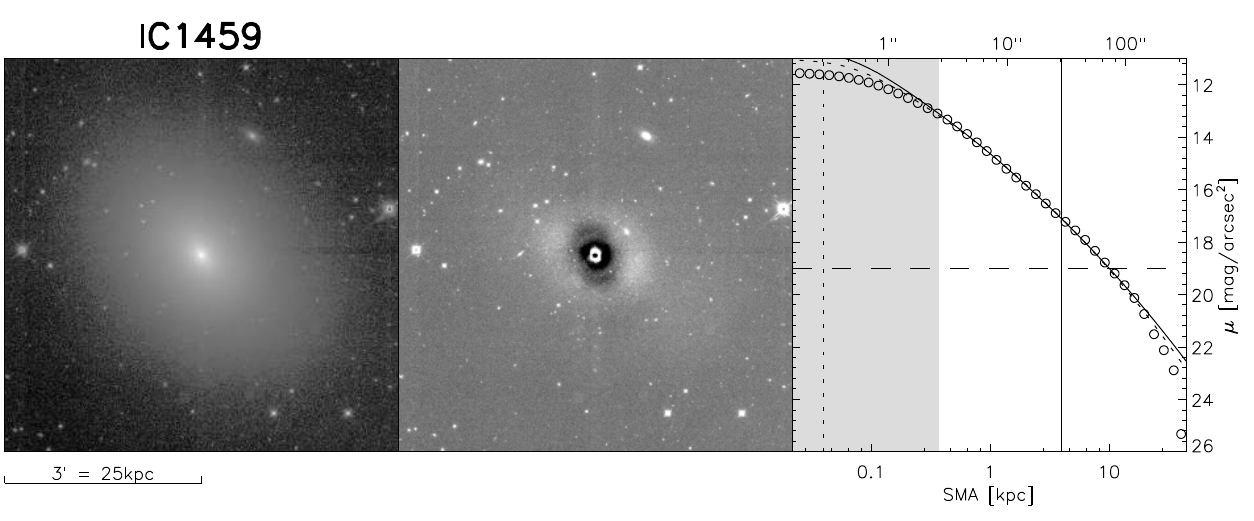}
\caption{Description as for Figure \ref{fig:CygA}, but now the \sersic\ model has been fitted after masking the core. The radial extent of the core mask is indicated by the shaded region in the figure on the right. For comparison, the profile from a model fitted without core masking is shown as a dotted curve.}
\label{fig:IC1459}
\end{center}
\end{figure*}

{\bf IC1459} (Fig. \ref{fig:IC1459}) does not show any clear sign of having a complex morphology, but the residual image from a single-component model reveals significant deviations from a 2D-\sersic\ profile. Apart from a clear core, which we accordingly mask for improved modeling, IC1459 exhibits a relatively strong isophotal twist ($\sim15\deg$), which might indicate that the galaxy is triaxial. Despite the central light deficit, the central surface brightness is notably higher than what we observe in other cored ellipcticals, and the residuals barely benefit from core-masking. Furthermore, beyond $R\sim100\arcsec$, the surface-brightness profile is mildly but consistently lower than that of the \sersic-model. However, attempted multi-component models proved unsustainable, in particular we cannot find a satisfactory fit when including an exponential disk. A disk is also not supported by the observed ellipticity profile and harmonic perturbations. The significant residuals might simply reflect the fact that the models do not account for the isophotal twist, and therefore we adopt the single-\sersic\ model for this galaxy. \\

\begin{figure*}
\begin{center}
\includegraphics[width=\linewidth]{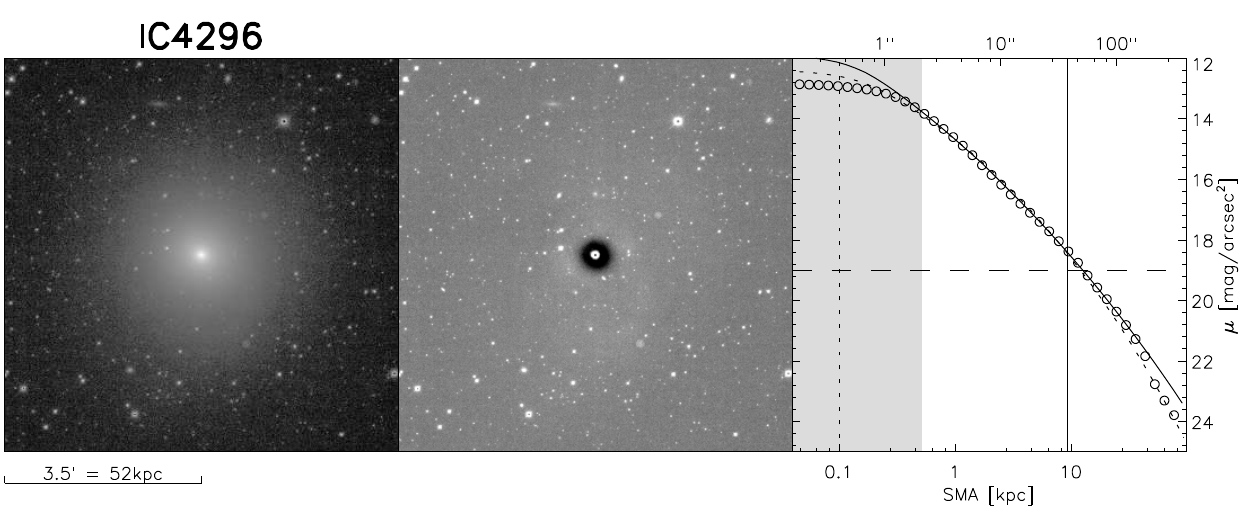}
\caption{As for Figure \ref{fig:IC1459}.}
\label{fig:IC4296}
\end{center}
\end{figure*}

{\bf IC4296} (A3565-BCG, Fig. \ref{fig:IC4296}) is an elliptical galaxy with very low flattening which, after masking the core, is very well fitted by a \sersic\ profile. \\

\begin{figure*}
\begin{center}
\includegraphics[width=\linewidth]{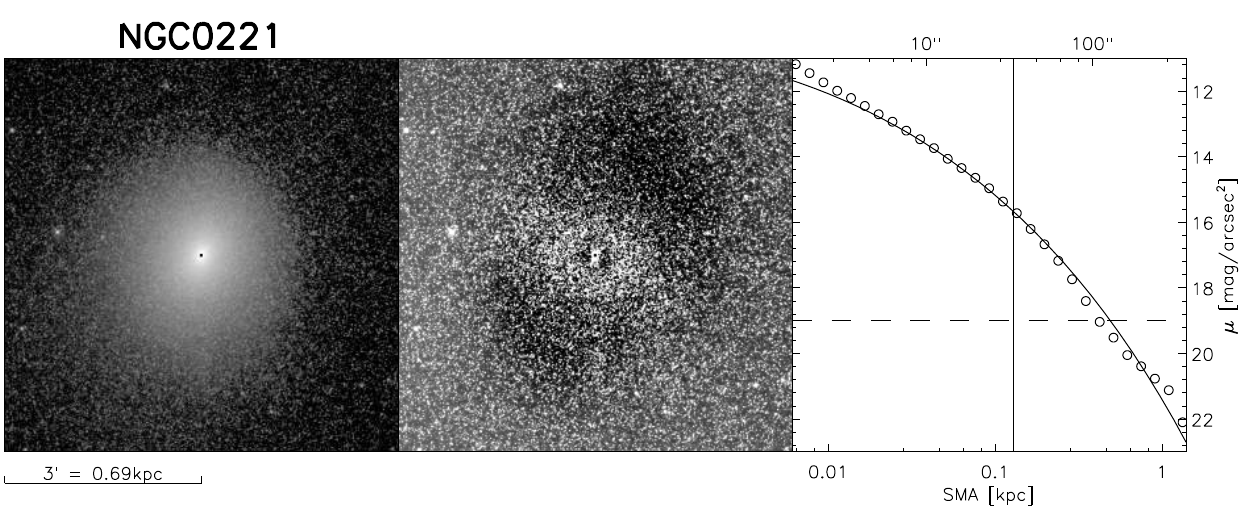}
\caption{As for Figure \ref{fig:CygA}.}
\label{fig:NGC0221}
\end{center}
\end{figure*}

{\bf NGC0221} (M32, Fig. \ref{fig:NGC0221}) overlaps with M31, which leads to unavoidable degeneracy. The background gradient due to M31 necessitated a specialized sky subtraction procedure, in which M31's diffuse (unresolved stellar) background is modeled from the co-added frame after the ``1st pass'' (see \S\ref{subsec:reduction}). After re-scaling and re-projection, but before performing the ``2nd-pass'' masking and sky subtraction, it is subtracted from the individual input frames. Additionally, both galaxies are partly resolved into stars, which should not be masked in order to not compromise M32's outer brightness profile. We therefore run \galfit\ without the source mask provided by \sex\, but account for M31's variable background of resolved stars by using \galfit's tilted ``Sky'' component. We find a good fit to M32 using a single-\sersic\ profile at almost all radii. M32 has a very bright center which saturates even in the shortest ($2.5\,\mathrm{s}$) exposures. As far inward as reliable data exists, the profile remains steep. The central resolved nucleus is almost entirely masked (at $R\lesssim1\farcs3$, inside the smallest radius plotted in Fig. \ref{fig:NGC0221}) and therefore does not affect our fit. \\

\begin{figure*}
\begin{center}
\includegraphics[width=\linewidth]{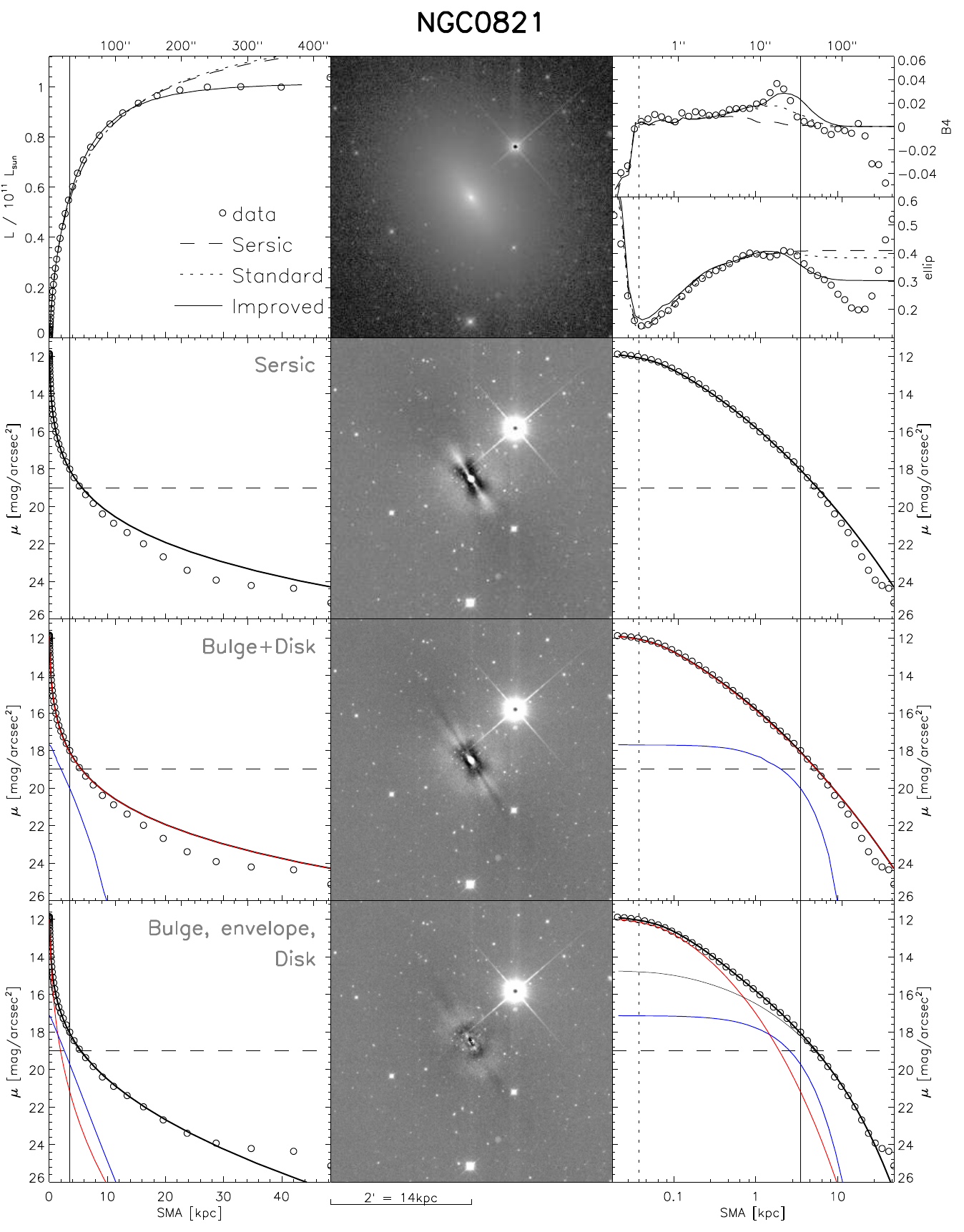}
\caption{As Figure \ref{fig:NGC1023}, with respect to which we have added the single-\sersic\ model (second row).}
\label{fig:NGC0821}
\end{center}
\end{figure*}

{\bf NGC0821} (Fig. \ref{fig:NGC0821}) is a typical lenticular galaxy hosting a highly-inclined disk (see \S\ref{subsec:beyond_b+d}). The disk is clearly visible by inspecting the images, and is reflected in the ellipticity and B4 harmonic coefficient profiles, as well as in the residuals of a single-\sersic\ model. While modeling this galaxy requires a disk component, the ``standard" bulge+disk configuration is fraught with problems. The solution is degenerate, as there exist two solutions with equivalent $\chi^2$, neither one of which appears plausible. Both have extreme disk flattening: in one case the disk is rounder than the bulge (and therefore discarded by us as unphysical), in the other case it is more flattened ($q=0.09$) than observed directly from the images. In the latter case, the disk component  deviates from an exponential profile and is modeled by a \sersic-profile with $n_d\sim0.5$. Finally, the ``standard" model bulge \sersic\ index is unusually high (7.2), leading to a significant overestimate of the enclosed flux at large radii, as shown by a comparison with the the curve-of-growth results.

Addition of an ''envelope''  with an $n=2.5$ \sersic-profile and larger axial ratio than bulge (and disk), addresses those issues and strongly reduces residuals. Although some residuals remain, we adopt this bulge+disk+envelope model, because there are no other discernible components apart from a possible weak nucleus, and it allows for a robust exponential disk extraction. More complex models that include a central point source or a second (inner) disk provide only minor reduction of residuals. It is not clear whether the envelope represents the outer part of the bulge, or a separate stellar component. Therefore, NGC821 is a case of an early-type galaxy in which the "\sersic-bulge + exponential-disk" picture is inadequate, and in which additionally the bulge may be photometrically ill-defined, depending on interpretation of the envelope component. \\

\begin{figure*}
\begin{center}
\includegraphics[width=\linewidth]{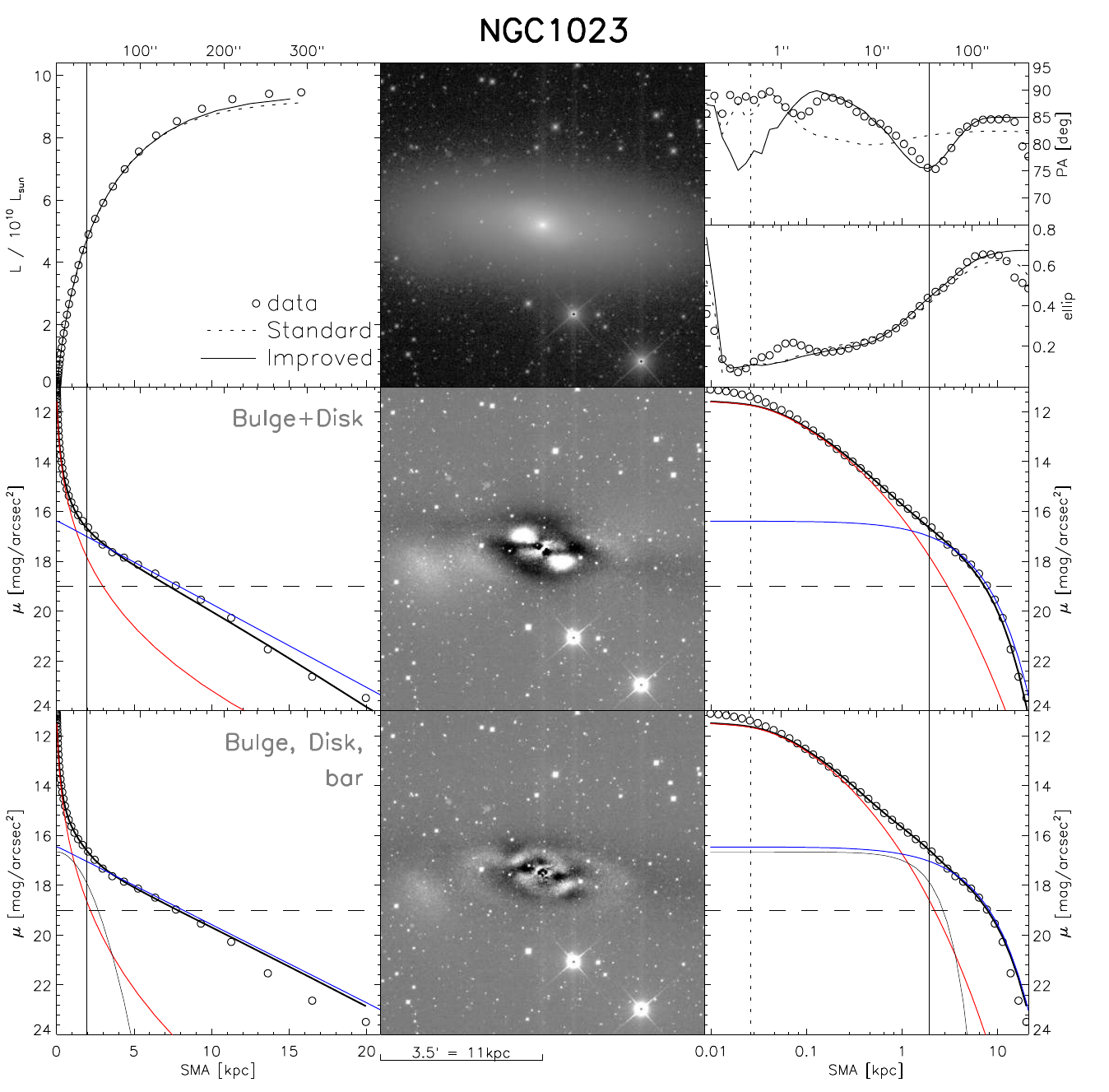}
\caption{Comparison of ``standard'' (bulge+disk) and improved models. The top row presents the data image (center), the curves-of-growth (left) and isophotal shape profiles (right figure). The middle and bottom rows show the residual image (center figure) and surface brightness profiles (left and right figure), respectively for both models. Image scaling, as well as vertical and horizontal lines in the surface brightness profiles, are defined as in Figure \ref{fig:CygA}. The surface brightness profile is shown on a linear radial scaling (left figures) in order to allow for comparison with an exponential decline (straight line). With respect to Figure \ref{fig:CygA}, the component profiles have been added (thick red and blue curves: bulge and disk, thin black curves: additional components). Components are listed in order of decreasing central surface brightness. Curves-of-growth (top-left figure) are shown as measured on data image (open circles) and model images (dotted line: bulge+disk, solid line: improved model).}
\label{fig:NGC1023}
\end{center}
\end{figure*}

{\bf NGC1023} (Fig. \ref{fig:NGC1023}) is an S0 that clearly contains a bar, which in this case is not aligned with the disk. Attempts to fit the isophotal twists with coordinate rotations in disk or bulge proved ineffective. A faint nuclear point source can be fitted, but we omit it since it has little effect on the bulge parameters. \\

\begin{figure*}
\begin{center}
\includegraphics[width=\linewidth]{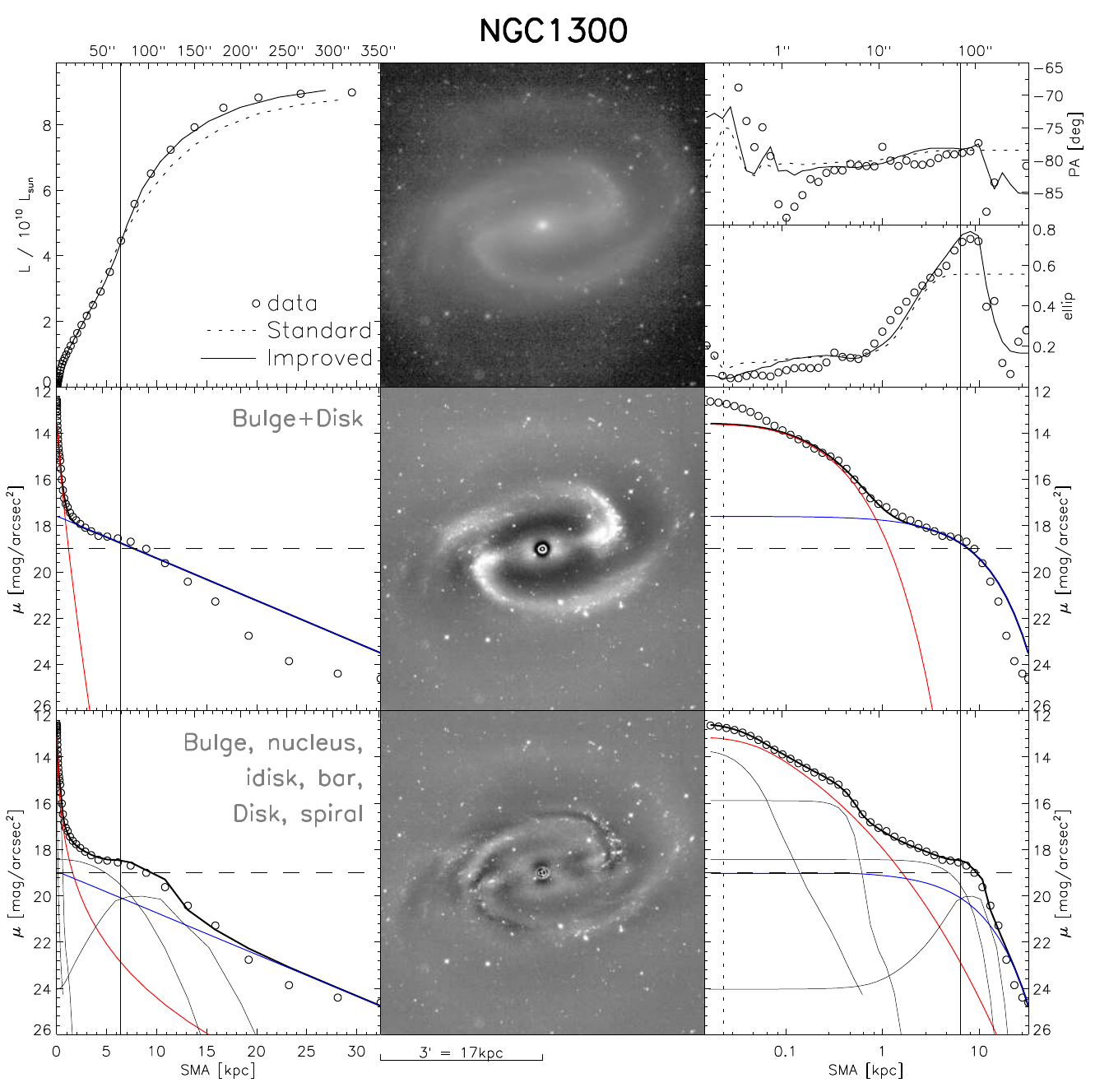}
\caption{As Figure \ref{fig:NGC1023}.}
\label{fig:NGC1300}
\end{center}
\end{figure*}

{\bf NGC1300} (Fig. \ref{fig:NGC1300}) is a grand-design spiral, with the bar and 2-arm structure being dominant. A conventional b+d model is created for the sake of completeness but is clearly inadequate. The most notable aspect of its mismatch with the data is that the ``disk" component of the best solution mostly follows the bar's light instead of the large-scale spiral+disk structure. 

This is the most complex galaxy in our sample. As a consequence, we require an ``improved model'' with 6 components for an adequate fit: bulge, disk, bar, spiral arms, inner disk and nucleus. The disk light is a superposition of an exponential profile (diffuse light) and the spiral arms. The ``maximum bulge" magnitude in this case is calculated by excluding the light of both the disk and spiral arms, while, as usual, the fluxes of bulge and all other additional components are summed in order obtain an upper limit to the bulge magnitude.

The spiral arms are modeled by employing coordinate rotation, and refined by bending and Fourier modes to account for their asymmetric strength and shape. Spiral arms and bar could not be modeled using a single component, due to a discontinuity in the surface-brightness profile, the different thickness of  the arms and bar, and the complexity of the rotational pattern (``kinks'', partial winding back onto itself). As a consequence, the spiral arms component requires an inner truncation.

Apart from bar, spiral and nucleus, modeling of the bright small-scale disk, whose sharp boundary creates a ring-like structure upon close visual inspection, is important. NGC1300 provides a good example of the \sersic\ index being \textit{under}estimated unless additional (and clearly visible at our resolution) components are included: it changes from 1.3 (standard) to 4.3 (improved model), despite the fact that the improved model includes a nuclear point source. Therefore, using a simple bulge+disk model, the bulge of this galaxy would be classified as a \textit{pseudo-bulge} (defined as having a \sersic\ index smaller than 2), while in fact it harbours a ``classical'' bulge \textit{and} an inner disk. \\

\begin{figure*}
\begin{center}
\includegraphics[width=\linewidth]{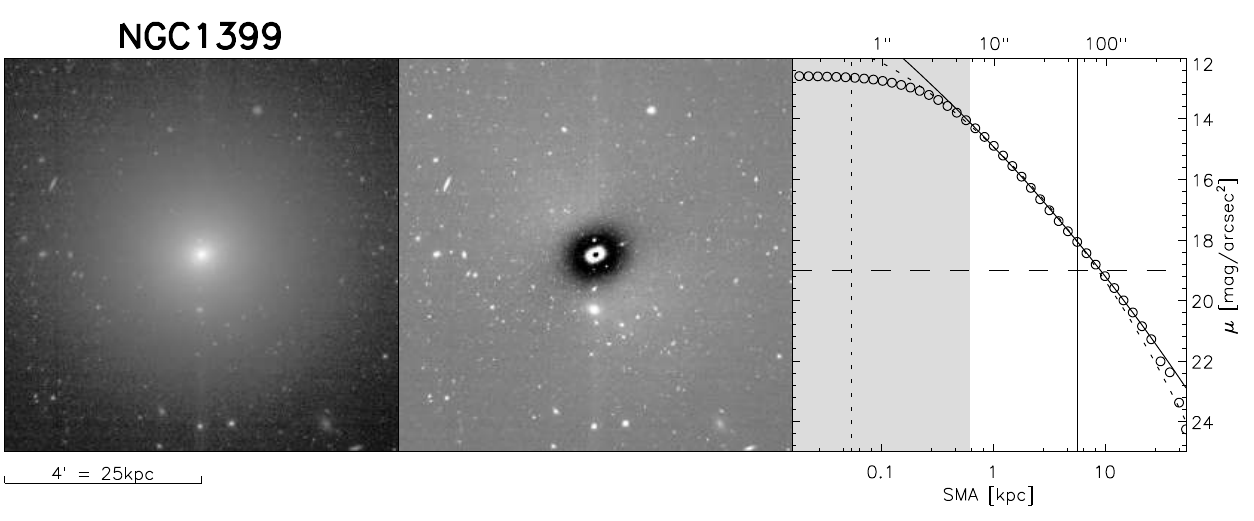}
\caption{As Figure \ref{fig:IC1459}.}
\label{fig:NGC1399}
\end{center}
\end{figure*}

{\bf NGC1399} (Fig. \ref{fig:NGC1399}) is a giant elliptical with an extended core ($R_c\sim 6\arcsec$) and a significant ellipticity gradient. Masking the core significantly changes the parameters of the \sersic\ model, including its magnitude. \\

\begin{figure*}
\begin{center}
\includegraphics[width=\linewidth]{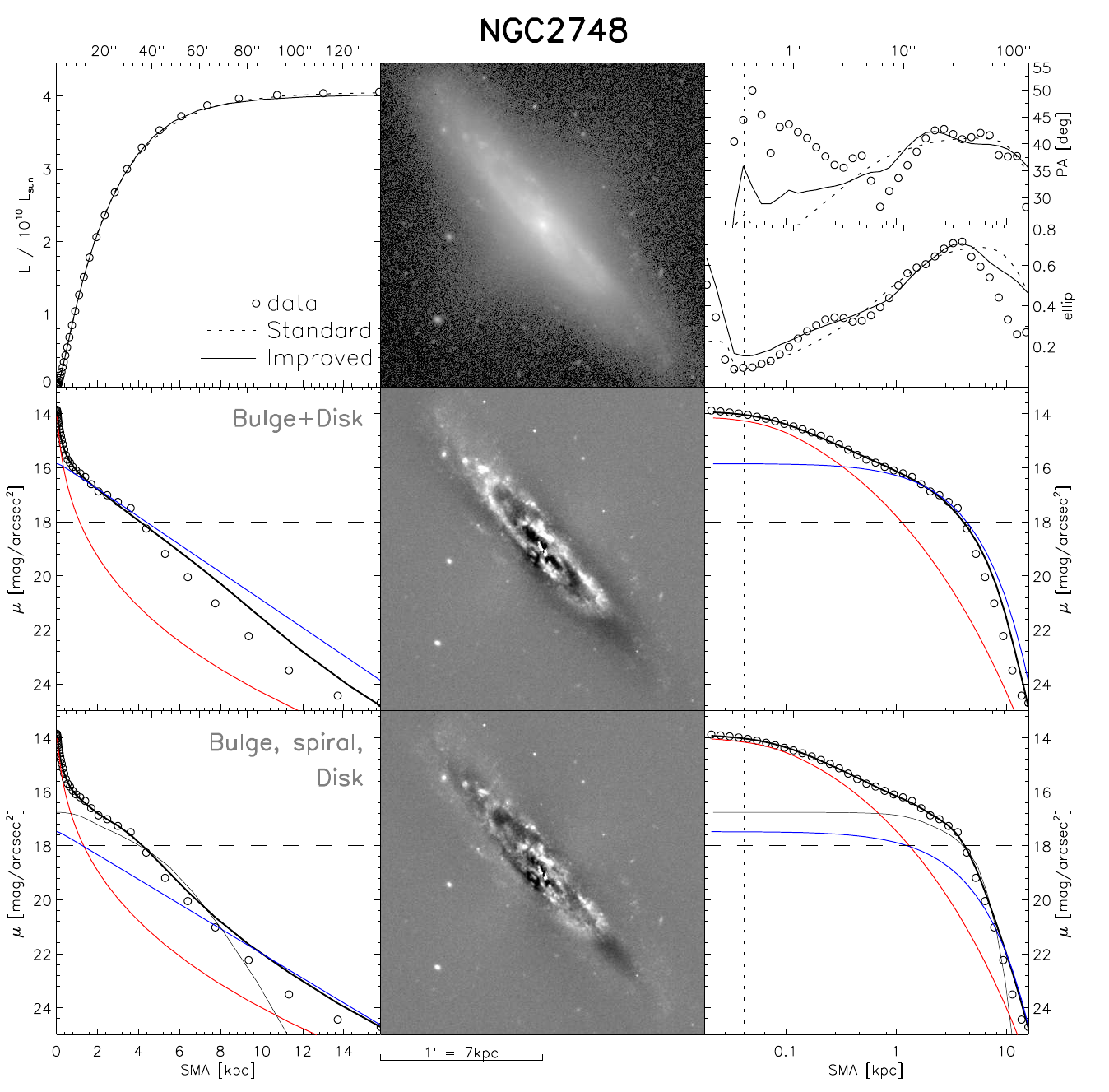}
\caption{As Figure \ref{fig:NGC1023}.}
\label{fig:NGC2748}
\end{center}
\end{figure*}

{\bf NGC2748} (Fig. \ref{fig:NGC2748}) is a spiral galaxy seen at high inclination. The spiral arms are tightly wound and can be traced at semi-major axis radii between $\sim6$ and $40\arcsec$ undergoing a full $\sim360\deg$ rotation. Inside $6\arcsec$ they are not well defined. The disk appears lopsided, and a faint elongated stellar feature is seen to extend from the center outwards along the minor axis. The radial profile is very close to exponential between $\sim6\arcsec$ and $120\arcsec$, except for a ``bump'' at $30\arcsec$ that can be ascribed to the spiral arms.

The bulge+disk model overestimates the disk's flux and scale radius, likely due to the (unaccounted) presence of the spiral arms. The disk profile, if modeled by a \sersic,  deviates from exponential, approaching $n_d\sim0.5$. The asymmetry in the disk contributes to the large residuals. Although generally not allowed in a ``standard'' model, we have tested adding perturbations by inclusion of Fourier mode F1 and bending mode B2 \citep[see][]{GF3}, and found that this has only a small impact on the standard-model parameters.

The evidently warranted improved model includes a spiral arms component. The spiral arms are difficult to model, because they are seen at high inclination and appear intermittent / flocculent. As a consequence, the arms are still  visible in the residual image, but their light contribution is approximately reproduced. There is some uncertainty as to the best way of accounting for the asymmetries. First, it is not clear whether it is the disk or the spiral component that should by modified. Second, asymmetries may be modeled by harmonic perturbations, bending modes, relative shifts of component centers, or any combination thereof. We tested all and find that the various configurations lead to different best-fit parameters, but often also slow down convergence or prevent a unique solution to be found. We eventually adopt a final ``best-fit'' solution in which all component centers are aligned, and asymmetric distortions are ascribed to the spiral component only, in the form of a first-order harmonic perturbation. Some models with higher-order perturbations or independent component centers provide improvements in $\chi^2$, but also impact the convergence or produce a non-exponential disk. We have not modeled the low surface brightness feature aligned with the minor axis, since it is very faint, it is unclear whether it can be described by a \sersic\ function, and would cause an unwarranted degree of degeneracy in the other components. 

Interestingly, despite \textit{adding} a component (the spiral arms), the improved model's bulge is \textit{brighter} than the standard model's. \\

\begin{figure*}
\begin{center}
\includegraphics[width=\linewidth]{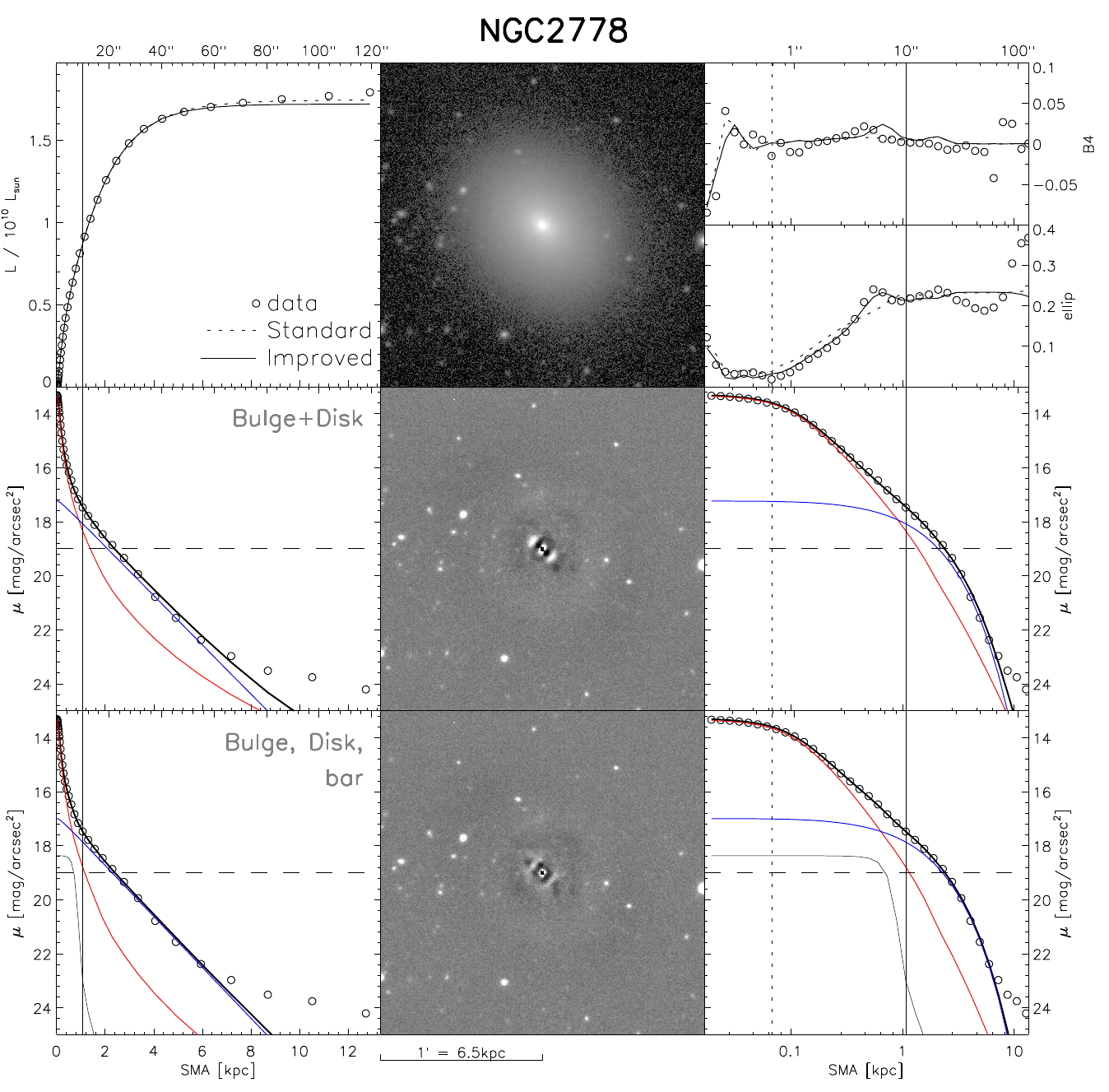}
\caption{As Figure \ref{fig:NGC1023}.}
\label{fig:NGC2778}
\end{center}
\end{figure*}

{\bf NGC2778} (Fig. \ref{fig:NGC2778}) is classified as an elliptical in the RC3  catalog, but we unambiguously identify it as an S0 with a large-scale disk. Furthermore, the 1D profiles (in particular the inflection in surface-brightness  and the peak in ellipticity at $\sim6\arcsec$) and residual image reveal a bar. Although the $\chi^2$ decreases only slightly by including the latter, we adopt the improved model on grounds that the bulge flux is overestimated by $\sim30\%$ when the bar is omitted. \\

\begin{figure*}
\begin{center}
\includegraphics[width=\linewidth]{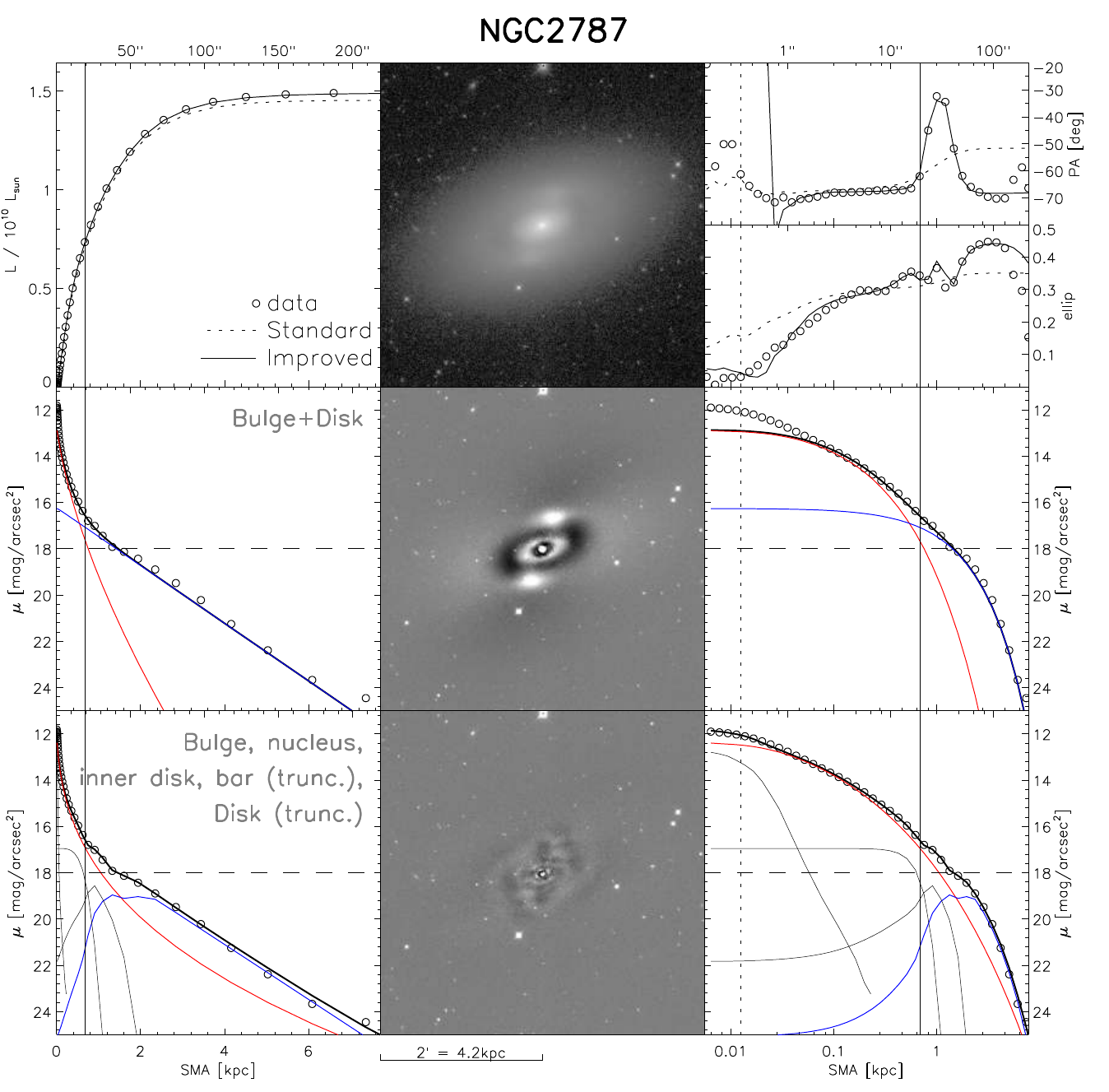}
\caption{As Figure \ref{fig:NGC1023}.}
\label{fig:NGC2787}
\end{center}
\end{figure*}

{\bf NGC2787} (Fig. \ref{fig:NGC2787}) is a lenticular galaxy which shows a strong ring and large bar, requiring an improved model. As suggested by a visual inspection of the image, comparing various models confirms that the bar does not extend all the way through the center and requires an inner truncation. The same applies to the (large-scale) disk in order to model the ring.

Inside the ring, an additional component needs to be introduced and may be interpreted as the inner continuation of the disk or a separate inner disk. Its flattening is close to that of the outer disk. Instead of adopting an exponential profile with outer truncation for the inner disk component, we find the best convergence and residuals by foregoing an exponential profile in favor of a \sersic\ profile. Its fit converges to $n\sim0.25$, in order to reproduce the rapid decline in surface brightness near and outside the ring.

NGC2787 also harbors a nucleus. Although our images marginally resolve the nucleus, we model it as a point source since models using a \sersic\ or King profile lead to a very similar magnitude and bulge parameters. \\

\begin{figure*}
\begin{center}
\includegraphics[width=\linewidth]{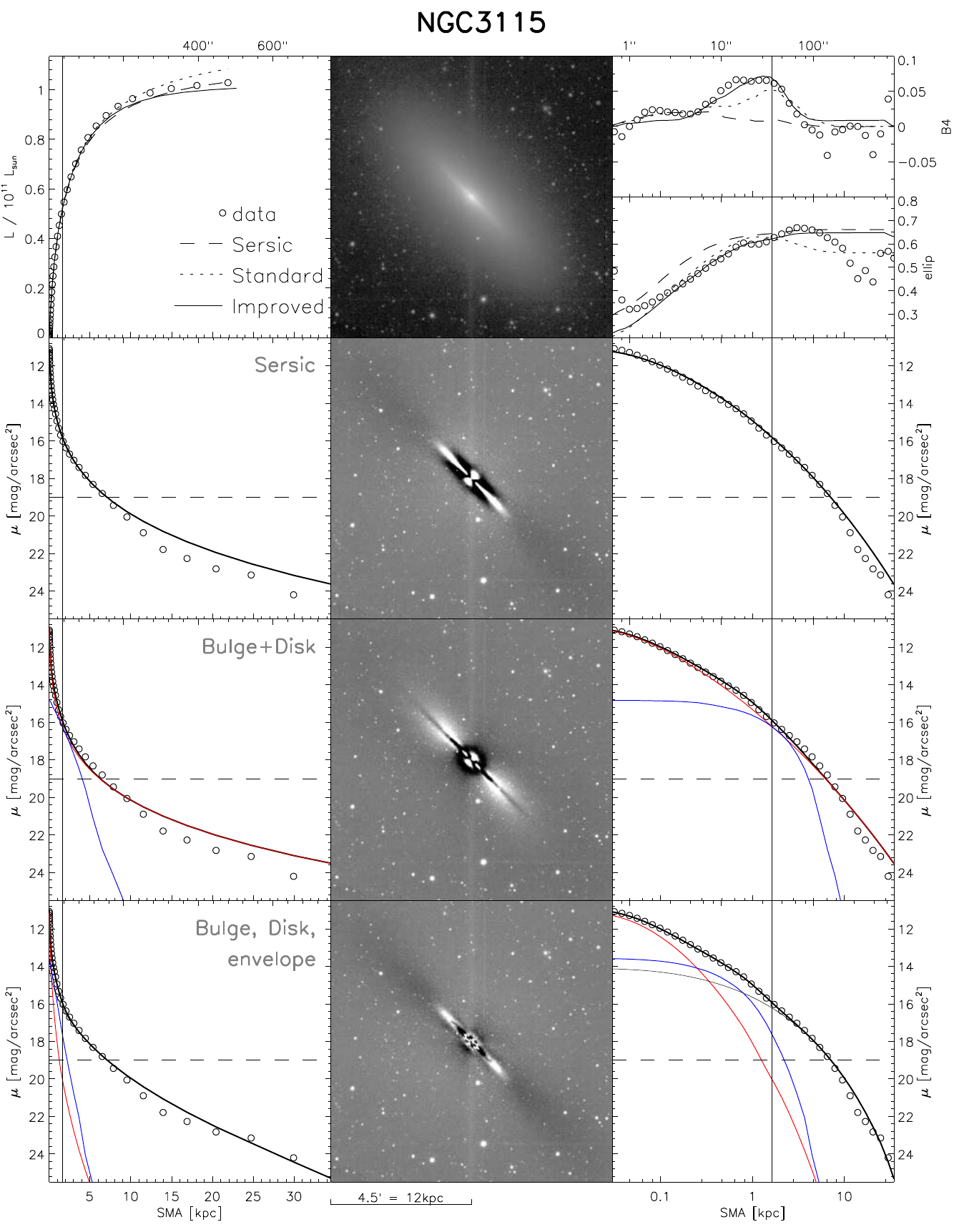}
\caption{As Figure \ref{fig:NGC0821}.}
\label{fig:NGC3115}
\end{center}
\end{figure*}

{\bf NGC3115} (Fig. \ref{fig:NGC3115}) is a rather flattened lenticular galaxy with a nearly edge-on embedded disk. The major-axis surface-brightness profile declines nearly exponentially between $\sim70\arcsec$ and $270\arcsec$. The ellipticity reaches its maximum ($e\sim0.6$) at $R\sim70\arcsec$, and decreases beyond this radius (in the exponential part of the surface brightness profile). There is another, weaker, ellipticity peak at $R\sim30\arcsec$ that may be due to a smaller inner disk.

NGC3115's disk is clearly visible but difficult to fit. A bulge+disk model fails, as is clearly seen from the radial profiles, residual image, and disk robustness (its profile deviates from exponential when the \sersic\ index is allowed to vary). The adopted 3-component model includes an envelope (\sersic\ $n=2.1$) and benefits from azimuthal Fourier modes (similar to diski-/boxiness) allowed for all components. The introduced envelope component, as in case of NGC4342, has flattening intermediate between bulge and disk and may thus be interpreted as a separate thick (``hot'') disk. The improved model much better reproduces the observed profiles, including ellipticity and diskiness. Even so, moderate residuals remain and hint at either profile mismatches or the presence of additional components (e.g. a secondary disk). We have however been unable to find a  4-component non-degenerate model. \\

\begin{figure*}
\begin{center}
\includegraphics[width=\linewidth]{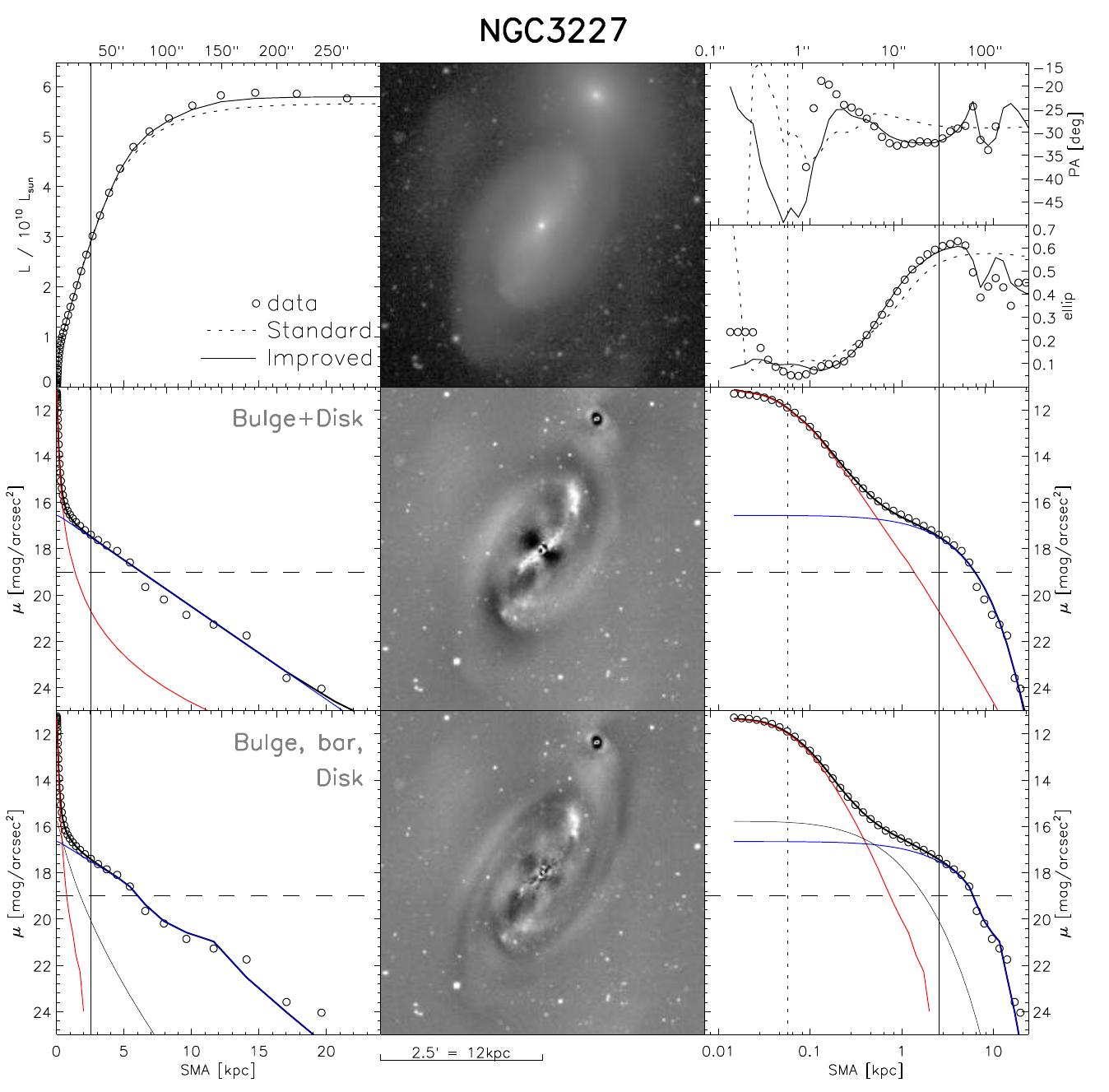}
\caption{As Figure \ref{fig:NGC1023}.}
\label{fig:NGC3227}
\end{center}
\end{figure*}

{\bf NGC3227} (Fig. \ref{fig:NGC3227}) poses a challenge to decomposition not only due to the presence of a two-armed spiral structure, but also because of a close neighbour, elliptical galaxy NGC3226, with which it probably interacts. The proximity, size and brightness of NGC3226 prevents masking, thus requiring its inclusion in the model (as a single-\sersic\ profile), unfortunately causing a degeneracy with NGC3227's outer disk. The disk profile consists of two parts, having approximately the same scale radius but different surface brightness. The transition between both portions of the disk profile occurs between $70\arcsec$ and $90\arcsec$, which marks the inner boundary of the discernible spiral. The latter has a largely smooth appearance with no prominent knots and might therefore be produced by tidal interactions, a view supported by the two faint extended features that resemble stripped material seen at $R\sim120\arcsec$ along the minor axis. However, mild star formation regions are apparent near both inner ends of the spiral, where they connect with the inner, rather regularly shaped part of the disk profile.

NGC3227's bulge has very high central surface brightness but is very small, dominating the surface brightness only inside $\sim20\arcsec$ (compared to $\sim300\arcsec$ out to which we can measure the disk profile). A bulge+disk model produces a best-fit bulge \sersic-index of almost 12, and naturally cannot reproduce the spiral morphology. The residual image also reveals a strongly elongated structure connecting the spiral arms through the galaxy centre. This may be interpreted as the spiral arms' continuation within the inner part of the disk, as a separate bar component, or as a superposition of both. Visual impression suggests a central point source, while the shape of the central brightness profile is inconclusive in this regard.

For these reasons we fit an improved model including a nucleus, a bar (or inner spiral arm) component, and coordinate rotation of the exponential disk component. We found that instead ascribing the rotation to the bar / inner spiral component lead to inferior residuals and fit convergence. This lends further support to the view that the outer spiral structure is caused by tidal interactions and distortion of the stellar disk. The \sersic\ index of the additional bar / inner spiral component fits best at $\sim1.4$, which we interpret as evidence that it does not exclusively represent the putative bar (since $n\lesssim1$ would be expected in this case), but includes flux of the inner spiral structure or the bulge. Perhaps surprisingly, the nuclear component in the best-fit model is rather faint, with central surface-brightness lower than the bulge component. Omitting it leaves the residuals and the bulge \sersic\ index almost unaltered at $\sim4$. Therefore, in our final model we exclude it as unnecessary. We caution though that given the small bulge effective radius of $\sim1\arcsec$, even a slight error in the point-spread-function may alter the bulge parameters, as well as the significance of a central point source component. \\

\begin{figure*}
\begin{center}
\includegraphics[width=\linewidth]{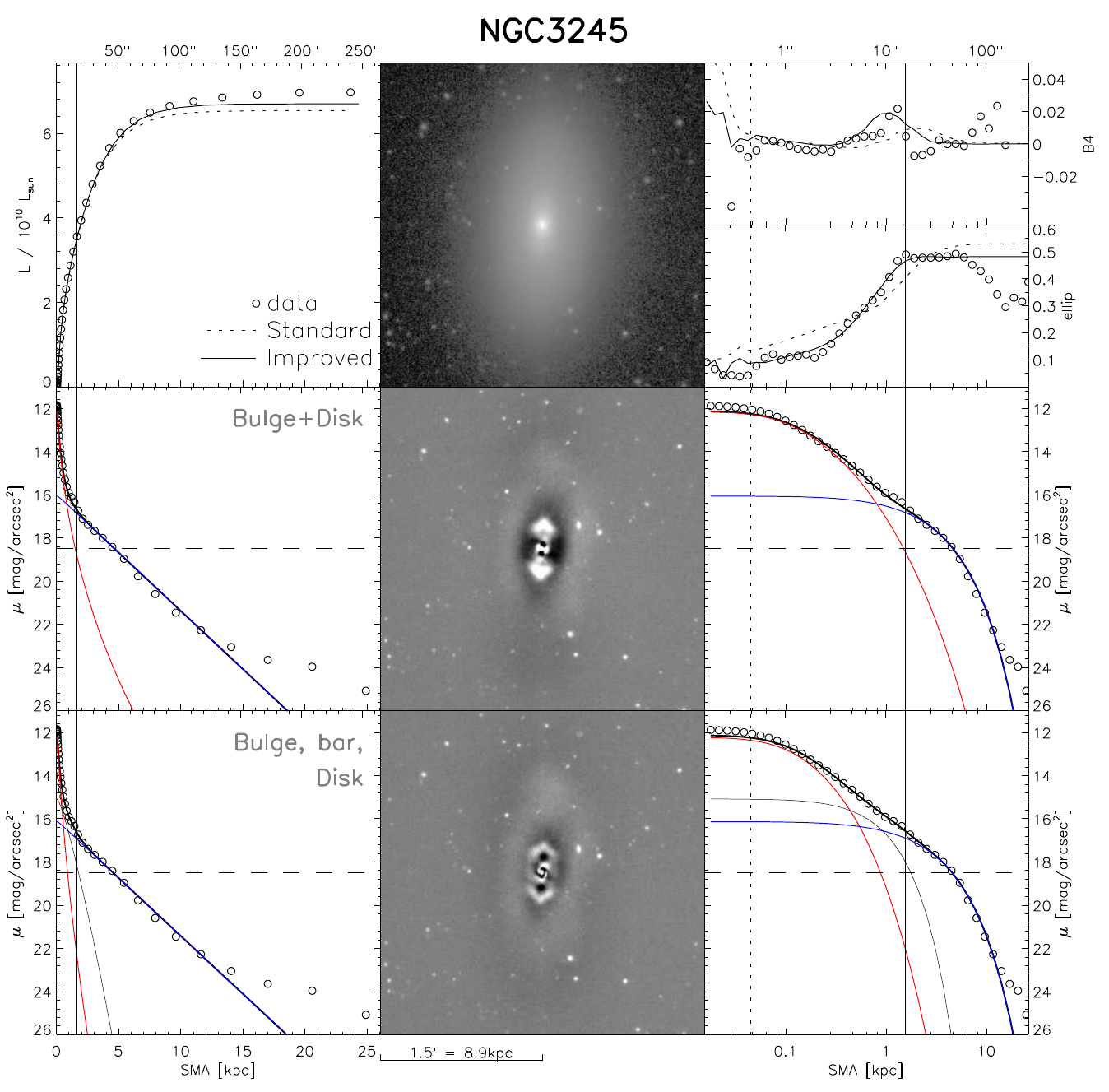}
\caption{As Figure \ref{fig:NGC1023}.}
\label{fig:NGC3245}
\end{center}
\end{figure*}

{\bf NGC3245} (Fig. \ref{fig:NGC3245}) has a clear disk component but it lacks any obvious spiral structure. Besides the bulge and the exponential disk, the radial profile shows that the disk is partially truncated beyond $\sim60\arcsec$, followed by a low-surface-brightness extended disk (or halo) beyond $\sim120\arcsec$. A small inflection at $\sim15\arcsec$ may suggest a bar.

The bulge+disk model produces significant residuals between $6\arcsec$ and $20\arcsec$ along the major axis, confirming the presence of a bar. The disk in the standard model converges to  $n\sim0.5$ if modeled with a \sersic\ profile.

Inclusion of a bar improves the residuals, but we could not establish a suitable improved model that accounts for the extended profile at the largest radii. Also, residuals in the central regions may be indicative of a nucleus or an inner disk, as well as a bulge isophotal twist or a spiral structure. Yet, we could not identify and fit any of these features unambiguously, and the corresponding models do not converge. This unsatisfactory situation arises not only because our resolution is still insufficient at these small spatial scales, but because the suitable profile and morphology of the putative additional component(s) and modifications are unknown and degenerate with other model parameters. \\

\begin{figure*}
\begin{center}
\includegraphics[width=\linewidth]{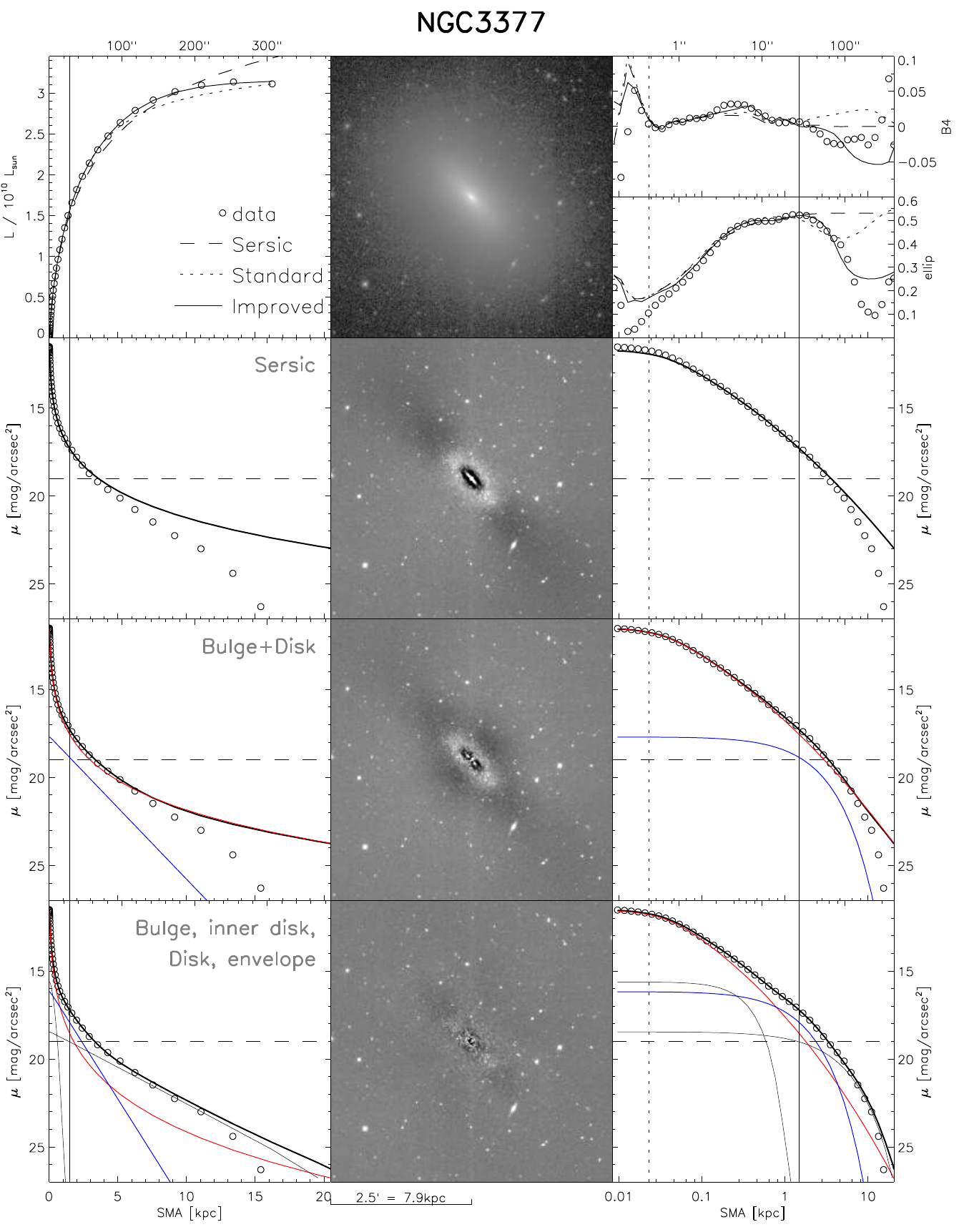}
\caption{As figure \ref{fig:NGC0821}.}
\label{fig:NGC3377}
\end{center}
\end{figure*}

{\bf NGC3377} (Fig. \ref{fig:NGC3377}) is classified as E5 in the RC3 catalogue. Yet, the surface-brightness profile exhibits an exponential decline at radii $>60\arcsec$, and closer inspection of the $\varepsilon$- and $B_4$-profiles reveal peaks at $R\sim8\arcsec$ and $R\sim30\arcsec$, respectively,  suggesting two embedded disks. A single-\sersic\ model not only produces quadrupole residuals, but also a  high \sersic\ index of 8.7. Its surface brightness profile strongly overestimates the flux at large radii, which we suggest is the result of the steep inner profile. The latter is caused by the bright small-scale inner disk which can be clearly seen in the residual image. Bulge+disk models produce similar results, independent of whether the disk component is made to fit the inner or the outer disk. Hence, we require an improved model with 4 components (bulge, disk, inner disk and envelope). 3-component models did not adequately fit this galaxy, because even with two disk components in place, the bulge component cannot account for the near-exponential decline and low ellipticity of the outer profile. \\

\begin{figure*}
\begin{center}
\includegraphics[width=\linewidth]{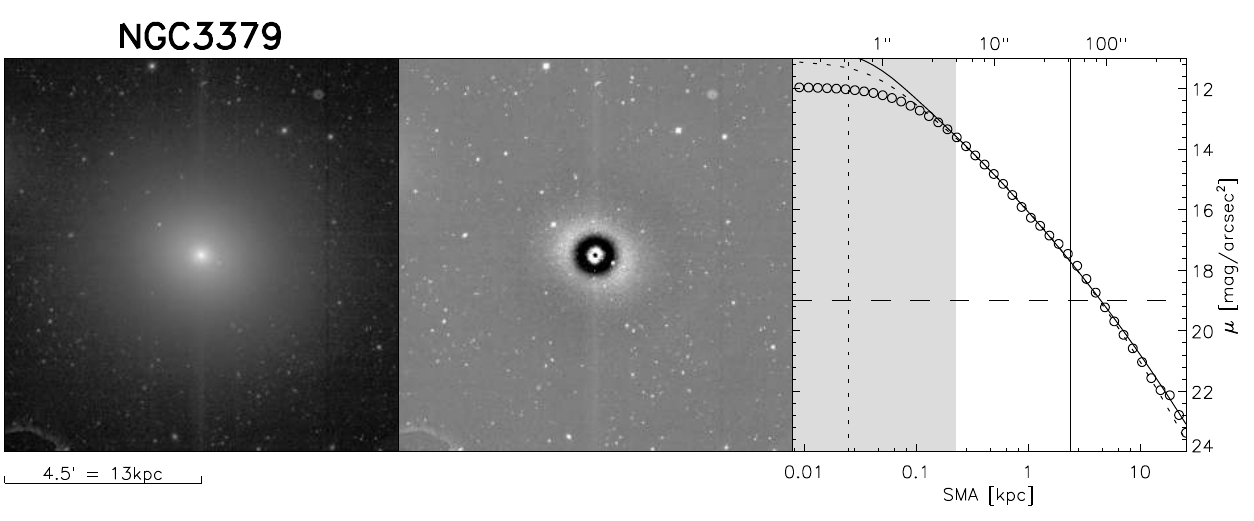}
\caption{As Figure \ref{fig:IC1459}.}
\label{fig:NGC3379}
\end{center}
\end{figure*}

{\bf NGC3379} (Fig. \ref{fig:NGC3379}) appears to be a typical giant elliptical, with the brightness profile showing a core ($R<3\arcsec$) and no evidence of a disk. We thus fit it with a single-\sersic\ model while masking the core. The ellipticity is nearly constant at $\sim0.1$. There is a mild ($\sim10\deg$) and smooth isophotal twist. The isophotes do not show deviations from pure ellipses. \\

\begin{figure*}
\begin{center}
\includegraphics[width=\linewidth]{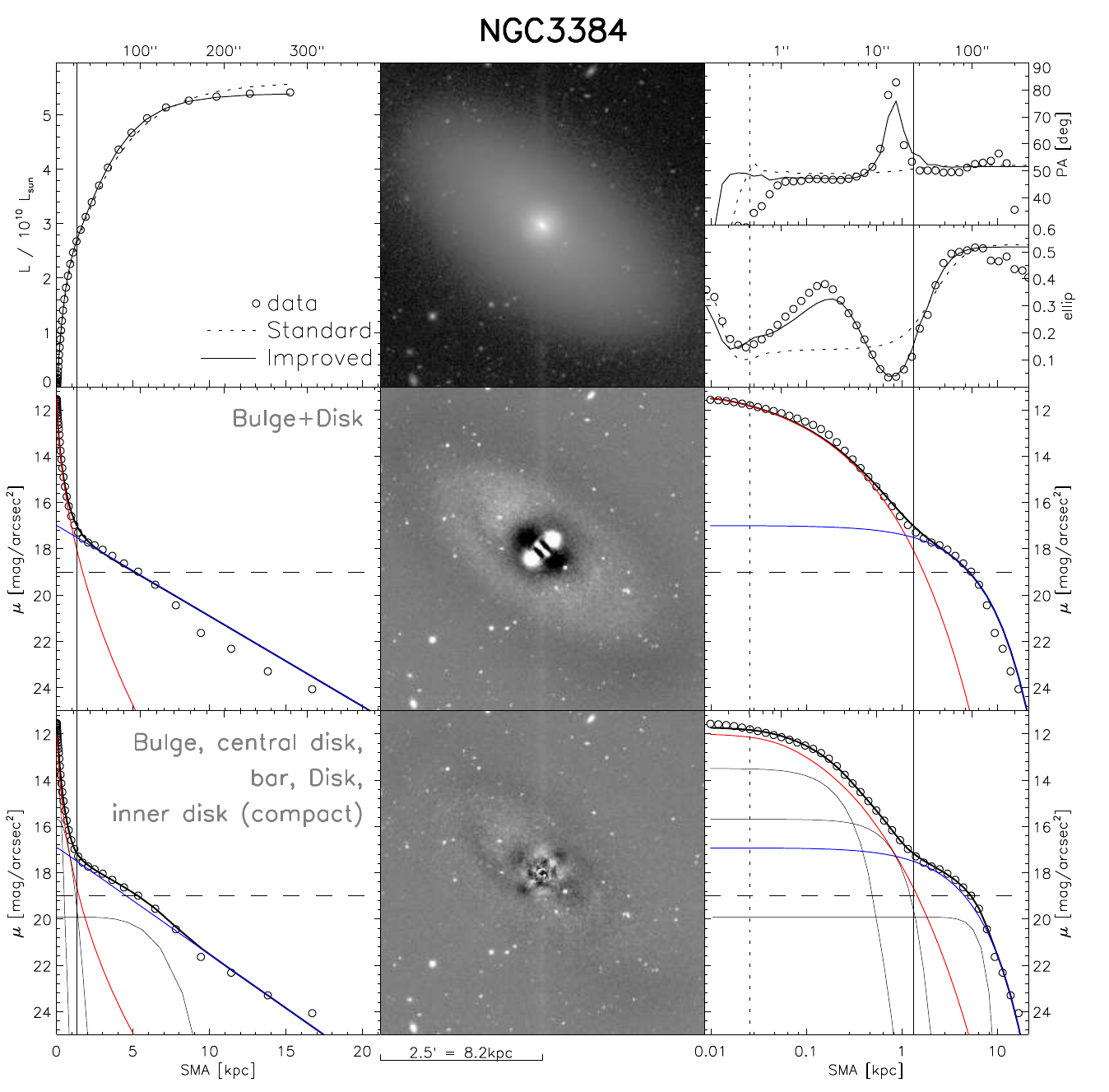}
\caption{As Figure \ref{fig:NGC1023}.}
\label{fig:NGC3384}
\end{center}
\end{figure*}

{\bf NGC3384}'s surface brightness profile (Fig. \ref{fig:NGC3384}) reveals a large-scale exponential disk showing a ``step" at $\sim150\arcsec$. Beyond, the scale radius may be somewhat larger than for the inner part, or the profile may not be exponential: the depth of our images is not sufficient to draw an unambiguous conclusion. As is the case for a few other galaxies in our sample, the disk may consist of two parts, or may be truncated with an additional envelope-like component responsible for the flux excess at the largest radii.

The standard model's disk is not exponential (if modeled by a \sersic\ profile, $n$ drops to almost 0.5). Furthermore, strong and characteristic residuals remain: a bright and highly flattened ($R\lesssim10\arcsec$) structure, aligned with the disk major axis and reminiscent of an inner edge-on disk, a  bar-like structure nearly aligned with the minor axis, and a dark ring at the radius of the above-mentioned step in the disk profile.

We thus fit an improved model that accounts for these features. Both the central disk and the bar are fitted by \sersic\ components. It is necessary to include a component for the bar before fitting the central disk. The latter is yet required for satisfactory residuals. Both of these additional components, as well as the bulge, benefit from introducing and fitting modified (disky/boxy) isophotal shapes.

The step in the disk's profile can be fitted by introducing an additional component that increases the surface brightness within $\sim150\arcsec$ above that of the underlying outer disk's. The solution is not unique, and similar $\chi^2$ can be achieved by using two different profiles: a \sersic\ with $n<1$, declining quickly at radii beyond $\sim150\arcsec$, or an exponential with outer truncation and scale radius constrained to be equal to that of the outer disk's. In both variants, the position angle and ellipticity of the two components are constrained to be the same: a test shows that even when allowed to vary independently, their axial ratios converges to the same value, confirming that they represent different parts of the same disk. We eventually adopt the configuration in which the inner part of the main disk is modeled with a \sersic\ profile, since it converges faster than a disk with truncation. Other configurations aimed at modeling the break in the disk profile proved to be relatively disadvantageous.

The adopted improved model is thus composed of a bulge, central disk, a bar and a two-component main disk. \\
 
\begin{figure*}
\begin{center}
\includegraphics[width=\linewidth]{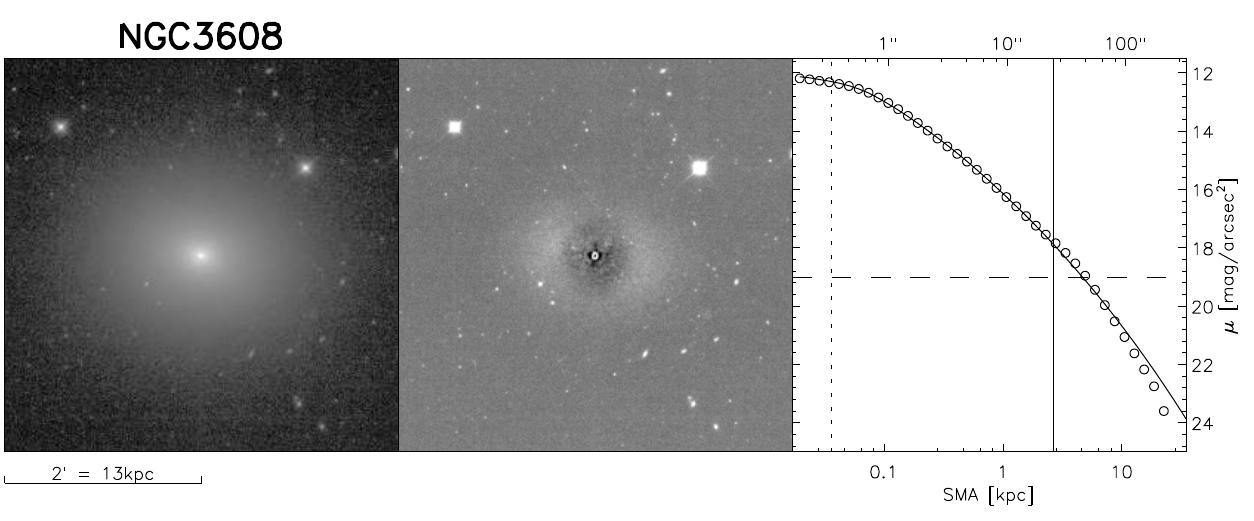}
\caption{As Figure \ref{fig:CygA}.}
\label{fig:NGC3608}
\end{center}
\end{figure*}

{\bf NGC3608}'s surface-brightness profile (see Fig. \ref{fig:NGC3608}) shows slight deviations from a \sersic\ at intermediate radii, but variations in ellipticity are very modest. We hence fit it with a single-\sersic\ component. In our image, we cannot discern a core. NGC3608 is a relatively close neighbor of another (larger) elliptical galaxy. As their light distributions overlap slightly, we simultaneously fit the neighbor, also using a \sersic\ profile. \\

\begin{figure*}
\begin{center}
\includegraphics[width=\linewidth]{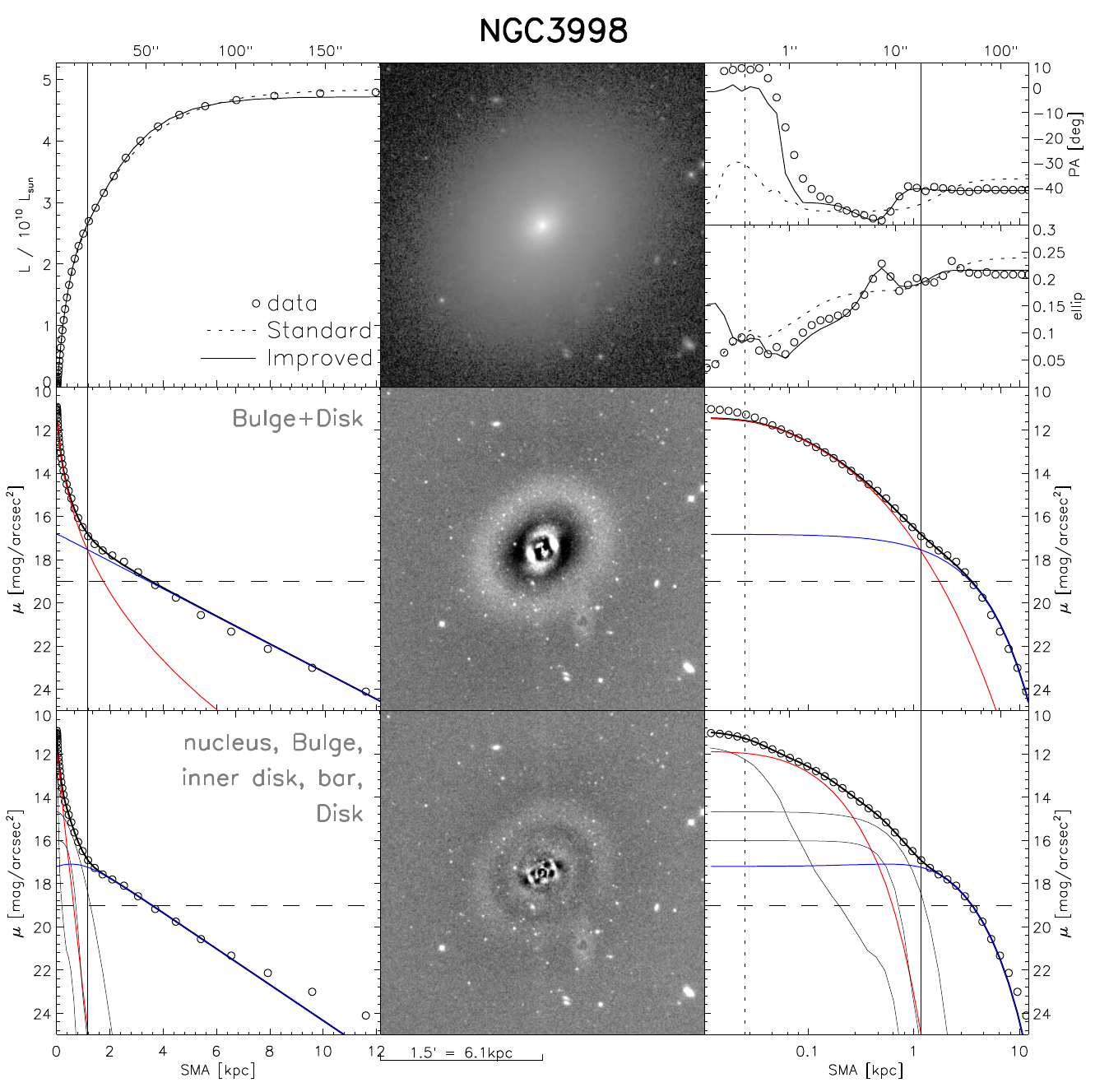}
\caption{As Figure \ref{fig:NGC1023}.}
\label{fig:NGC3998}
\end{center}
\end{figure*}

{\bf NGC3998} (Fig. \ref{fig:NGC3998}) appears to be a lenticular galaxy, but the radial profile indicates deviations from a simple bulge+disk composition.

The surface brightness profile exhibits an inflection between $\sim10\arcsec$ and $40\arcsec$ (best seen when scaling the radial coordinate logarithmically), corresponding to the location of a ring seen in the image. This inflection is located in between the bulge- and disk-dominated regions, and its surface brightness is lower than the inward extrapolation of the outer, exponential part of the profile. Additionally, beyond $\sim150\arcsec$, there is a light excess with respect to an exponential profile, perhaps indicative of an extended outer disk \citep[see e.g.][]{Minchev12}. The ellipticity increases from the center outwards, reaching a maximum at $\sim8\arcsec$ where the residual image reveals the presence of a small-scale bar, and then levels off beyond $\sim15\arcsec$, where the disk dominates. The position angle profile shows an extremum at the same radius, as well as some isophotal twist in the innermost (bulge-dominated) regions.

Relatively strong residuals, due to the ring, bar, as well a a central nucleus, are evident when fitting a standard bulge+disk model. The improved model includes the bulge, a bar, an inner disk that is best-fitted by a compact ($n\sim0.6$) \sersic\ profile, the main (outer) disk, and a central point source. The main disk is modified by an inner truncation, with the best-fit truncation parameters (``softening length'') corresponding to a partial truncation. The inner disk component may thus be interpreted as either a continuation of the outer disk which is,  with the main disk truncation, required to reproduce the ring, or as a separate stellar component. Models that include \textit{both} components (inner disk \textit{and} and a continuation of the outer disk in the region inside the ring) were not supported by our data, producing degenerate results and best-fit parameters sensitive to initial values. However, models without an inner disk component could be ruled out. \\

\begin{figure*}
\begin{center}
\includegraphics[width=\linewidth]{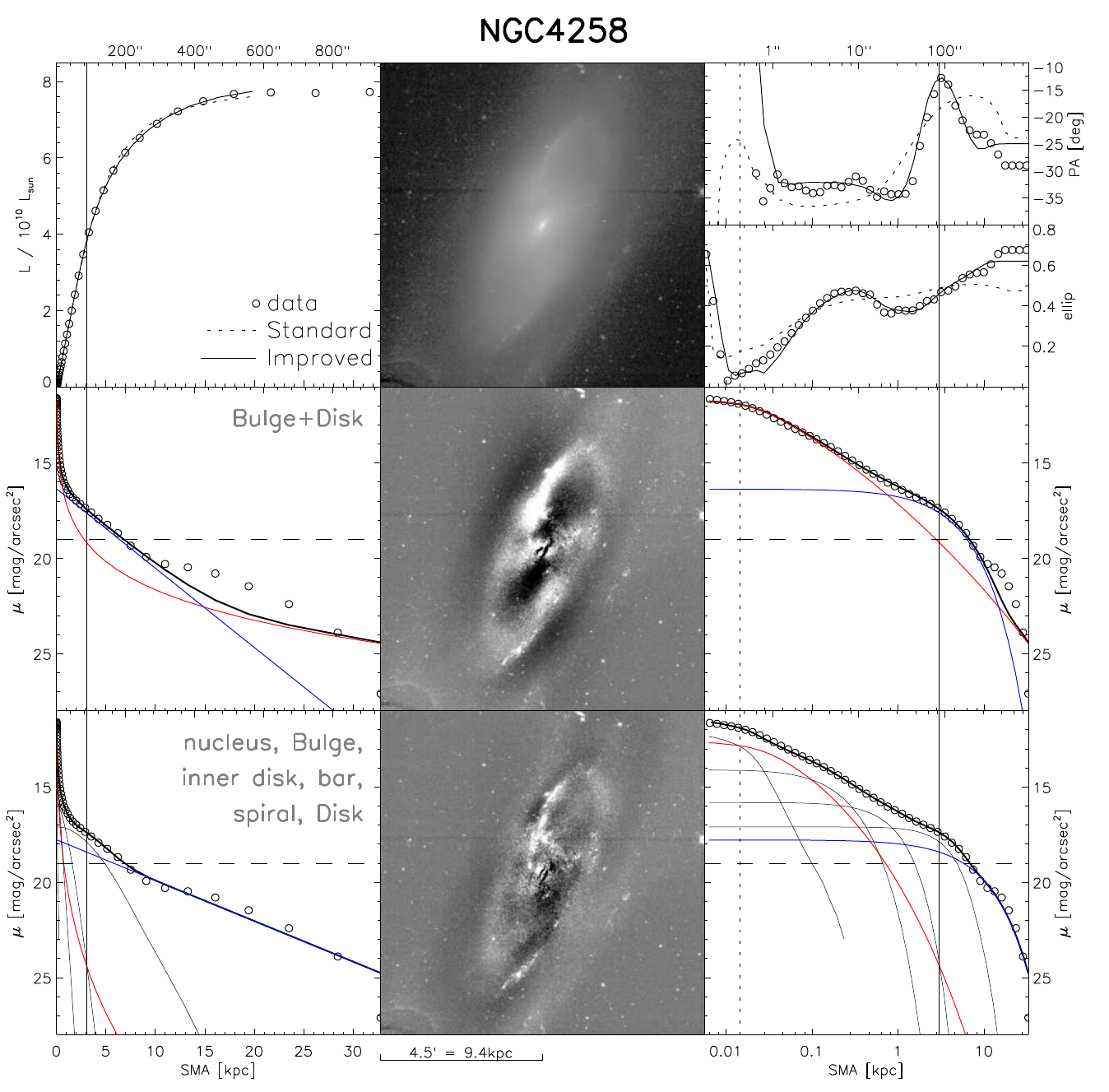}
\caption{As Figure \ref{fig:NGC1023}.}
\label{fig:NGC4258}
\end{center}
\end{figure*}

{\bf NGC4258} (M106, Fig. \ref{fig:NGC4258}) is, after M32, the most nearby galaxy in our sample and has one of the largest apparent sizes.  

The disk profile is divided into an inner disk that dominates within $\sim200\arcsec$, i.e. the same regions where the spiral arms are present, and an outer disk with apparently larger scale radius, visible\footnote{not in Figure \ref{fig:NGC4258} due to the smaller size of the cutout, which covers mostly the main (inner) part of the disk} out to $R\sim800\arcsec$ and showing some warping as well as modest asymmetry. Both disks appear truncated at $\sim200\arcsec$ and $\sim800\arcsec$, respectively.

The spiral arms are not well defined at all radii, and can hardly be traced inside $\sim100\arcsec$, where they weaken and broaden. The knotty (presumably star-forming) regions are located mainly in two narrow regions near the edge of the main disk. The pair of (main) spiral arms is asymmetric, with the northern arm brighter than the southern arm. Both data and residual image suggest an additional pair of (weaker) minor-axis arms, but this structure may also result from a bar. 

Besides a nucleus, there appears to be a small, bright and elongated structure near the galaxy centre. It resembles a nuclear disk except for its relatively large physical size. It can be seen in the image, and it affects the ellipticity and isophotal shapes, which show enhancements between $\sim5\arcsec$ and $15\arcsec$ ($\sim100$ and $300\,\mathrm{pc}$).

A standard bulge+disk model does not account for any of the above-mentioned features. Additionally, the best-fit  \sersic\ index for the bulge is very high ($\sim8.7$) and, as a consequence, the bulge  dominates the surface brightness in the outer region. The fitted bulge is likely biased by the bright central disk, and overestimates the true bulge magnitude.

The main challenges in building the improved model are the interplay of main disk and spiral arms, the treatment of the arms' asymmetry and ambiguous morphology in the inner parts, and the degeneracy of bulge, spiral arms, main disk and edge-on central disk. The configuration we eventually adopt is not unique, and the bulge parameters may vary greatly if other component configurations are applied.

We account for the spiral arms and part of the main disk using a compact \sersic\ ($n\sim0.5$) profile component, modified by rotation and 1st-order bending modes to reproduce the main arms, as well as 2nd- and 4th-order Fourier modes to reproduce the minor-axis extension. An exponential profile component  accounts for the remaining main disk light and the outer/extended disk. We note that this way of modeling spiral arm and disk structure only approximately accounts for the observed truncation, but forcing a  truncation function led to a degenerate model, and was not applied. Similarly, we forego uneven Fourier modes or independent component centres with which we attempted to model the asymmetries, as they lead to undesirable degeneracy with bulge and other central components.

To model the inner part of NGC4258, we employ 4 components, of which we have already mentioned the bulge, the nucleus, and the central flat disk. When fitting only those three, the central flat component has higher central surface brightness than the bulge, and the latter's profile is almost exponential. The best-fit parameters of this configuration produce a data-model profile mismatch inside $\sim3\arcsec$, that is, a relatively strong light excess of the model in the very center and a deficit around it. Analysis of the components' profiles reveals that the point source is overrepresented as the fitting routine attempts to account for the extended (resolved) spheroidal flux, and that the central disk itself has unexpectedly high axial ratio -- almost as high as the axial ratio of the ``bulge'', and higher than observed from the image -- indicating that it, too, is probably accounting for part of the bulge. We therefore introduce another central \sersic\ component, which converges to $n\sim3.3$ and high axial ratio. We accordingly identify this as the bulge component, but unfortunately now the interpretation of the first \sersic\ component (the earlier model's ``bulge'') is unclear. It is not aligned with the galaxy major axis, but rotated towards the minor-axis structure in the disk. It may represent a bar, but equally likely a part of the spiral structure. We also suspect that it is still degenerate to some degree with the underlying main disk and even the bulge, as its axial ratio is higher than the disk's. In summary, we cannot draw reliable conclusions with respect to the nature of this component, but tentatively term it ``bar'' to avoid confusion with the other, more unequivocally identified, components. \\

\begin{figure*}
\begin{center}
\includegraphics[width=\linewidth]{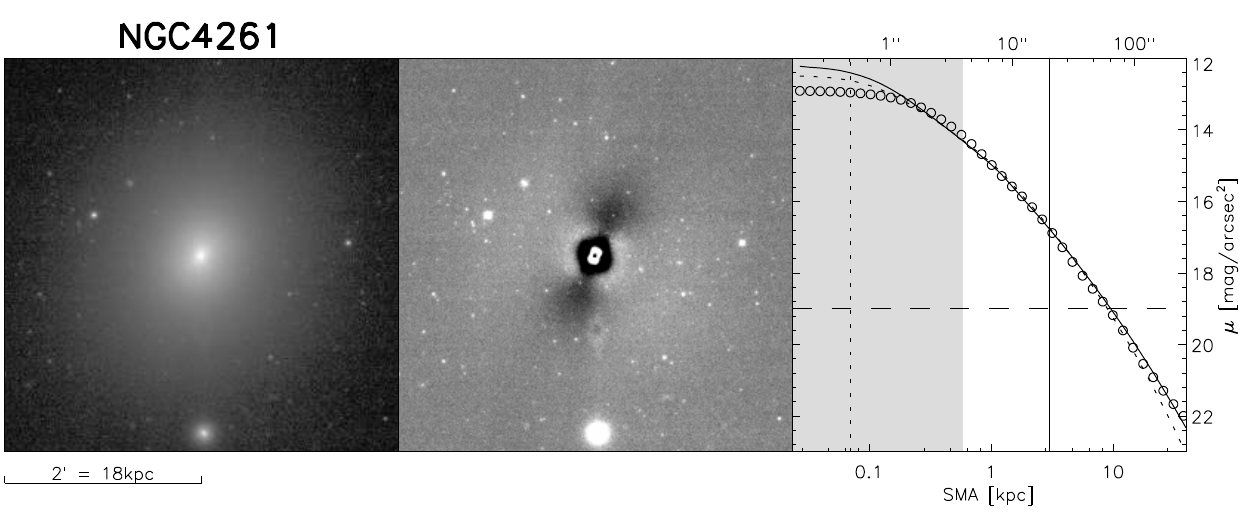}
\caption{As Figure \ref{fig:IC1459}.}
\label{fig:NGC4261}
\end{center}
\end{figure*}

{\bf NGC4261} (Fig. \ref{fig:NGC4261}) is a giant elliptical galaxy. The residual image shows at least two peculiarities: a high-ellipticity region immediately surrounding the core (between $\sim3\arcsec$ and $15\arcsec$), and strong boxiness between $\sim3\arcsec$ and $45\arcsec$, while the isophotes are perfectly elliptical at all other radii. Interestingly, the boxiness is most pronounced at the same radius where the ellipticity peaks. The central structure in the residual image resembles a bar. We tested various multi-component models, but none provided an improvement over a single-\sersic\ model (with the core masked), which we therefore adopt. \\

\begin{figure*}
\begin{center}
\includegraphics[width=\linewidth]{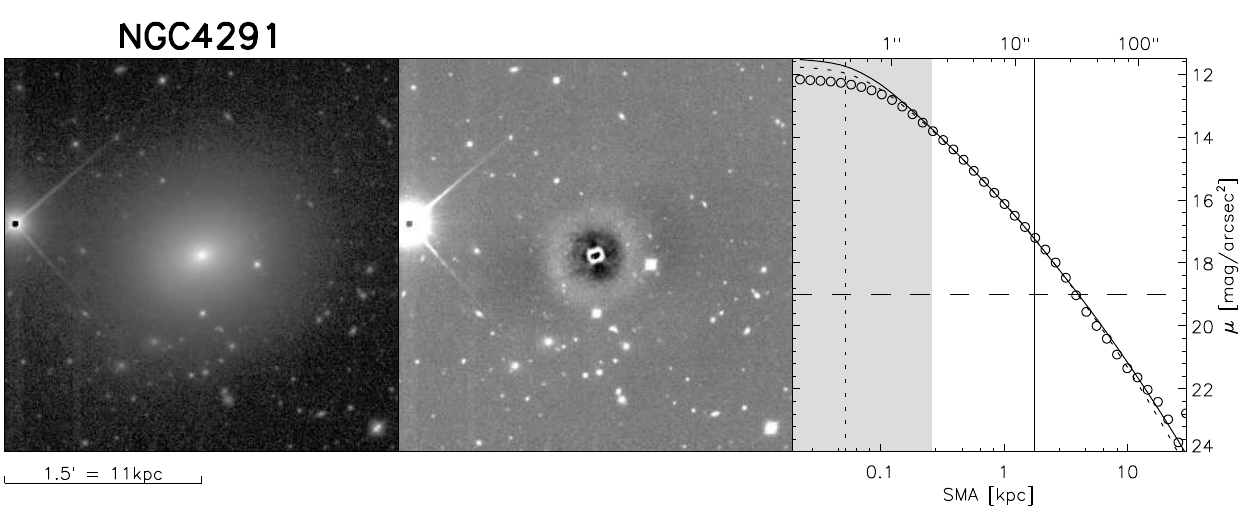}
\caption{As Figure \ref{fig:IC1459}.}
\label{fig:NGC4291}
\end{center}
\end{figure*}

{\bf NGC4291} (Fig. \ref{fig:NGC4291}) is a relatively small ($R_e\sim2\kpc$) elliptical with a pronounced  core. The ellipticity changes and peaks at $\sim15\arcsec$, followed by a minimum at $\sim45\arcsec$, to then increase again at larger radii. The second ellipticity peak may be caused by a nearby bright star, despite masking. The isophotes are very closely elliptical. As was the case for NGC4261, the residuals from a single-\sersic\ model produce some peculiar structure, in this case a ring between $\sim15\arcsec$ and $20\arcsec$, which lessens, but does not entirely disappear, when the core is masked. We tested various multi-component models, but none provided an improvement over a single-\sersic\ model (with the core masked), which we therefore adopt. \\

\begin{figure*}
\begin{center}
\includegraphics[width=\linewidth]{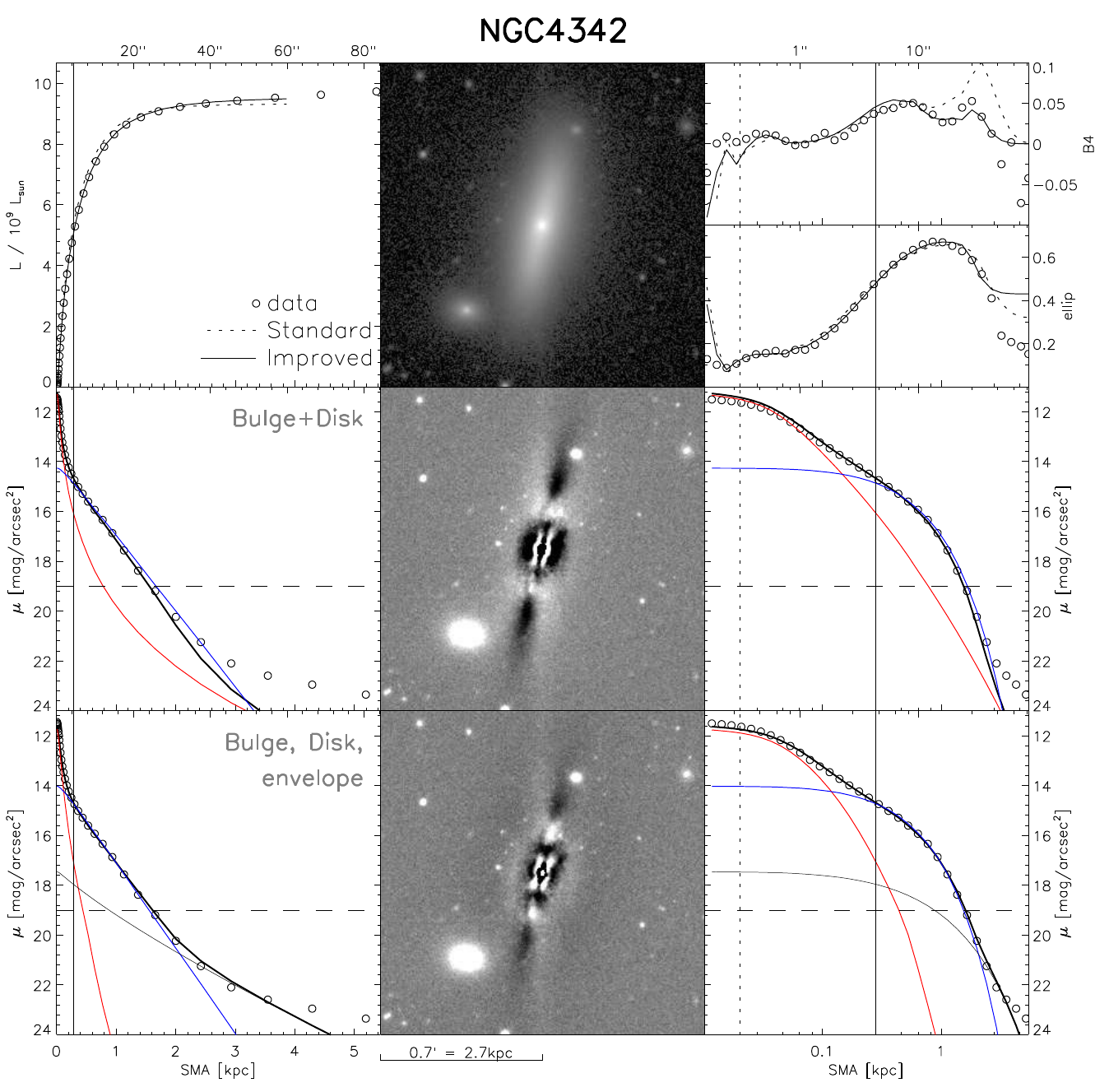}
\caption{As Figure \ref{fig:NGC0821}.}
\label{fig:NGC4342}
\end{center}
\end{figure*}

{\bf NGC4342} (Fig. \ref{fig:NGC4342}) is a lenticular galaxy seen at high inclination. As other edge-on lenticulars, it presents challenges for the fitting routine, further compounded in this case by the small apparent size of the galaxy and the resulting relative lack of information in the data, which hinders a robust fit of the multiple additional components that are required to obtain acceptable residuals.

Between $\sim7\arcsec$ and $35\arcsec$, NGC4342's profile is largely exponential, while the isophotes are disky at most radii, signaling the presence of a large-scale disk. The surface brightness profile may also suggest the presence of an outer envelope or halo. The residuals from a standard bulge+disk model are very pronounced, suggesting that the best fit disk is too flattened, and that the flux in the central regions is overestimated. Using a \sersic\ instead of an exponential profile for the disk produces a better fit. 

We tried to account for the mismatch by introducing a (\sersic) envelope (see \S\ref{subsec:beyond_b+d}), a nucleus, an inner truncation of the disk, and experimented with implementing a bulge ellipticity gradient by means of symmetric truncation functions.
                                                                                                                                                                                                                                                                                                                                                                                                                                                                                                                                                                                                                                                                                                                      
As for other highly flattened E/S0 galaxies, the ``envelope" seems to be required if the disk is assumed to be exponential (see the bottom-left panel of Figure \ref{fig:NGC4342}). Additionally,  we add a nuclear component to account for the mismatch of the model in the innermost arcsecond. This significantly complicates the analysis: if the nuclear component is fit with a \sersic\ with size and axial ratio that are allowed to vary freely, it converges to  values typical of a bulge, while the component that represented the bulge in the Bulge+Disk+Envelope model converges to parameters intermediate between a bar and a disk. We allow generalized ellipses in the models, and find boxy isophotes to fit the best. In spite of this, the octupole residuals seen in the 3-component model at intermediate radii ($\sim4\arcsec$)  are unaffected. 

Reducing the residuals even further might require the addition of a bar and/or a (secondary) embedded disk.
However, doing so produces fits that are highly degenerate and do not converge. While these components  may not be spurious, their interpretation is problematic. We therefore feel that the Bulge+Disk+Envelope model is to be preferred, although it should be noted that the bulge magnitude, as well as other parameters, is likely to be very uncertain. \\

\begin{figure*}
\begin{center}
\includegraphics[width=\linewidth]{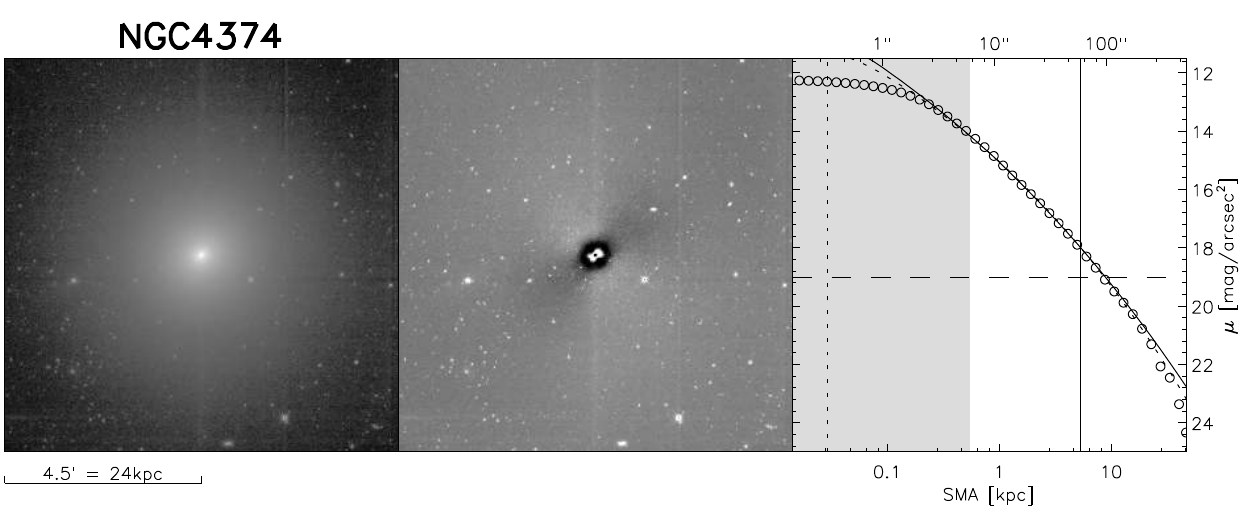}
\caption{As Figure \ref{fig:IC1459}.}
\label{fig:NGC4374}
\end{center}
\end{figure*}

{\bf NGC4374} (M84, Fig. \ref{fig:NGC4374}) is a giant elliptical galaxy with a large core. The ellipticity is low ($\sim0.2$) at small radii and decreases to almost zero outside $\sim100\arcsec$. The isophotes are boxy throughout. There are no signs of any sub-components, and we therefore model this galaxy with a single-\sersic\ profile, after masking the core. \\

\begin{figure*}
\begin{center}
\includegraphics[width=\linewidth]{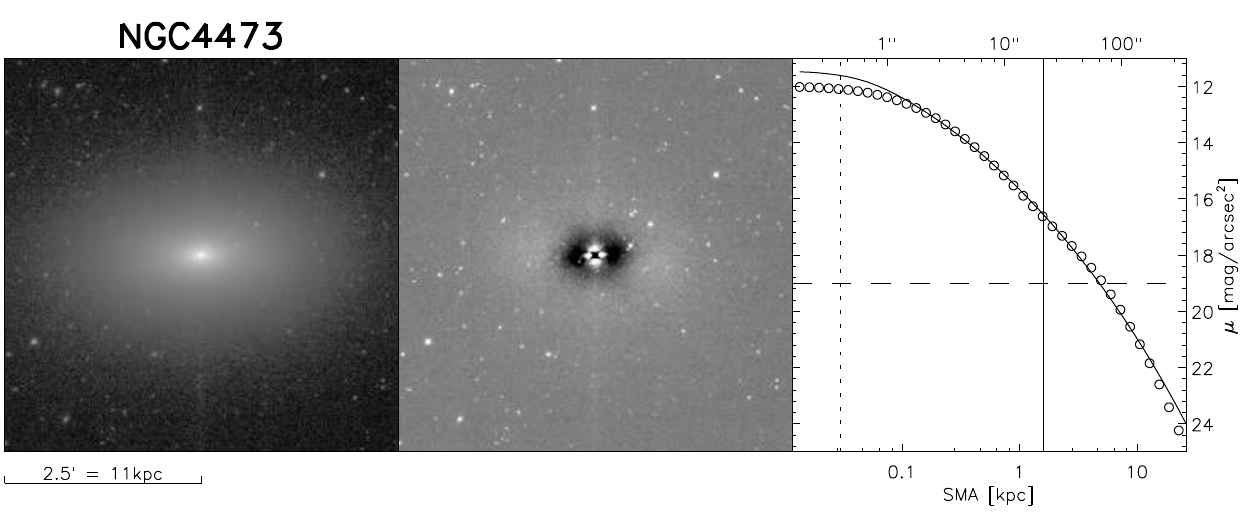}
\caption{As figure \ref{fig:CygA}.}
\label{fig:NGC4473}
\end{center}
\end{figure*}

{\bf NGC4473} (Fig. \ref{fig:NGC4473}) is a flattened ($\epsilon\approx0.5$) early-type galaxy with disky isophotes. Its surface-brightness profile is not perfectly fit by a \sersic\ model, and the ellipticity profile peaks twice, at $\sim3\arcsec$ and $\sim45\arcsec$, possibly suggesting the presence of two disks (we note that a disk was also suggested by \citealt{Ferrarese_etal06b} based on HST images). An inner disk is in fact clearly seen in the residuals from a single-\sersic\ model. Yet, we were unable to fit a model that includes an \textit{exponential} disk, whether only in conjunction with a bulge or a bulge plus a secondary disk. We therefore adopt a single-\sersic\ model as the best model for this galaxy, although this overestimates the flux in  the central regions, likely due to the presence of  the (unaccounted-for) substructure mentioned above.\\

\begin{figure*}
\begin{center}
\includegraphics[width=\linewidth]{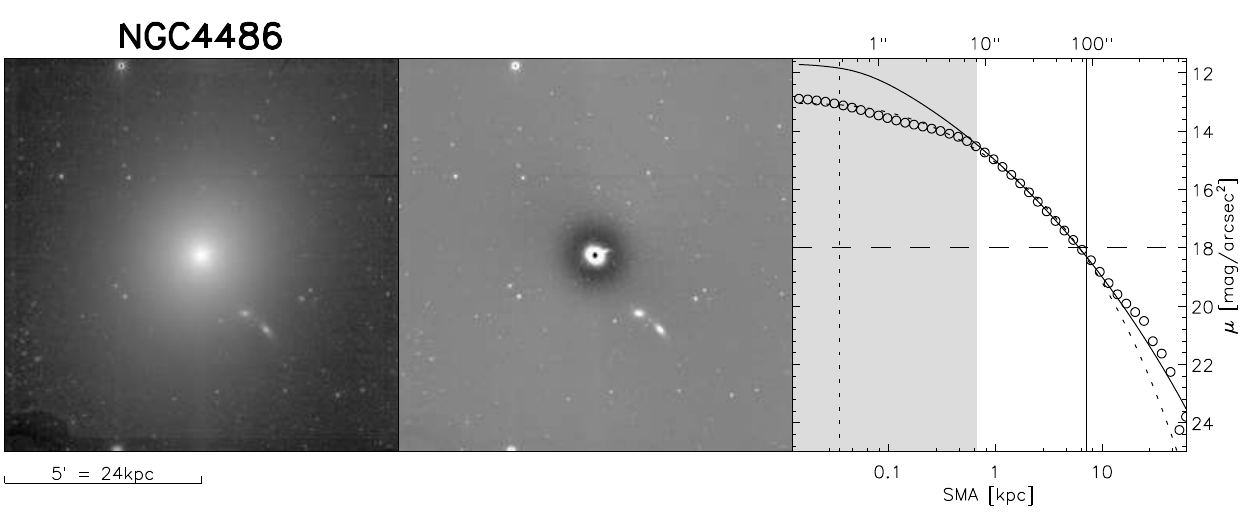}
\caption{As Figure \ref{fig:IC1459}.}
\label{fig:NGC4486}
\end{center}
\end{figure*}

{\bf NGC4486} (M87, Fig. \ref{fig:NGC4486}) is a cD galaxy with a large ($\sim8\arcsec$) core. Within our sample, it is also the galaxy with the largest (apparent) effective radius, $122\arcsec$ as measured from the improved model in which the core is  masked. The non-thermal jet as well as central AGN are clearly visible, and were also masked during the analysis.  The radial surface brightness profile might have a ``knee'' with subsequent steeper decline at very large radii ($\approx400\arcsec$), but this observation is tentative since the profile becomes noisy and  background uncertainties are significant in this region. Regardless, since this possible change in the profile is present at very large radii and low surface brightness, it does not affect the fit. \\

\begin{figure*}
\begin{center}
\includegraphics[width=\linewidth]{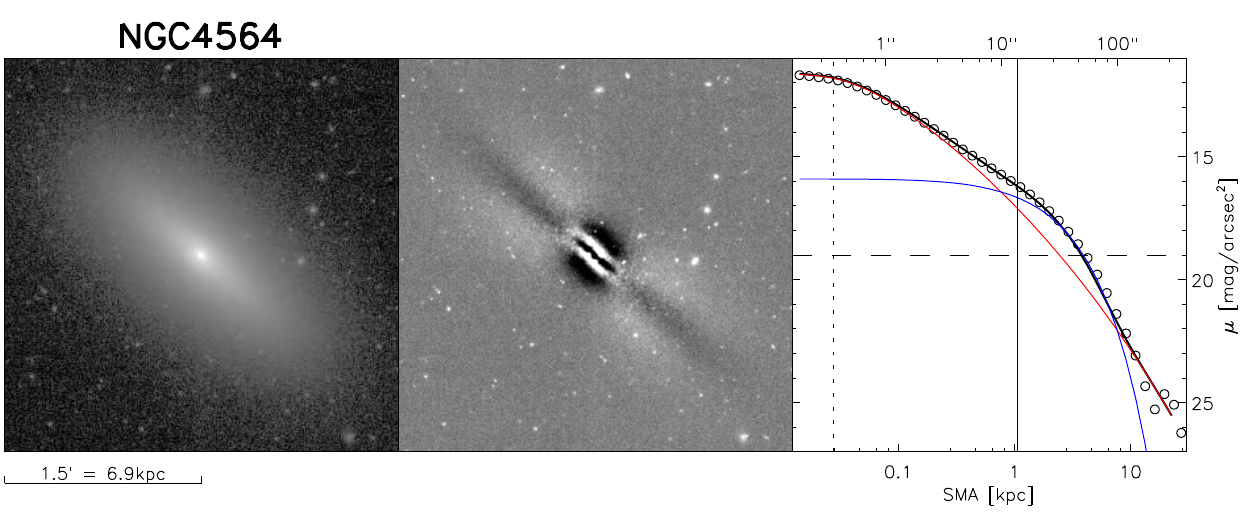}
\caption{As figure \ref{fig:NGC1023}.}
\label{fig:NGC4564}
\end{center}
\end{figure*}

{\bf NGC4564} (Fig. \ref{fig:NGC4564}) is an early-type galaxy with high average ellipticity. The presence of a disk is suggested by  the exponential decline in the surface brightness profile between $\sim30\arcsec$ and $\sim75\arcsec$, combined with the increase in ellipticity and diskiness of the isophotes in the same region. However, like other elongated early-types in our sample, it is not perfectly fit by a simple bulge plus exponential disk model: such model does not fully account for the ellipticity gradient, and a bar-like structure can be seen in the residuals in the inner regions. Tests  using \galfit's ``edge disk" profile produce little improvement, as do 2nd- and 4th-order harmonic modes added to the ellipsoidal shape. If the \sersic\ index of the disk is allowed to vary, it converges to $n_d \sim 1.2$, and residuals are relatively moderate. Furthermore, introduction of a third component (a bar  or a large-scale``envelope") reduces the residuals only slightly. We therefore  adopt the standard bulge+disk model for this galaxy. \\
 
\begin{figure*}
\begin{center}
\includegraphics[width=\linewidth]{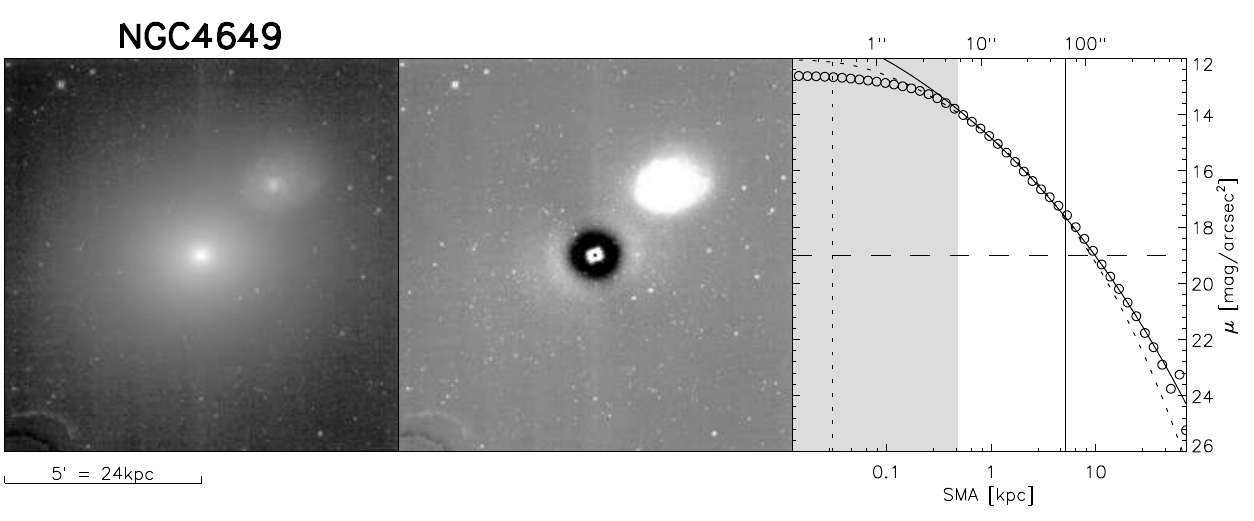}
\caption{As Figure \ref{fig:IC1459}.}
\label{fig:NGC4649}
\end{center}
\end{figure*}

{\bf NGC4649} (M60, Fig. \ref{fig:NGC4649}) is an elliptical with a core, which was masked during the fits. The companion spiral galaxy, NGC 4647, was also masked; residual (unmasked) light from the outer disk of NGC4647 is unlikely to bias the fit given its low surface brightness compared to NGC4649.  \\

{\bf NGC4697} (Fig. \ref{fig:CygA}) is an early-type galaxy with a very elongated embedded disk. The ellipticity and diskiness  of the isophotes show two distinct peaks. The diskiness of the isophotes at small and intermediate radii is also clearly visible in the images.

The residuals from a single-\sersic\ model support the existence of two disks. The \sersic\ model is biased to a large ellipticity by the embedded disk, making it too elongated to match the data at large radii. There is a similar small, but bright, residual structure at the smallest radii ($<10\arcsec$).

If an exponential disk is added, it converges to high ellipticity ($\epsilon\sim0.93$) and low total magnitude ($m_d=11.34\mg$ compared to $m_b=5.69\mg$). When the \sersic\ index of the disk is allowed as a free parameter, it  converges to $n \sim 5.9$. Our best improved model therefore features a second, inner, \sersic\ component (which converges to ellipticity intermediate  between disk and bulge and has smaller size -- $R_e\sim 7"$ --  than both the disk  -- $R_{e,d}\sim 38\arcsec$ -- and the bulge -- $R_{e,b}\sim 130\arcsec$), and a nuclear point source (which has the effect of reducing the \sersic\ index of the previous component, which is otherwise biased to $n \sim 3.1$ while trying to absorb some of the nuclear light). The inner \sersic\ component might be a small-scale disk due to its high ellipticity, although its profile ($n\sim2.1$) is not exponential. Alternatively, it might be part of the bulge, and we therefore take this possibility into account by including it in our estimate of the  ``maximal bulge" magnitude, $M_{b,max}$. \\

\begin{figure*}
\begin{center}
\includegraphics[width=\linewidth]{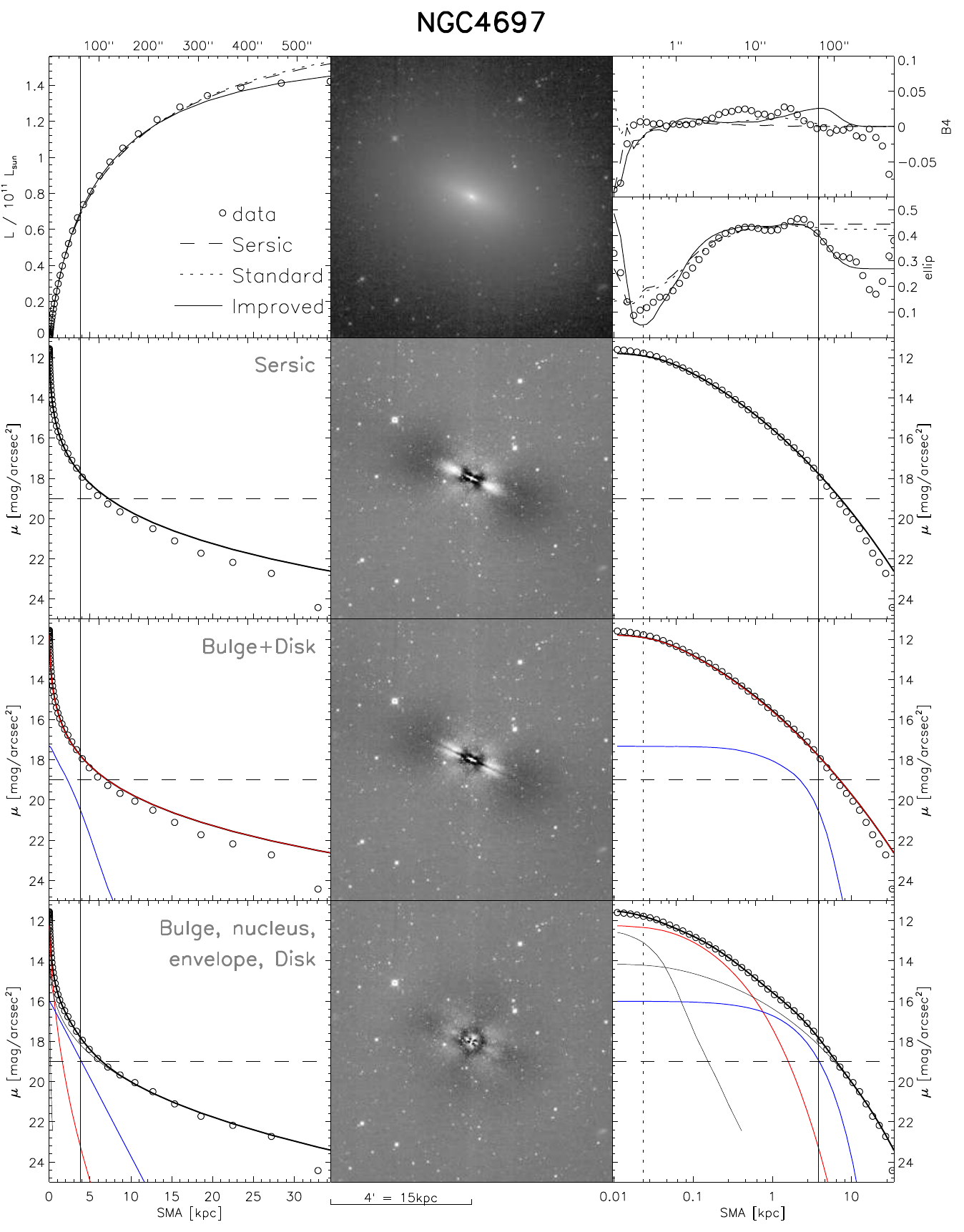}
\caption{As Figure \ref{fig:NGC0821}.}
\label{fig:NGC4697}
\end{center}
\end{figure*}

\begin{figure*}
\begin{center}
\includegraphics[width=\linewidth]{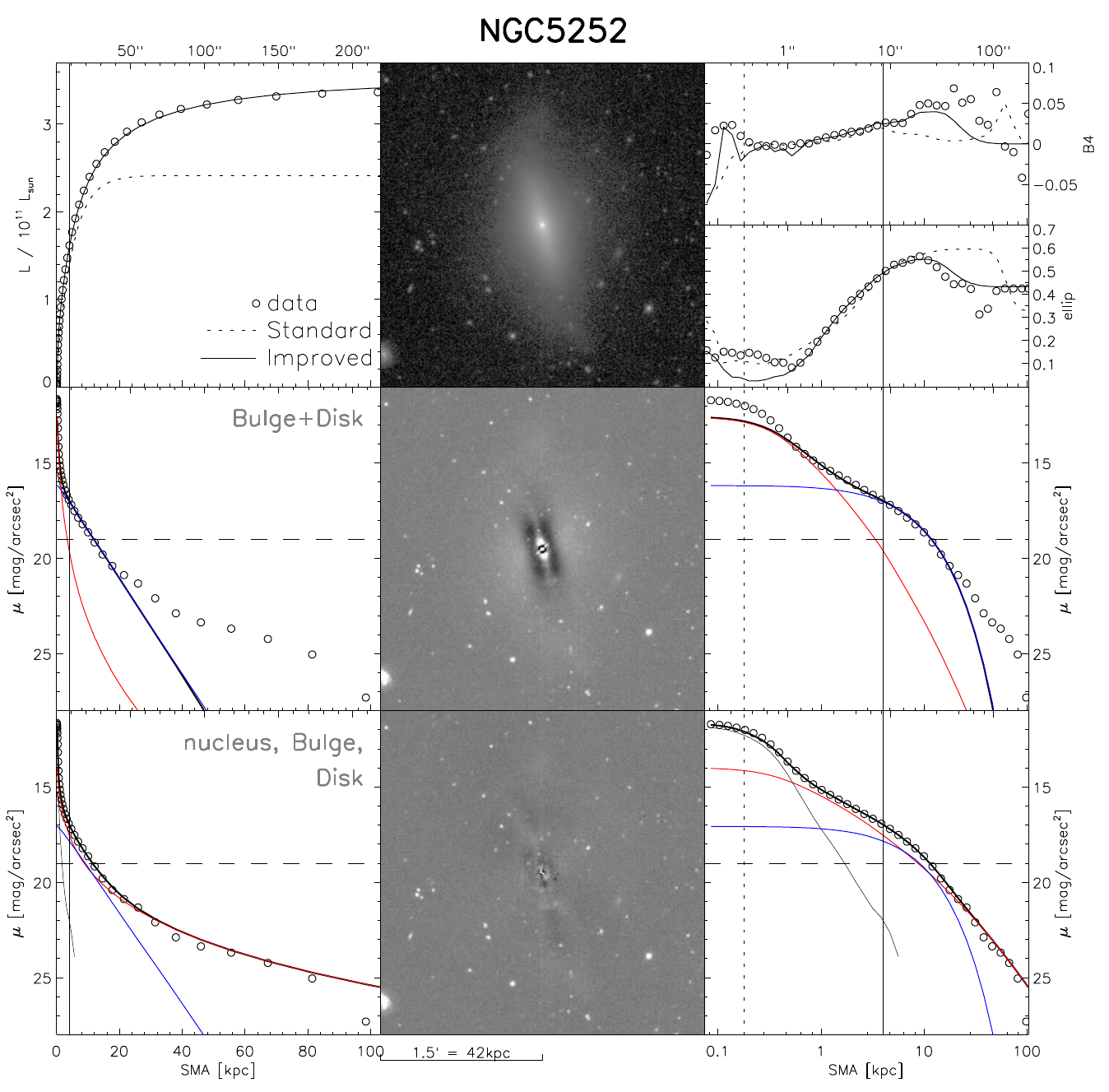}
\caption{As Figure \ref{fig:NGC1023}.}
\label{fig:NGC5252}
\end{center}
\end{figure*}

{\bf NGC5252} (Fig. \ref{fig:NGC5252}) is a lenticular galaxy that harbours a prominent AGN. If unaccounted for, the AGN significantly biases the parameters of the standard  bulge+disk models, to the point that the \sersic\ index of the bulge becomes larger than 10. For this reason, in the standard model we fix the \sersic\ index of the bulge to $n=4$, although this results in a best-fit with an unrealistically small bulge component. The unresolved nucleus is included in the improved model; doing so produces a very good fit. \\
 
\begin{figure*}
\begin{center}
\includegraphics[width=\linewidth]{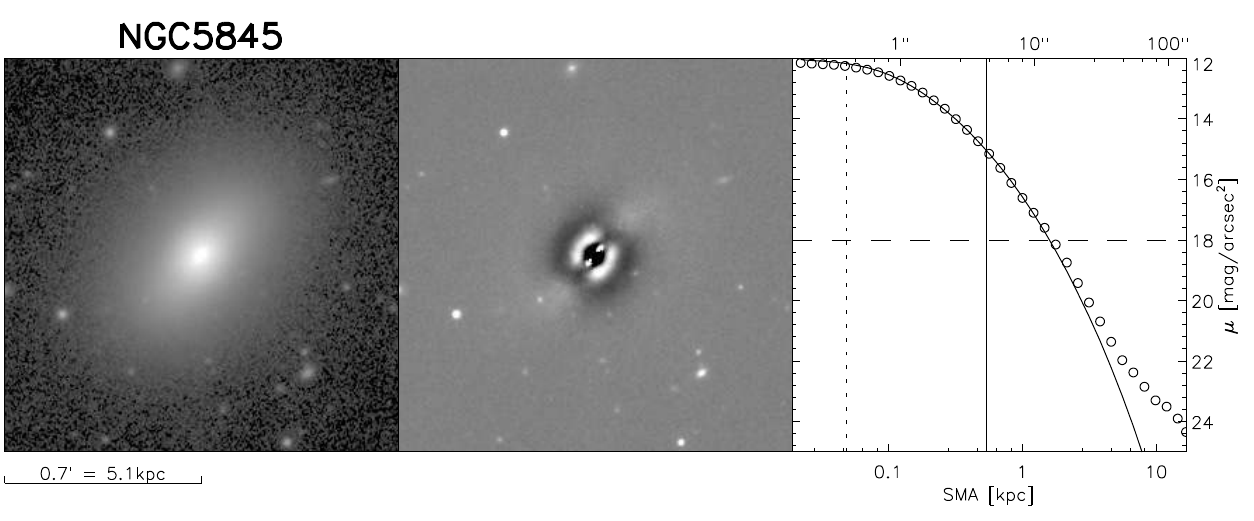}
\caption{As Figure \ref{fig:CygA}.}
\label{fig:NGC5845}
\end{center}
\end{figure*}

{\bf NGC5845} (Fig. \ref{fig:NGC5845}) is a difficult galaxy to classify and model, both because it  is one of the smallest in our sample, implying relatively low resolution, and because a large and bright nearby elliptical galaxy (NGC5846) requires a separate model. NGC5845's profile shows a distinct inflection at $\sim35\arcsec$, where the surface brightness is low ($\sim22\magarcsec$), and declines slowly beyond this radius. This extended outer profile can \textit{not} be attributed to NGC5846, as it is still present after modeling the latter. The ellipticity, isophotal shape, and position angle  show numerous fluctuations with radius, at least some of which are likely due to real sub-structure.

The inflection at $\sim35\arcsec$ may be caused by several, not mutually exclusive, components: an inner embedded disk (a thesis supported by the large ellipticity and diskiness at smaller radii);  a large-scale ring; or a distinct outer envelope/halo. None of these scenarios, however, can account for the isophotal twist, which may result from a triaxial spheroid or a bar (for which we however do otherwise not find conclusive evidence). 

Although the residuals  from both a single-\sersic\ and a bulge+disk model suggest a small-scale edge-on disk, the data do not allow us to discriminate between the different scenarios. A bulge+disk model produces large residuals that are barely an improvement over those resulting from a single-\sersic\ model. The disk component does not converge to exponential when the \sersic\ index is allowed to vary, and has low ellipticity (lower than the bulge). Introducing additional components did not allow for unambiguous physical interpretation, either.

In conclusion, while this galaxy shows clear evidence of sub-structure, and the residuals from a single-\sersic\ model are significant, we did not succeed in finding a unique solution with exponential disk, low residuals and a clear interpretation of all components. We therefore adopt a single-\sersic\ model for this galaxy, but  caution that such a model underestimates the flux at radii outside $\sim30\arcsec$. \\

\begin{figure*}
\begin{center}
\includegraphics[width=\linewidth]{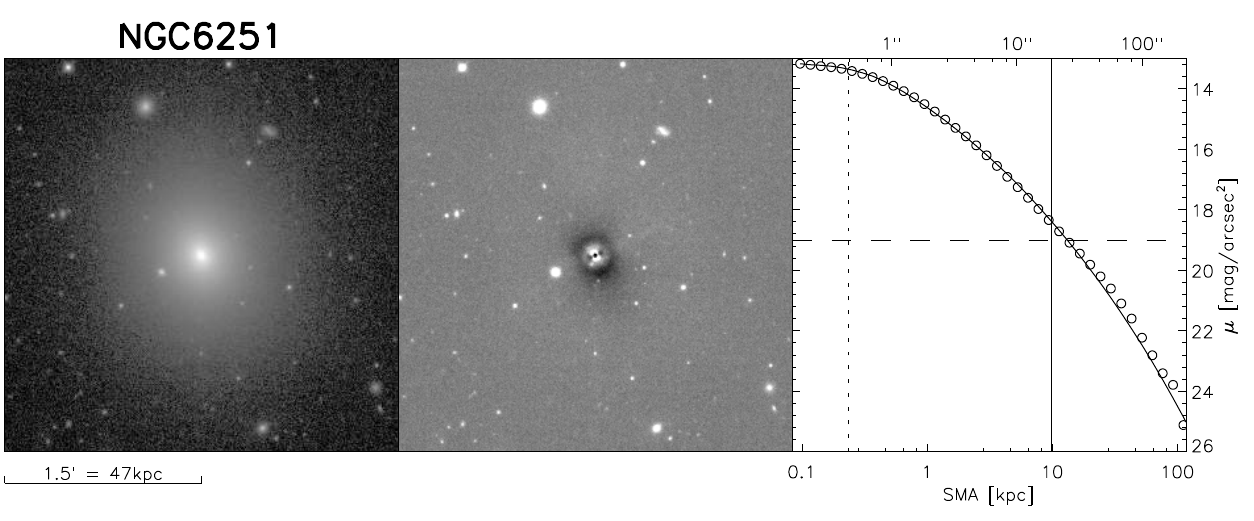}
\caption{As Figure \ref{fig:CygA}.}
\label{fig:NGC6251}
\end{center}
\end{figure*}

{\bf NGC6251} (Fig. \ref{fig:NGC6251}) is a luminous early-type galaxy, although its core cannot be resolved in our data. The slight ``knee" in the profile at $\sim50\arcsec$ may indicate an outer disk, and a fit that includes a disk indeed produces a decrease in the $\chi^2$. Yet, neither the ellipticity nor the isophotal shape hint at the presence of a disk, and the residual likewise not suggest significant mismatch at intermediate or large radii. We conclude that the data do not justify introduction of a disk component and adopt a single-\sersic\ model for this galaxy. \\

\begin{figure*}
\begin{center}
\includegraphics[width=\linewidth]{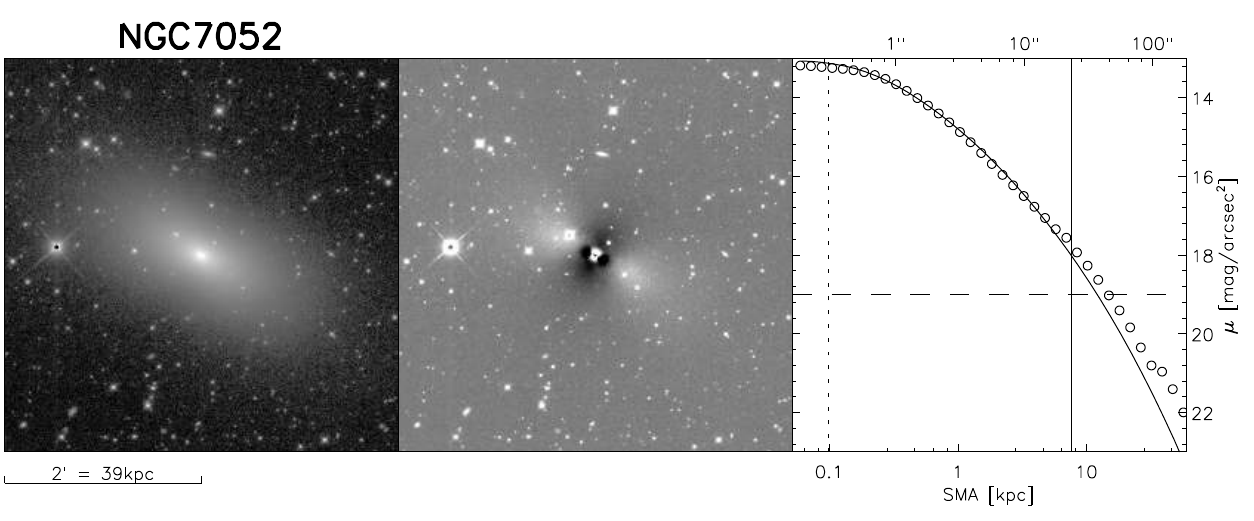}
\caption{As Figure \ref{fig:CygA}.}
\label{fig:NGC7052}
\end{center}
\end{figure*}

{\bf NGC7052} (Fig. \ref{fig:NGC7052}) is a very elongated elliptical galaxy.  Stellar foreground contamination is severe: as was the case for Cyg A, creating the mask required a ``second-pass" using the residual image. The galaxy shows a strong ellipticity gradient ($e\approx0.2$ in the center, and $\approx0.6$ beyond $\sim30\arcsec$). Unusually, despite the relatively high ellipticity, the isophotes are predominantly boxy, and very  strongly so  outside $\sim30\arcsec$. A disk is not visible in the images, but the residuals decrease significantly when a second component is added. The \sersic\ index of the latter is $>2$, therefore this component is not a classical exponential disk. Indeed whether it is {\it real} is subject to debate: the residuals seen in the single-\sersic\ model could easily be caused by the ellipticity gradient [we tried to account for this by using a truncation function to join two \sersic\ components with shared parameters, except for ellipticity and isophotal shape, but the model converged very slowly and did not significantly improve the residuals]. Therefore, we choose to adopt as the best fit the one obtained using  a single-\sersic\ profile. \\
 
\begin{figure*}
\begin{center}
\includegraphics[width=\linewidth]{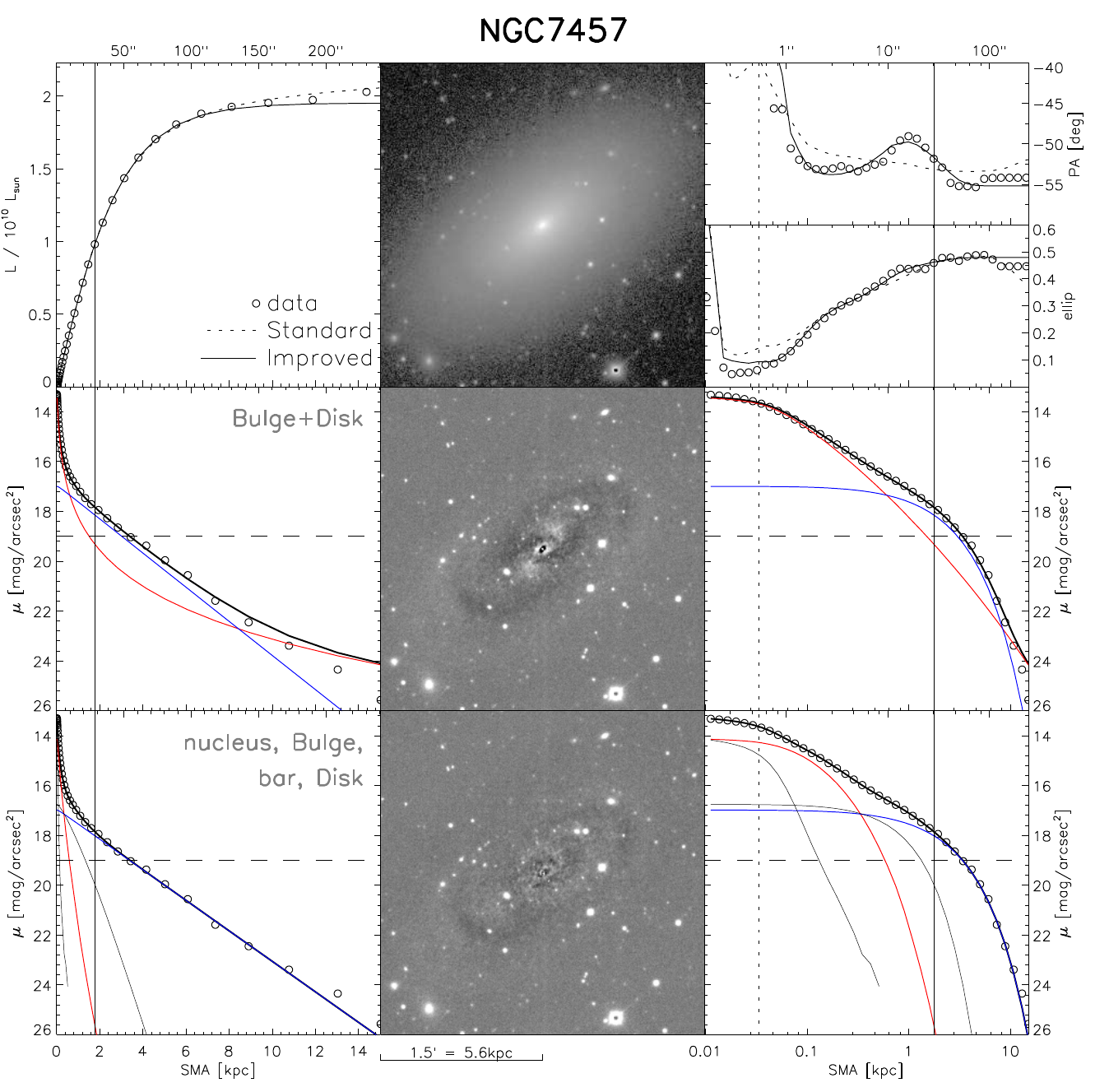}
\caption{As Figure \ref{fig:NGC1023}.}
\label{fig:NGC7457}
\end{center}
\end{figure*}

{\bf NGC7457} (Fig. \ref{fig:NGC7457}) is a lenticular galaxy with a prominent large-scale disk. The residuals from a standard model are relatively good, but show signs of a bar, which is not aligned with the disk nor the bulge, and a nucleus. Additional justification for attempting a more complex model is given by the fact that the best fit bulge in a standard model is too extended and over-predicts the data at large radii
($\gtrsim120\arcsec$), where it dominates the surface brightness. On the contrary, the data shows a roughly exponential profile at these radii. The high \sersic\ index of the bulge in this model, $n=7.7$, is  likely biased by the presence of  the nucleus.

The improved model, which includes a central point source as well as a bar, converges to a bulge with a  lower \sersic\ index, and an exponential disk that dominates the light at large radii. The bar component in this model is substantial. Adding it allows us to reproduce the isophotal twist at intermediate radii, and leads to a disk profile that is close to exponential. The bar may be equally well modeled by a modified Ferrer profile or a \sersic\ profile. Interestingly, while the resulting $\chi^2$ is virtually the same in both cases, the  bulge parameters differ: when the bar is fitted with a Ferrer profile, the bulge is brighter and has \sersic\ index $n>2$, instead of $\sim1.5$ as in case of a bar with a \sersic\ profile. This highlights our overall conclusion that bulge parameters are quite sensitive to the details of the decomposition. For consistency with the other galaxies, we  adopt a \sersic\ profile for the bar. \\
 
\begin{figure*}
\begin{center}
\includegraphics[width=\linewidth]{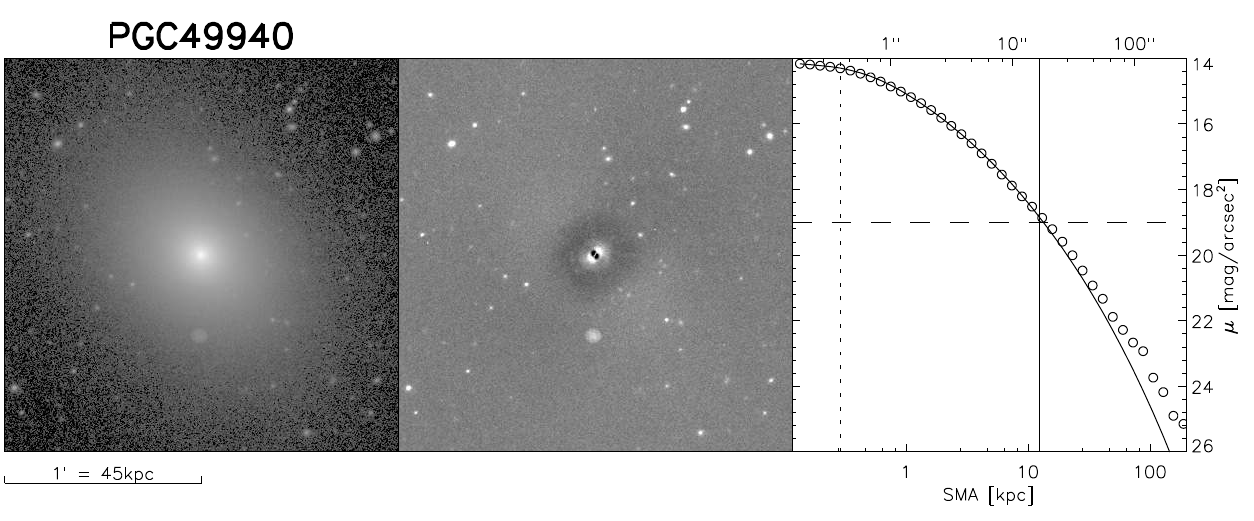}
\caption{As figure \ref{fig:CygA}.}
\label{fig:PGC49940}
\end{center}
\end{figure*}

{\bf PGC49940} (Fig. \ref{fig:PGC49940}) is a luminous elliptical galaxy at relatively large distance ($153\Mpc$). A core, if present, is not resolved. Although the \sersic\ profile mildly underestimates the flux at $R\gtrsim30\arcsec$, where $\mu\lesssim22\magarcsec$, we find no justification for introducing a large-scale disk.

\label{lastpage}

\end{document}

%% file: Mbh-Lbt_abbrev.tex
\newcommand{\galfit}{\textsc{galfit}}
\newcommand{\gfthree}{\textsc{galfit3}}
\newcommand{\sersic}{S\'{e}rsic}
\newcommand{\iraf}{\textsc{IRAF}}
\newcommand{\sex}{\textsc{SExtractor}}
\newcommand{\scamp}{\textsc{Scamp}}
\newcommand{\swarp}{\textsc{SWarp}}

\newcommand{\twomass}{\textsc{2MASS}}      
\newcommand{\ukidss}{\textsc{UKIDSS}}      

\newcommand{\smbh}{SMBH}
\newcommand{\mbh}{M_\bullet} 

\newcommand{\eps}{\epsilon} 
\newcommand{\mpc}{\,\mathrm{Mpc}}
\newcommand{\magarcsec}{\,\mathrm{mag\,arcsec^{-2}}}
\newcommand{\mg}{\,\mathrm{mag}}
\newcommand{\kpc}{\,\mathrm{kpc}}
\newcommand{\Mpc}{\,\mathrm{Mpc}}

\newcommand{\lsun}{\,L_\odot}

\newcommand{\bul}{_\mathrm{bul}}
\newcommand{\bstd}{_\mathrm{b,std}}

\newcommand{\bmin}{_\mathrm{b,min}}
\newcommand{\bmax}{_\mathrm{b,max}}
\newcommand{\sph}{_\mathrm{sph}}
\newcommand{\tot}{_\mathrm{tot}}
\newcommand{\ser}{_\mathrm{ser}}
\newcommand{\tstd}{_\mathrm{t,std}}
\newcommand{\timp}{_\mathrm{t,imp}}
\newcommand{\iso}{_\mathrm{24}}
\newcommand{\twom}{_\mathrm{t,2M}}
\newcommand{\mh}{_\mathrm{MH03}}
\newcommand{\vika}{_\mathrm{V12}}

\newcommand{\lbul}{L\bul}
\newcommand{\lbstd}{L\bstd}

\newcommand{\ltot}{L\tot}

\newcommand{\liso}{L\iso}

\newcommand{\mbstd}{m\bstd}

\newcommand{\mtstd}{m\tstd}

\newcommand{\miso}{m\iso}

\newcommand{\Mbul}{M\bul}
\newcommand{\Mbstd}{M\bstd}

\newcommand{\Mbmin}{M\bmin}
\newcommand{\Mbmax}{M\bmax}
\newcommand{\Msph}{M\sph}
\newcommand{\Mtot}{M\tot}
\newcommand{\Mser}{M\ser}
\newcommand{\Mtstd}{M\tstd}
\newcommand{\Mtimp}{M\timp}
\newcommand{\Miso}{M\iso}
\newcommand{\Mtwom}{M\twom}
\newcommand{\Mmh}{M\mh}
\newcommand{\Mv}{M\vika}

%% file: table_obs.tex
CygA & E & Ell & $36.88\pm0.24$ & 2 & 0.140 & 112$\times$18s & 2016s & no & 26.0 & 0.90 & 7.8 \\
IC1459 & E3 & Ell & $32.27\pm0.28$ & 1 & 0.006 & 20$\times$24s & 480s & yes & 25.0 & 0.57 & 2.1 \\
IC4296 & E & Ell & $33.53\pm0.16$ & 7 & 0.023 & 24$\times$18s & 432s & yes & 25.0 & 0.82 & 4.3 \\
NGC0221 & cE2 & Ell & $24.49\pm0.08$ & 1 & 0.023 & 16$\times$24s & 384s & yes & 24.0 & 0.68 & 4.3 \\
NGC0821 & E6 & Len & $31.85\pm0.17$ & 1 & 0.040 & 24$\times$24s & 576s & yes & 25.4 & 0.66 & 2.1 \\
NGC1023 & S0 & Len,bar & $30.23\pm0.16$ & 1 & 0.022 & 24$\times$24s & 576s & yes & 25.4 & 0.96 & 4.6 \\
NGC1300 & SBbc & Spi,bar & $31.39\pm0.24$ & 2 & 0.011 & 48$\times$24s & 1152s & yes & 25.9 & 0.49 & 2.0 \\
NGC1399 & E1pec & Ell & $31.63\pm0.06$ & 8 & 0.005 & 24$\times$24s & 576s & yes & 24.7 & 1.05 & 3.8 \\
NGC2748 & SAbc & Spi & $31.90\pm0.24$ & 2 & 0.010 & 40$\times$20s & 800s & no & 25.5 & 0.72 & 4.1 \\
NGC2778 & S0 & Len,bar & $31.74\pm0.30$ & 1 & 0.008 & 52$\times$24s & 1248s & no & 25.8 & 1.26 & 2.5 \\
NGC2787 & SB(r)0 & Len,bar & $29.31\pm0.26$ & 1 & 0.048 & 20$\times$20s & 400s & no & 25.2 & 0.70 & 4.9 \\
NGC3115 & S0 & Len & $29.87\pm0.09$ & 1 & 0.017 & 24$\times$24s & 576s & yes & 24.8 & 1.08 & 2.1 \\
NGC3227 & SAB(s)pec & Spi,bar & $31.13\pm0.24$ & 2 & 0.008 & 54$\times$24s & 1296s & yes & 25.6 & 1.39 & 2.4 \\
NGC3245 & SA0(r)? & Len,bar & $31.54\pm0.20$ & 1 & 0.009 & 48$\times$24s & 1152s & no & 25.7 & 0.90 & 2.0 \\
NGC3377 & E5 & Len & $30.19\pm0.09$ & 1 & 0.013 & 24$\times$24s & 576s & yes & 25.0 & 0.87 & 2.7 \\
NGC3379 & E1 & Ell & $30.06\pm0.11$ & 1 & 0.009 & 24$\times$24s & 576s & yes & 25.0 & 0.99 & 2.6 \\
NGC3384 & SB(s)0- & Len,bar & $30.26\pm0.14$ & 1 & 0.010 & 24$\times$24s & 576s & yes & 25.0 & 0.94 & 2.3 \\
NGC3608 & E2 & Ell & $31.74\pm0.14$ & 1 & 0.008 & 48$\times$18s & 864s & no & 25.5 & 0.67 & 2.2 \\
NGC3998 & SA(r)0 & Len,bar & $30.71\pm0.19$ & 1 & 0.006 & 24$\times$18s & 432s & yes & 24.9 & 0.75 & 2.9 \\
NGC4258 & SAB(s)bc & Spi,bar & $29.29\pm0.09$ & 4 & 0.006 & 24$\times$18s & 432s & yes & 24.5 & 0.83 & 3.5 \\
NGC4261 & E2 & Ell & $32.44\pm0.19$ & 1 & 0.007 & 24$\times$18s & 432s & yes & 24.8 & 0.93 & 3.7 \\
NGC4291 & E3 & Ell & $32.03\pm0.32$ & 1 & 0.013 & 48$\times$18s & 864s & no & 25.6 & 0.84 & 3.8 \\
NGC4342 & S0 & Len & $30.62\pm0.25$ & 2 & 0.008 & 48$\times$18s & 864s & no & 25.7 & 0.62 & 2.0 \\
NGC4374 & E1 & Ell & $31.34\pm0.07$ & 5 & 0.015 & 24$\times$18s & 432s & yes & 24.7 & 0.65 & 4.0 \\
NGC4473 & E5 & Ell & $30.92\pm0.07$ & 5 & 0.010 & 24$\times$18s & 432s & yes & 24.8 & 0.80 & 4.2 \\
NGC4486 & E0pec & Ell & $31.11\pm0.08$ & 5 & 0.008 & 24$\times$18s & 432s & yes & 24.8 & 0.87 & 2.5 \\
NGC4564 & S0 & Len & $31.01\pm0.07$ & 5 & 0.013 & 24$\times$18s & 432s & yes & 25.2 & 0.71 & 2.4 \\
NGC4649 & E2 & Ell & $31.08\pm0.08$ & 5 & 0.010 & 24$\times$18s & 432s & yes & 24.8 & 0.76 & 5.7 \\
NGC4697 & E6 & Len & $30.49\pm0.06$ & 1 & 0.011 & 26$\times$18s & 468s & yes & 24.8 & 0.76 & 1.7 \\
NGC5252 & S0 & Len & $34.94\pm0.24$ & 2 & 0.012 & 48$\times$18s & 864s & no & 25.4 & 0.76 & 2.1 \\
NGC5845 & E* & Ell & $32.01\pm0.21$ & 1 & 0.020 & 48$\times$18s & 864s & no & 25.7 & 0.80 & 2.0 \\
NGC6251 & E & Ell & $35.15\pm0.24$ & 2 & 0.032 & 48$\times$18s & 864s & no & 25.5 & 0.90 & 3.3 \\
NGC7052 & E & Ell & $34.15\pm0.24$ & 2 & 0.044 & 53$\times$18s & 954s & no & 25.6 & 0.60 & 2.3 \\
NGC7457 & SA(rs)0- & Len,bar & $30.55\pm0.21$ & 1 & 0.019 & 40$\times$20s & 800s & no & 25.5 & 1.07 & 3.5 \\
PGC49940 & E & Ell & $35.93\pm0.24$ & 2 & 0.024 & 96$\times$18s & 1728s & no & 26.1 & 0.78 & 3.2 \\

%% file: table_galfits.tex
CygA & E & 9.91 & -- & 18.4 & -- & 2.36 & -- & -- & -- & -- & -- & -- & -- & -- & -- \\
IC1459 & E3 & 6.43 & 6.27 & 46.2 & 62.4 & 6.5 & 8.25 & -- & -- & -- & -- & -- & -- & -- & -- \\
IC4296 & E & 7.13 & 6.68 & 39.9 & 97.8 & 5 & 8.24 & -- & -- & -- & -- & -- & -- & -- & -- \\
NGC0221 & cE2 & 5.01 & -- & 35.7 & -- & 3.4 & -- & -- & -- & -- & -- & -- & -- & -- & -- \\
NGC0821 & E6 & 7.36 & 9.41 & 56.1 & 3.81 & 7.15 & 3.13 & 12.44 & 11.29 & 13.1 & 12.2 & \parbox[c][0.9cm][c]{1cm}{ 7.89 \\ (env)} & -- & -- & -- \\
NGC1023 & S0 & 6.87 & 7.31 & 16.6 & 9.6 & 3.55 & 3.1 & 6.82 & 6.65 & 60 & 64.2 & \parbox[c][0.9cm][c]{1cm}{ 8.65 \\ (bar)} & -- & -- & -- \\
NGC1300 & SBbc & 9.86 & 9.55 & 4.08 & 10.4 & 1.34 & 4.3 & 7.40 & 8.03 & 64.8 & 65.4 & \parbox[c][0.9cm][c]{1cm}{13.86 \\ (psf)} & \parbox[c][0.9cm][c]{1cm}{11.30 \\ (idisk)} & \parbox[c][0.9cm][c]{1cm}{ 9.63 \\ (bar)} & \parbox[c][0.9cm][c]{1cm}{ 8.74 \\ (spir)} \\
NGC1399 & E1pec & 5.95 & 5.41 & 51.6 & 154 & 5.25 & 11.1 & -- & -- & -- & -- & -- & -- & -- & -- \\
NGC2748 & SAbc & 10.08 & 9.79 & 10.5 & 12.6 & 3.19 & 3.09 & 8.99 & 9.80 & 18.1 & 18.8 & \parbox[c][0.9cm][c]{1cm}{ 9.98 \\ (spir)} & -- & -- & -- \\
NGC2778 & S0 & 10.18 & 10.46 & 4.14 & 2.75 & 4.6 & 3.98 & 10.16 & 9.99 & 11 & 10.7 & \parbox[c][0.9cm][c]{1cm}{13.83 \\ (bar)} & -- & -- & -- \\
NGC2787 & SB(r)0 & 8.26 & 7.65 & 7.8 & 14.3 & 1.53 & 2.77 & 7.69 & 8.68 & 24.7 & 25.9 & \parbox[c][0.9cm][c]{1cm}{10.30 \\ (bar)} & \parbox[c][0.9cm][c]{1cm}{10.22 \\ (idisk)} & \parbox[c][0.9cm][c]{1cm}{12.69 \\ (psf)} & -- \\
NGC3115 & S0 & 5.52 & 7.92 & 75 & 3.9 & 6.81 & 3.01 & 8.34 & 7.56 & 23.1 & 10.3 & \parbox[c][0.9cm][c]{1cm}{ 6.01 \\ (env)} & -- & -- & -- \\
NGC3227 & SAB(s)pec & 9.02 & 9.54 & 2.04 & 0.717 & 11.7 & 4.08 & 7.83 & 7.78 & 33.3 & 34.5 & \parbox[c][0.9cm][c]{1cm}{10.32 \\ (bar)} & -- & -- & -- \\
NGC3245 & SA0(r)? & 8.94 & 9.50 & 3.54 & 1.95 & 2.27 & 1.6 & 8.24 & 8.20 & 20.4 & 20.5 & \parbox[c][0.9cm][c]{1cm}{ 9.91 \\ (bar)} & -- & -- & -- \\
NGC3377 & E5 & 7.13 & 8.34 & 56.7 & 10.1 & 5.27 & 6.04 & 10.49 & 8.87 & 1.12 & 16.5 & \parbox[c][0.9cm][c]{1cm}{11.75 \\ (idisk)} & \parbox[c][0.9cm][c]{1cm}{ 8.19 \\ (env)} & -- & -- \\
NGC3379 & E1 & 5.78 & 5.52 & 57.3 & 96.3 & 6.45 & 9.28 & -- & -- & -- & -- & -- & -- & -- & -- \\
NGC3384 & SB(s)0- & 7.60 & 8.08 & 7.38 & 5.88 & 2.03 & 2.46 & 7.26 & 7.51 & 50.7 & 42.9 & \parbox[c][0.9cm][c]{1cm}{ 9.20 \\ (idisk)} & \parbox[c][0.9cm][c]{1cm}{ 9.18 \\ (bar)} & \parbox[c][0.9cm][c]{1cm}{10.25 \\ (cdisk)} & -- \\
NGC3608 & E2 & 7.40 & -- & 48.9 & -- & 6.6 & -- & -- & -- & -- & -- & -- & -- & -- & -- \\
NGC3998 & SA(r)0 & 7.99 & 9.14 & 5.37 & 2.02 & 2.6 & 1.14 & 8.08 & 7.98 & 25.2 & 19.4 & \parbox[c][0.9cm][c]{1cm}{11.09 \\ (bar)} & \parbox[c][0.9cm][c]{1cm}{ 8.90 \\ (idisk)} & \parbox[c][0.9cm][c]{1cm}{11.75 \\ (psf)} & -- \\
NGC4258 & SAB(s)bc & 6.26 & 9.24 & 182 & 6.27 & 8.74 & 3.26 & 5.80 & 6.00 & 75 & 146 & \parbox[c][0.9cm][c]{1cm}{12.02 \\ (psf)} & \parbox[c][0.9cm][c]{1cm}{ 9.82 \\ (idisk)} & \parbox[c][0.9cm][c]{1cm}{ 8.23 \\ (bar)} & \parbox[c][0.9cm][c]{1cm}{ 6.47 \\ (spir)} \\
NGC4261 & E2 & 6.95 & 6.65 & 37.5 & 68.4 & 4.67 & 6.49 & -- & -- & -- & -- & -- & -- & -- & -- \\
NGC4291 & E3 & 8.09 & 7.96 & 17 & 21.3 & 6.79 & 8.55 & -- & -- & -- & -- & -- & -- & -- & -- \\
NGC4342 & S0 & 9.69 & 10.21 & 1.59 & 0.99 & 5.25 & 1.94 & 9.76 & 9.64 & 5.61 & 4.98 & \parbox[c][0.9cm][c]{1cm}{10.95 \\ (env)} & -- & -- & -- \\
NGC4374 & E1 & 5.65 & 5.40 & 84 & 139 & 6.28 & 8.3 & -- & -- & -- & -- & -- & -- & -- & -- \\
NGC4473 & E5 & 6.97 & -- & 27.9 & -- & 5.11 & -- & -- & -- & -- & -- & -- & -- & -- & -- \\
NGC4486 & E0pec & 5.48 & 5.03 & 61.2 & 122 & 2.76 & 5.6 & -- & -- & -- & -- & -- & -- & -- & -- \\
NGC4564 & S0 & 8.22 & -- & 13.5 & -- & 6.1 & -- & 9.18 & -- & 17.1 & -- & -- & -- & -- & -- \\
NGC4649 & E2 & 5.54 & 5.18 & 51.9 & 95.7 & 3.41 & 5.81 & -- & -- & -- & -- & -- & -- & -- & -- \\
NGC4697 & E6 & 5.59 & 8.57 & 154 & 6.27 & 6.32 & 2.11 & 11.24 & 8.24 & 20.6 & 22.9 & \parbox[c][0.9cm][c]{1cm}{12.38 \\ (psf)} & \parbox[c][0.9cm][c]{1cm}{ 6.01 \\ (env)} & -- & -- \\
NGC5252 & S0 & 10.99 & 9.74 & 1.09 & 21.2 & 4 & 5.04 & 10.19 & 11.48 & 9.15 & 9.72 & \parbox[c][0.9cm][c]{1cm}{11.41 \\ (psf)} & -- & -- & -- \\
NGC5845 & E* & 9.08 & -- & 3.48 & -- & 2.77 & -- & -- & -- & -- & -- & -- & -- & -- & -- \\
NGC6251 & E & 8.67 & -- & 20.6 & -- & 4.95 & -- & -- & -- & -- & -- & -- & -- & -- & -- \\
NGC7052 & E & 8.26 & -- & 26.6 & -- & 4.15 & -- & -- & -- & -- & -- & -- & -- & -- & -- \\
NGC7457 & SA(rs)0- & 8.71 & 10.95 & 51.9 & 3 & 7.7 & 1.55 & 8.68 & 8.35 & 25.4 & 28.4 & \parbox[c][0.9cm][c]{1cm}{10.36 \\ (bar)} & \parbox[c][0.9cm][c]{1cm}{13.40 \\ (psf)} & -- & -- \\
PGC49940 & E & 9.52 & -- & 16.7 & -- & 3.74 & -- & -- & -- & -- & -- & -- & -- & -- & -- \\

%% file: table_magabs.tex
CygA & no & no & no & -26.74 & -27.02 & -26.97 & -26.97 & -26.97 & -26.97 & -26.97 & -26.97 & -26.97 & -27.28 \\
IC1459 & yes & no & yes & -25.47 & -25.64 & -25.84 & -25.84 & -26.00 & -25.84 & -26.00 & -26.00 & -26.00 & -25.84 \\
IC4296 & yes & no & yes & -26.05 & -26.39 & -26.40 & -26.40 & -26.85 & -26.40 & -26.85 & -26.85 & -26.85 & -- \\
NGC0221 & no & no & no & -19.42 & -19.53 & -19.48 & -19.48 & -19.48 & -19.48 & -19.48 & -19.48 & -19.48 & -19.77 \\
NGC0821 & no & yes & yes & -23.99 & -24.24 & -24.48 & -24.50 & -24.24 & -24.49 & -22.44 & -24.20 & -24.20 & -24.74 \\
NGC1023 & no & yes & yes & -24.01 & -24.16 & -24.38 & -24.13 & -24.15 & -23.36 & -22.92 & -23.19 & -22.92 & -23.45 \\
NGC1300 & no & yes & yes & -23.84 & -24.09 & -23.57 & -24.10 & -24.14 & -21.53 & -21.84 & -22.67 & -21.84 & -- \\
NGC1399 & yes & no & yes & -25.32 & -25.70 & -25.68 & -25.68 & -26.22 & -25.68 & -26.22 & -26.22 & -26.22 & -- \\
NGC2748 & no & yes & yes & -23.19 & -23.23 & -23.40 & -23.25 & -23.24 & -21.82 & -22.11 & -22.11 & -22.11 & -- \\
NGC2778 & no & yes & yes & -22.24 & -22.32 & -22.79 & -22.33 & -22.32 & -21.56 & -21.28 & -21.33 & -21.28 & -22.94 \\
NGC2787 & no & yes & yes & -22.10 & -22.15 & -22.26 & -22.12 & -22.15 & -21.05 & -21.66 & -21.85 & -21.66 & -21.23 \\
NGC3115 & no & yes & yes & -24.01 & -24.25 & -24.32 & -24.43 & -24.24 & -24.35 & -21.95 & -24.03 & -24.03 & -24.34 \\
NGC3227 & no & yes & yes & -23.50 & -23.64 & -24.78 & -23.61 & -23.63 & -22.11 & -21.59 & -22.02 & -21.59 & -- \\
NGC3245 & no & yes & yes & -23.69 & -23.82 & -24.20 & -23.76 & -23.79 & -22.60 & -22.04 & -22.61 & -22.04 & -23.24 \\
NGC3377 & no & yes & yes & -22.76 & -22.96 & -23.35 & -23.11 & -22.97 & -23.06 & -21.85 & -22.70 & -22.68 & -23.54 \\
NGC3379 & yes & no & yes & -23.80 & -24.20 & -24.28 & -24.28 & -24.54 & -24.28 & -24.54 & -24.54 & -24.54 & -24.13 \\
NGC3384 & no & yes & yes & -23.52 & -23.55 & -23.52 & -23.59 & -23.55 & -22.66 & -22.18 & -22.85 & -22.18 & -22.54 \\
NGC3608 & no & no & no & -23.65 & -24.05 & -24.34 & -24.34 & -24.34 & -24.34 & -24.34 & -24.34 & -24.34 & -24.04 \\
NGC3998 & no & yes & yes & -23.36 & -23.43 & -23.51 & -23.43 & -23.41 & -22.72 & -21.57 & -22.57 & -21.57 & -- \\
NGC4258 & no & yes & yes & -23.84 & -23.93 & -24.45 & -24.04 & -23.97 & -23.03 & -20.05 & -21.61 & -20.05 & -22.40 \\
NGC4261 & yes & no & yes & -25.19 & -25.64 & -25.49 & -25.49 & -25.79 & -25.49 & -25.79 & -25.79 & -25.79 & -25.54 \\
NGC4291 & yes & no & yes & -23.62 & -23.88 & -23.94 & -23.94 & -24.07 & -23.94 & -24.07 & -24.07 & -24.07 & -23.84 \\
NGC4342 & no & yes & yes & -21.61 & -21.73 & -21.54 & -21.65 & -21.67 & -20.93 & -20.41 & -20.85 & -20.85 & -21.04 \\
NGC4374 & yes & no & yes & -25.13 & -25.53 & -25.69 & -25.69 & -25.94 & -25.69 & -25.94 & -25.94 & -25.94 & -25.72 \\
NGC4473 & no & no & no & -23.77 & -23.88 & -23.95 & -23.95 & -23.95 & -23.95 & -23.95 & -23.95 & -23.95 & -23.74 \\
NGC4486 & yes & no & yes & -25.31 & -25.86 & -25.63 & -25.63 & -26.08 & -25.63 & -26.08 & -26.08 & -26.08 & -25.68 \\
NGC4564 & no & yes & no & -23.08 & -23.14 & -23.23 & -23.16 & -23.16 & -22.79 & -22.79 & -22.79 & -22.79 & -23.53 \\
NGC4649 & yes & no & yes & -25.35 & -25.71 & -25.54 & -25.54 & -25.90 & -25.54 & -25.90 & -25.90 & -25.90 & -25.75 \\
NGC4697 & no & yes & yes & -24.13 & -24.59 & -24.90 & -24.91 & -24.70 & -24.90 & -21.92 & -24.58 & -24.58 & -24.75 \\
NGC5252 & no & yes & yes & -25.18 & -25.48 & -25.80 & -25.18 & -25.58 & -23.95 & -25.20 & -25.42 & -25.20 & -25.61 \\
NGC5845 & no & no & no & -22.92 & -23.02 & -22.93 & -22.93 & -22.93 & -22.93 & -22.93 & -22.93 & -22.93 & -22.94 \\
NGC6251 & no & no & no & -26.15 & -26.46 & -26.48 & -26.48 & -26.48 & -26.48 & -26.48 & -26.48 & -26.48 & -26.60 \\
NGC7052 & no & no & no & -25.62 & -25.92 & -25.89 & -25.89 & -25.89 & -25.89 & -25.89 & -25.89 & -25.89 & -25.98 \\
NGC7457 & no & yes & yes & -22.38 & -22.50 & -23.30 & -22.60 & -22.45 & -21.84 & -19.60 & -20.73 & -19.60 & -21.75 \\
PGC49940 & no & no & no & -25.96 & -26.43 & -26.41 & -26.41 & -26.41 & -26.41 & -26.41 & -26.41 & -26.41 & -- \\

%% file: table_magcomp.tex
(1) & $\Mbmin$ & $\Mbstd$ & -24.23 & 0.36 & 0.14 & 1.01 \\
(2) & $\Msph$ & $\Mbstd$ & -24.23 & 0.13 & 0.13 & 0.73 \\
(3) & $\Mbstd$ & $\Mmh$ & -24.10 & 0.17 & 0.02 & 0.48 \\
(4) & $\Msph$ & $\Mv$ & -23.88 & -0.54 & 0.00 & 0.84 \\
(5) & $\Mtimp$ & $\Miso$ & -24.23 & -0.10 & 0.05 & 0.18 \\
\hline
(6) & $\Mtimp$ & $\Mtwom$ & -24.23 & -0.32 & 0.11 & 0.21 \\

%% file: Mbh-Lbt_I.bbl
\begin{thebibliography}{}
\expandafter\ifx\csname natexlab\endcsname\relax\def\natexlab#1{#1}\fi

\bibitem[{{Bell} \& {de Jong}(2001)}]{BelldeJong01}
{Bell}, E.~F., \& {de Jong}, R.~S. 2001, \apj, 550, 212

\bibitem[{{Bertin}(2006)}]{Scamp}
{Bertin}, E. 2006, in Astronomical Society of the Pacific Conference Series,
  Vol. 351, Astronomical Data Analysis Software and Systems XV, ed.
  C.~{Gabriel}, C.~{Arviset}, D.~{Ponz}, \& S.~{Enrique}, 112

\bibitem[{{Bertin} \& {Arnouts}(1996)}]{SEx}
{Bertin}, E., \& {Arnouts}, S. 1996, \aaps, 117, 393

\bibitem[{{Bertin} {et~al.}(2002){Bertin}, {Mellier}, {Radovich}, {Missonnier},
  {Didelon}, \& {Morin}}]{SWarp}
{Bertin}, E., {Mellier}, Y., {Radovich}, M., {et~al.} 2002, in Astronomical
  Society of the Pacific Conference Series, Vol. 281, Astronomical Data
  Analysis Software and Systems XI, ed. D.~A. {Bohlender}, D.~{Durand}, \&
  T.~H. {Handley}, 228

\bibitem[{{Blakeslee} {et~al.}(2009){Blakeslee}, {Jord{\'a}n}, {Mei},
  {C{\^o}t{\'e}}, {Ferrarese}, {Infante}, {Peng}, {Tonry}, \&
  {West}}]{Blakeslee_etal09}
{Blakeslee}, J.~P., {Jord{\'a}n}, A., {Mei}, S., {et~al.} 2009, \apj, 694, 556

\bibitem[{{Cole} {et~al.}(2001){Cole}, {Norberg}, {Baugh}, {Frenk},
  {Bland-Hawthorn}, {Bridges}, {Cannon}, {Colless}, {Collins}, {Couch},
  {Cross}, {Dalton}, {De Propris}, {Driver}, {Efstathiou}, {Ellis},
  {Glazebrook}, {Jackson}, {Lahav}, {Lewis}, {Lumsden}, {Maddox}, {Madgwick},
  {Peacock}, {Peterson}, {Sutherland}, \& {Taylor}}]{Cole_etal01}
{Cole}, S., {Norberg}, P., {Baugh}, C.~M., {et~al.} 2001, \mnras, 326, 255

\bibitem[{{Croton} {et~al.}(2006){Croton}, {Springel}, {White}, {De Lucia},
  {Frenk}, {Gao}, {Jenkins}, {Kauffmann}, {Navarro}, \&
  {Yoshida}}]{Croton_etal06}
{Croton}, D.~J., {Springel}, V., {White}, S.~D.~M., {et~al.} 2006, \mnras, 365,
  11

\bibitem[{{Ferrarese} {et~al.}(2006){Ferrarese}, {C{\^o}t{\'e}}, {Jord{\'a}n},
  {Peng}, {Blakeslee}, {Piatek}, {Mei}, {Merritt}, {Milosavljevi{\'c}},
  {Tonry}, \& {West}}]{Ferrarese_etal06b}
{Ferrarese}, L., {C{\^o}t{\'e}}, P., {Jord{\'a}n}, A., {et~al.} 2006, \apjs,
  164, 334

\bibitem[{{Granato} {et~al.}(2004){Granato}, {De Zotti}, {Silva}, {Bressan}, \&
  {Danese}}]{Granato_etal04}
{Granato}, G.~L., {De Zotti}, G., {Silva}, L., {Bressan}, A., \& {Danese}, L.
  2004, \apj, 600, 580

\bibitem[{{Greene} {et~al.}(2008){Greene}, {Ho}, \& {Barth}}]{GreeneHoBarth08}
{Greene}, J.~E., {Ho}, L.~C., \& {Barth}, A.~J. 2008, \apj, 688, 159

\bibitem[{{Greene} {et~al.}(2010){Greene}, {Peng}, {Kim}, {Kuo}, {Braatz},
  {Impellizzeri}, {Condon}, {Lo}, {Henkel}, \& {Reid}}]{Greene_etal10}
{Greene}, J.~E., {Peng}, C.~Y., {Kim}, M., {et~al.} 2010, \apj, 721, 26

\bibitem[{{H{\"a}ring} \& {Rix}(2004)}]{HR04}
{H{\"a}ring}, N., \& {Rix}, H.-W. 2004, \apjl, 604, L89

\bibitem[{{Herrnstein} {et~al.}(1999){Herrnstein}, {Moran}, {Greenhill},
  {Diamond}, {Inoue}, {Nakai}, {Miyoshi}, {Henkel}, \&
  {Riess}}]{Herrnstein_etal99}
{Herrnstein}, J.~R., {Moran}, J.~M., {Greenhill}, L.~J., {et~al.} 1999, \nat,
  400, 539

\bibitem[{{Hopkins} {et~al.}(2006){Hopkins}, {Hernquist}, {Cox}, {Di Matteo},
  {Robertson}, \& {Springel}}]{Hopkins_etal06}
{Hopkins}, P.~F., {Hernquist}, L., {Cox}, T.~J., {et~al.} 2006, \apjs, 163, 1

\bibitem[{{Hu}(2008)}]{Hu08}
{Hu}, J. 2008, \mnras, 386, 2242

\bibitem[{{Jahnke} \& {Macci{\`o}}(2011)}]{JahnkeMaccio11}
{Jahnke}, K., \& {Macci{\`o}}, A.~V. 2011, \apj, 734, 92

\bibitem[{{Jensen} {et~al.}(2003){Jensen}, {Tonry}, {Barris}, {Thompson},
  {Liu}, {Rieke}, {Ajhar}, \& {Blakeslee}}]{Jensen_etal03}
{Jensen}, J.~B., {Tonry}, J.~L., {Barris}, B.~J., {et~al.} 2003, \apj, 583, 712

\bibitem[{{Jun} \& {Im}(2008)}]{JunIm08}
{Jun}, H.~D., \& {Im}, M. 2008, \apjl, 678, L97

\bibitem[{{Kormendy} {et~al.}(2011){Kormendy}, {Bender}, \& {Cornell}}]{K11}
{Kormendy}, J., {Bender}, R., \& {Cornell}, M.~E. 2011, \nat, 469, 374

\bibitem[{{Marconi} \& {Hunt}(2003)}]{MH03}
{Marconi}, A., \& {Hunt}, L.~K. 2003, \apjl, 589, L21

\bibitem[{{Marconi} {et~al.}(2004){Marconi}, {Risaliti}, {Gilli}, {Hunt},
  {Maiolino}, \& {Salvati}}]{Marconi_etal04}
{Marconi}, A., {Risaliti}, G., {Gilli}, R., {et~al.} 2004, \mnras, 351, 169

\bibitem[{{McLure} \& {Dunlop}(2004)}]{McD04}
{McLure}, R.~J., \& {Dunlop}, J.~S. 2004, \mnras, 352, 1390

\bibitem[{{Mei} {et~al.}(2007){Mei}, {Blakeslee}, {C{\^o}t{\'e}}, {Tonry},
  {West}, {Ferrarese}, {Jord{\'a}n}, {Peng}, {Anthony}, \&
  {Merritt}}]{Mei_etal07}
{Mei}, S., {Blakeslee}, J.~P., {C{\^o}t{\'e}}, P., {et~al.} 2007, \apj, 655,
  144

\bibitem[{{Minchev} {et~al.}(2012){Minchev}, {Famaey}, {Quillen}, {Di Matteo},
  {Combes}, {Vlajic}, {Erwin}, \& {Bland-Hawthorn}}]{Minchev12}
{Minchev}, I., {Famaey}, B., {Quillen}, A.~C., {et~al.} 2012, ArXiv e-prints,
  arXiv:1203.2621

\bibitem[{{Nowak} {et~al.}(2010){Nowak}, {Thomas}, {Erwin}, {Saglia}, {Bender},
  \& {Davies}}]{Nowak_etal10}
{Nowak}, N., {Thomas}, J., {Erwin}, P., {et~al.} 2010, \mnras, 403, 646

\bibitem[{{Peng} {et~al.}(2010){Peng}, {Ho}, {Impey}, \& {Rix}}]{GF3}
{Peng}, C.~Y., {Ho}, L.~C., {Impey}, C.~D., \& {Rix}, H.-W. 2010, \aj, 139,
  2097

\bibitem[{{Puget} {et~al.}(2004){Puget}, {Stadler}, {Doyon}, {Gigan},
  {Thibault}, {Luppino}, {Barrick}, {Benedict}, {Forveille}, {Rambold},
  {Thomas}, {Vermeulen}, {Ward}, {Beuzit}, {Feautrier}, {Magnard}, {Mella},
  {Preis}, {Vallee}, {Wang}, {Lin}, {Hall}, \& {Hodapp}}]{WIRCam}
{Puget}, P., {Stadler}, E., {Doyon}, R., {et~al.} 2004, in Society of
  Photo-Optical Instrumentation Engineers (SPIE) Conference Series, Vol. 5492,
  Society of Photo-Optical Instrumentation Engineers (SPIE) Conference Series,
  ed. A.~F.~M. {Moorwood} \& M.~{Iye}, 978--987

\bibitem[{{Sani} {et~al.}(2011){Sani}, {Marconi}, {Hunt}, \&
  {Risaliti}}]{Sani_etal11}
{Sani}, E., {Marconi}, A., {Hunt}, L.~K., \& {Risaliti}, G. 2011, \mnras, 413,
  1479

\bibitem[{{S{\'e}rsic}(1963)}]{Sersic63}
{S{\'e}rsic}, J.~L. 1963, Boletin de la Asociacion Argentina de Astronomia La
  Plata Argentina, 6, 41

\bibitem[{{Shankar} {et~al.}(2004){Shankar}, {Salucci}, {Granato}, {De Zotti},
  \& {Danese}}]{Shankar_etal04}
{Shankar}, F., {Salucci}, P., {Granato}, G.~L., {De Zotti}, G., \& {Danese}, L.
  2004, \mnras, 354, 1020

\bibitem[{{Shen} {et~al.}(2003){Shen}, {Mo}, {White}, {Blanton}, {Kauffmann},
  {Voges}, {Brinkmann}, \& {Csabai}}]{Shen_etal03}
{Shen}, S., {Mo}, H.~J., {White}, S.~D.~M., {et~al.} 2003, \mnras, 343, 978

\bibitem[{{Silk} \& {Rees}(1998)}]{SilkRees98}
{Silk}, J., \& {Rees}, M.~J. 1998, \aap, 331, L1

\bibitem[{{Tonry} {et~al.}(2001){Tonry}, {Dressler}, {Blakeslee}, {Ajhar},
  {Fletcher}, {Luppino}, {Metzger}, \& {Moore}}]{Tonry_etal01}
{Tonry}, J.~L., {Dressler}, A., {Blakeslee}, J.~P., {et~al.} 2001, \apj, 546,
  681

\bibitem[{{Tundo} {et~al.}(2007){Tundo}, {Bernardi}, {Hyde}, {Sheth}, \&
  {Pizzella}}]{Tundo_etal07}
{Tundo}, E., {Bernardi}, M., {Hyde}, J.~B., {Sheth}, R.~K., \& {Pizzella}, A.
  2007, \apj, 663, 53

\bibitem[{{Vika} {et~al.}(2012){Vika}, {Driver}, {Cameron}, {Kelvin}, \&
  {Robotham}}]{V12}
{Vika}, M., {Driver}, S.~P., {Cameron}, E., {Kelvin}, L., \& {Robotham}, A.
  2012, \mnras, 419, 2264

\bibitem[{{Volonteri} {et~al.}(2011){Volonteri}, {Natarajan}, \&
  {G{\"u}ltekin}}]{VolNatGul11b}
{Volonteri}, M., {Natarajan}, P., \& {G{\"u}ltekin}, K. 2011, \apj, 737, 50

\end{thebibliography}
